\documentclass[lettersize,journal]{IEEEtran}
\usepackage{amsmath,amsfonts}
\usepackage{algorithm}
\usepackage{algpseudocode}
\usepackage{array}
\usepackage[caption=false,font=normalsize,labelfont=sf,textfont=sf]{subfig}
\usepackage{textcomp}
\usepackage{stfloats}
\usepackage{url}
\usepackage{verbatim}
\usepackage{graphicx}
\usepackage{cite}
\usepackage[dvipsnames]{xcolor}
\usepackage{tikz}
\usetikzlibrary{backgrounds}
\usepackage{multicol}
\usepackage{multirow}

\usepackage[bookmarks=true]{hyperref}
\hypersetup{
    colorlinks=true,
    linkcolor=blue,
    filecolor=magenta,      
    urlcolor=blue,
    pdfpagemode=FullScreen,
    }

\usepackage{bm}
\usepackage{bbm}
\usepackage{dsfont}
\usepackage{stmaryrd}

\usepackage{caption}

\usepackage{diagbox}

\newtheorem{definition}{Definition}
\newtheorem{remark}{Remark}
\newtheorem{assumption}{Assumption}
\newtheorem{problem}{Problem}

\newcommand{\T}{^\mathrm{T}}

\newcommand{\ba}{{\bm a}}
\newcommand{\bb}{{\bm b}}
\newcommand{\bc}{{\bm c}}
\newcommand{\bd}{{\bm d}}

\newcommand{\bff}{{\bm f}}
\newcommand{\bg}{{\bm g}}
\newcommand{\bh}{{\bm h}}

\newcommand{\bk}{{\bm k}}

\newcommand{\bp}{{\bm p}}

\newcommand{\bs}{{\bm s}}

\newcommand{\bu}{{\bm u}}
\newcommand{\bv}{{\bm v}}
\newcommand{\bw}{{\bm w}}
\newcommand{\bx}{{\bm x}}
\newcommand{\by}{{\bm y}}
\newcommand{\bz}{{\bm z}}

\newcommand{\bA}{{\bm A}}
\newcommand{\bB}{{\bm B}}
\newcommand{\bC}{{\bm C}}

\newcommand{\bI}{{\bm I}}

\newcommand{\bK}{{\bm K}}

\newcommand{\bM}{{\bm M}}

\newcommand{\bP}{{\bm P}}
\newcommand{\bQ}{{\bm Q}}
\newcommand{\bR}{{\bm R}}

\newcommand{\bT}{{\bm T}}

\newcommand{\bW}{{\bm W}}

\usepackage{calrsfs}
\DeclareMathAlphabet{\pazocal}{OMS}{zplm}{m}{n}

\newcommand{\calC}{\pazocal{C}}

\newcommand{\calI}{\pazocal{I}}

\newcommand{\calL}{\pazocal{L}}

\newcommand{\calN}{\pazocal{N}}

\newcommand{\calO}{\pazocal{O}}

\newcommand{\calP}{\pazocal{P}}

\newcommand{\calV}{\pazocal{V}}

\newcommand{\calX}{\pazocal{X}}

\newcommand{\Rb}{\mathbb{R}}

\newcommand{\Zb}{\mathbb{Z}}

\newcommand{\beeta}{{\bm \eta}}
\newcommand{\blambda}{{\bm \lambda}}
\newcommand{\bnu}{{\bm \nu}}
\newcommand{\bxi}{{\bm \xi}}

\DeclareMathOperator*{\argmin}{arg\,min}

\newcommand{\diag}{{\mathrm{diag}}}
\newcommand{\bdiag}{{\mathrm{bdiag}}}

\begin{document}

\setlength{\abovedisplayskip}{4pt}
\setlength{\belowdisplayskip}{4pt}
\setlength{\textfloatsep}{12pt}

\title{Distributed Differential Dynamic Programming Architectures for Large-Scale Multi-Agent Control
}

\author{Augustinos D. Saravanos, Yuichiro Aoyama, Hongchang Zhu and Evangelos A. Theodorou
\thanks{This work was supported by the ARO Award $\#$W911NF2010151 and the Georgia COVID relief fund. Augustinos D. Saravanos acknowledges financial support by the A. Onassis Foundation Scholarship. \textit{(Corresponding author: Augustinos D. Saravanos)}}
\thanks{The authors are with the Daniel Guggenheim School of Aerospace Engineering, Georgia Institute of Technology, North Avenue, Atlanta 30332, GA, USA 
(e-mail: asaravanos@gatech.edu; yaoyama3@gatech.edu; hozhu@gatech.edu; evangelos.theodorou@gatech.edu).
}
\vspace{-0.4cm}
}



\maketitle

\begin{abstract}
In this paper, we propose two novel decentralized optimization frameworks for multi-agent nonlinear optimal control problems in robotics. The aim of this work is to suggest architectures that inherit the computational efficiency and scalability of Differential Dynamic Programming (DDP) and the distributed nature of the Alternating Direction Method of Multipliers (ADMM). In this direction, two frameworks are introduced. The first one called Nested Distributed DDP (ND-DDP), is a three-level architecture which employs ADMM for enforcing a consensus between all agents, an Augmented Lagrangian layer for satisfying local constraints and DDP as each agent’s optimizer. In the second approach, both consensus and local constraints are handled with ADMM, yielding a two-level architecture called Merged Distributed DDP (MD-DDP), which further reduces computational complexity. Both frameworks are \textit{fully decentralized} since all computations are parallelizable among the agents and only local communication is necessary. Simulation results that scale up to thousands of vehicles and hundreds of drones verify the effectiveness of the methods. Superior scalability to large-scale systems against centralized DDP and centralized/decentralized sequential quadratic programming is also illustrated. Finally, hardware experiments on a multi-robot platform demonstrate the applicability of the proposed algorithms, while highlighting the importance of optimizing for feedback policies to increase robustness against uncertainty.  
A video with all results is available here \url{https://youtu.be/tluvENcWldw}.
\end{abstract}


\section{Introduction}

Multi-agent problems are emerging increasingly often in the robotics and control fields, with numerous significant applications such as guiding fleets of autonomous cars \cite{chalaki_malikopoulos2021time, xiao_cassandras2021decentralized}, navigating formations of unmanned aerial vehicles (UAVs) \cite{kabore2021distributed}, multi-robot motion coordination \cite{yu_dimarogonas2022distributed}, air traffic management \cite{Tomlin_AirTrafficManagement2011}, swarm robotics \cite{turgut2008self} and so forth. There is a tendency that the size of such systems increases from a handful to large-scale teams of hundreds or even thousands of robots that are required to cooperatively achieve a common objective \cite{zhu_ferrari2021adaptive, abdulghafoor_bakolas2022multi, zhou2022anovel}.
It is also desirable that these problems are solved optimally regarding the performance and control effort of the agents while ensuring their safe operation within their environment. The latter specifications can be captured intrinsically with a multi-agent optimal control setting where a set of agents must collectively minimize a global cost function while satisfying dynamics, actuation and safety constraints.  


A straightforward solution to address multi-agent optimal control problems is through centralized methods, where a powerful controller is assumed to perform all computations and communicate with all agents \cite{riegger2016centralized}. Unfortunately, this is often prohibitive for large-scale systems due to poor scalability with the increase of agents and potential communication limitations. As a consequence, there is a great necessity for designing fully decentralized optimal control approaches that enjoy computational and communication efficiency along with scalability to large-scale systems.

%
%
%


\begin{figure}[t]
\centering
\includegraphics[width=0.485\textwidth, trim={13.6cm 7.5cm 8.75cm 12.4cm},clip]{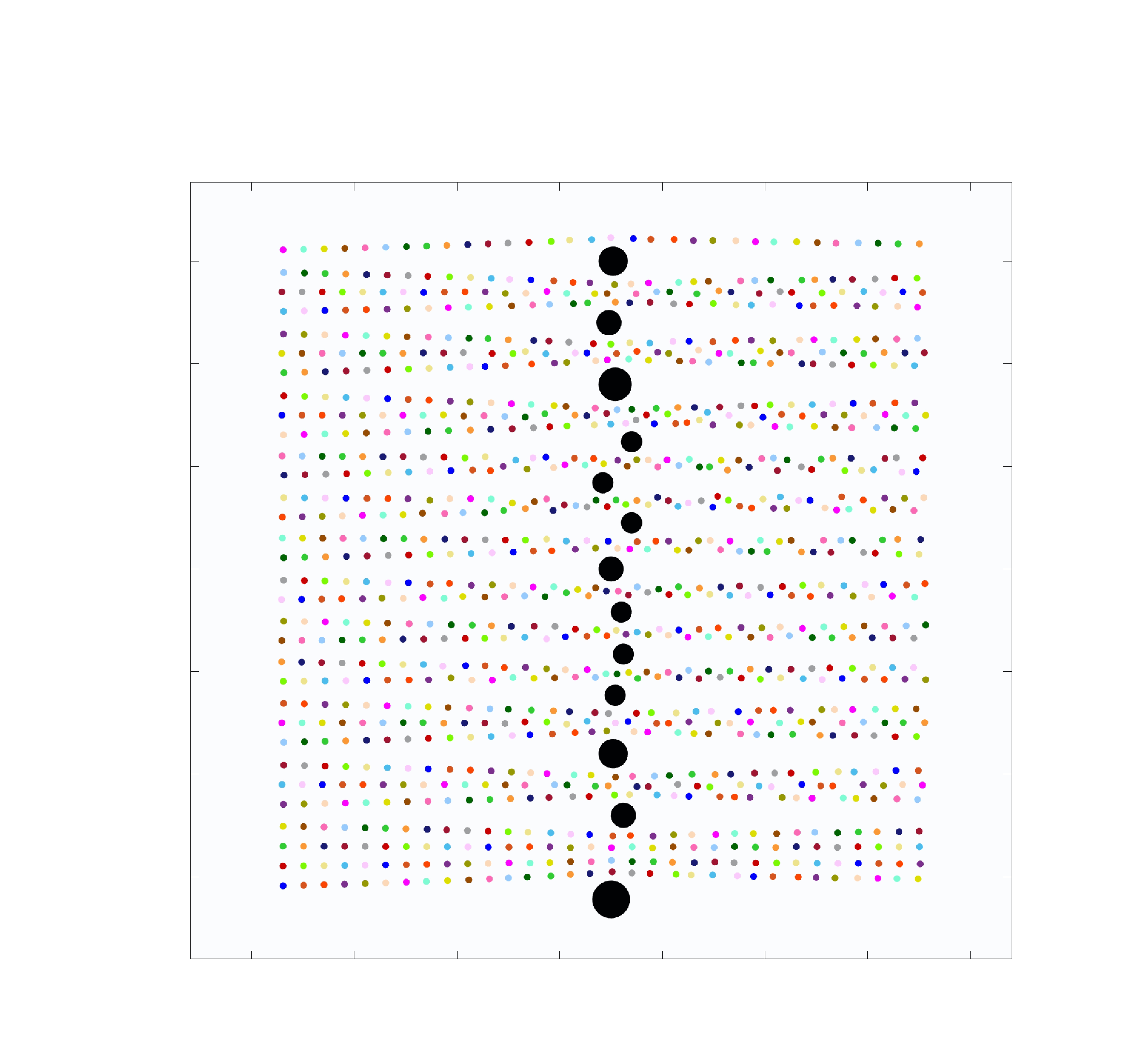}\caption{A  $1{,}024$ cars fleet navigation problem in a multi-obstacle environment solved with MD-DDP.}
\end{figure}

There exist two main classes of methods for solving nonlinear optimal control problems. The first one, referred to as direct trajectory optimization, relies on representing both the states and controls as optimization variables and solving the problem
with an off-the-shelf nonlinear programming (NLP) algorithm such as sequential quadratic programming (SQP) \cite{diehl2009nonlinearmpc, mombaur2009using}. The second class entails the indirect methods, where only the controls are parametrized while the states are obtained through forward integration of the dynamics, alleviating the need for explicitly imposing them as constraints \cite{von1992direct}.

Differential Dynamic Programming (DDP) is a well-known second-order indirect method that was first proposed by Jacobson and Mayne in 1970 \cite{jacobson1970differential} and has found numerous successful robotics applications \cite{morimoto2003minimax, kumar2016optimal, budhiraja_carpentier2018differential, houghton2022path}. Its solution yields an optimal state trajectory along with locally optimal feed-forward and feedback control policies. It also includes the iterative linear quadratic regulator (iLQR) algorithm \cite{todorov2005iLQR} as a special case if only the first-order approximations of the dynamics are considered. Some appealing features of DDP include its global convergence guarantees with a quadratic rate and its low computational and memory complexities \cite{liao1991convergence, yakowitz1984computational}. These characteristics have established DDP as one of the most scalable nonlinear optimal control methods \cite{yakowitz1984computational, liao1992advantages}, available in the literature. Several extensions for handling control and/or state constraints with DDP can be found in \cite{murray1979constrained, tassa2014controlddp, xie2017differential, pavlov2021interior, jallet2022constrained, howell2019altro,  aoyama2020constrainedDDPicra, almubarak2022safety}. 
Game theoretic DDP-based methods for multi-agent settings have been presented in \cite{kavuncu2021potential, so2022multimodal}.

Although DDP is considered to be one of the most scalable optimal control approaches, yet its scalability would be limited if directly applied on high-dimensional control problems in a centralized fashion. Therefore, it is highly desirable to develop new distributed algorithms that retain the advantages of DDP and are inherently scalable for large-scale multi-robot control.
Along this direction, but in a non-robotics setting, a preconditioned distributed DDP-based method was presented in \cite{wang2020preconditioned}, where inter-agent interactions only appear through the dynamics but not through inequality constraints that are typical in multi-robot systems.

An attractive approach adopted in this paper in order to overcome the scalability issue is to combine DDP with the Alternating Direction Method of Multipliers (ADMM).
ADMM is an optimization method that has recently received significant attention because it can lead to algorithms with a distributed structure \cite{boyd2011distributed}. Several ADMM-based decentralized control methods have been proposed recently \cite{saravanos2021distributed, pereira2022decentralized, rey_lygeros2018fully, shorinwa_macschwager2020scalable, halsted2021survey, tang2021fast}, demonstrating a significantly improved computational efficiency against centralized methods. Nevertheless, there has been very little work in proposing schemes that attempt to merge the well-appreciated attributes of both DDP and ADMM. In \cite{zhang2021semi, zhang2021parallel}, the authors provide multi-agent ADMM-based DDP algorithms that are still semi-centralized, while scalability results are limited to a few vehicles. In \cite{park2019distributed}, a semi-centralized multi-target tracking method using iLQR and ADMM was proposed, yet being distributed only among the targets and not the agents. While these methods have shown promising results, they still require a central node for computational and communication purposes and have only demonstrated low scalability in practice. 

The scope of the present work is to propose distributed architectures that thoroughly exploit the capabilities of combining DDP and ADMM, leading to fully decentralized algorithms that are applicable to large-scale multi-robot systems. Two new schemes are proposed, termed thereafter as: Nested Distributed DDP (ND-DDP) and Merged Distributed DDP (MD-DDP). Both methods are extensively tested in simulation on various multi-vehicle and multi-UAV problems of an increasing scale. Hardware experiments on a multi-robot platform also illustrate their applicability on real systems. 
The specific contributions of this work can be listed as follows:
\begin{enumerate}
\item A three-level decentralized architecture, named ND-DDP is proposed, that utilizes ADMM for enforcing consensus between all agents, an AL layer for satisfying neighborhood constraints locally and DDP as each agent's optimizer. This framework has a general form where any constrained DDP method can be used instead of the AL-DDP combination. 
\item A two-level decentralized architecture, named MD-DDP is presented, which exploits the fact that both consensus and local constraints can be treated in the same level through ADMM, which further reduces computational complexity.
\item Two improvements are proposed to further enhance the performance of each framework. First, a Nesterov acceleration technique is utilized, leading to faster versions of the methods. Subsequently, a decentralized adaptation scheme for the penalty parameters induced by ADMM, that is particularly tailored for multi-agent optimal control problems, is also introduced. 
\item Extensive simulation results on teams of cars and drones showcase the effectiveness of both frameworks. ND-DDP is successfully applied to systems with up to 256 cars (1,024 states) and 64 drones (768 states). With MD-DDP an even higher scalability is achieved, solving problems with up to 4,096 cars (16,384 states) and 256 drones (3,072 states). To our best knowledge, ND-DDP and MD-DDP are the first fully decentralized DDP/iLQR-based methods that have been shown to scale for multi-robot systems of such a large scale.
\item An ablative analysis that illustrates the advantages of the improved versions of the architectures, as well as a computational time comparison that verifies the superior scalability of ND-DDP and MD-DDP against centralized DDP and centralized/decantrized SQP are also provided.
\item Hardware results on the Robotarium platform \cite{wilson2020robotarium} verify the applicability of the methods on actual systems. These experiments also underline the advantage of DDP as a dynamic optimizer to provide control policies that consist of feed-forward and feedback terms, thus enhancing the robustness of the methods against model uncertainty.
\end{enumerate}

The rest of the paper is organized as follows. In Section \ref{sec:Preliminaries}, a brief overview of DDP and ADMM is provided. The multi-agent optimal control problem is formulated in Section \ref{sec:Multi-AgentFormulation}. In Section \ref{sec:Nested_DDP}, the ND-DDP architecture is proposed, while the MD-DDP framework is introduced in Section \ref{sec:Merged_DistributedDDP}. In Section \ref{sec:improvements}, the improved versions of the architectures are presented, along with additional details on the methods. In Section \ref{sec:Simulation_Results}, the performance of the algorithms on multi-car and multi-UAV systems is demonstrated through simulations. Hardware experiments on a multi-robot platform are presented in Section \ref{sec:Hardware_Experiments}. Finally, a conclusion and an overview of future research directions are provided in Section \ref{sec:conclusion}.
\IEEEpubidadjcol
\section{Preliminaries}
\label{sec:Preliminaries}
\subsection{Notation}
Throughout this paper, non-bold symbols are used for scalars $a \in \Rb$ and bold lowercase and uppercase symbols for vectors $\ba \in \Rb^n$ and matrices $\bA \in \Rb^{n \times m}$, respectively. With $\llbracket a,b \rrbracket$, the integer interval $[a,b] \cap \Zb$ is defined. Given vectors $\ba_i, \ i \in \llbracket 1, n \rrbracket$, we refer to their vertical concatenation as $\ba = 
\big[ \ba_1 ; \dots ; \ba_n \big] = 
\big[ \{ \ba_i \}_{i \in \llbracket 1, n \rrbracket} \big]$. A diagonal matrix made up of scalars $a_i$, $i \in \llbracket 1,n \rrbracket$, is denoted with $\mathrm{diag}(a_1,\dots, a_n) \in \Rb^{n \times n}$, while a block diagonal matrix constructed by $\bA_i$, $i \in \llbracket 1,n \rrbracket$, is denoted with $\bdiag(\bA_1, \dots, \bA_n)$. The $\ell_2$-norm of $\ba = \big[ a_1 \ \dots \ a_n \big] \in \Rb^n$ is $\| \ba \|_2 = \sqrt{\ba\T \ba} = \sqrt{\sum_{i=1}^n a_i^2}$. We also define the weighted norm $\| \ba \|_{\bW} = \| \bW^{1/2} \ba \|_2 = \sqrt{\ba\T \bW \ba}$ for any symmetric positive definite matrix $\bW$. Furthermore, given a set $\calX$, its cardinality is provided by $|\calX|$. Finally, the projection of a vector $\bx$ onto a set $\calC$ is denoted with $\Pi_{\calC}(\bx)$, while $\calI_{\calC}(\bx)$ defines an indicator function such that $\calI_{\calC}(\bx) = 0$ if $\bx \in \calC$ and $\calI_{\calC}(\bx) = + \infty$, otherwise.

\subsection{Differential Dynamic Programming}

A brief introduction to DDP is initially provided in its well-known centralized/single-agent unconstrained form \cite{jacobson1970differential}. Let us consider the following discrete-time nonlinear dynamics
\begin{equation}
\bx_{k+1} = \bff(\bx_k, \bu_k), \ \ \bx_0: \text{given},
\label{Nonlinear Dynamics}
\end{equation}
where $\bx_k \in \Rb^p$, $\bu_k \in \Rb^{q}$ denote the state and control at time instant $k$, respectively, and $\bff: \Rb^p \times \Rb^q \rightarrow \Rb^p
$ is the dynamics function. Given a finite time horizon $K$, the state and control trajectories are denoted with $\bx = \big[ \bx_0; \dots; \bx_K \big]$ and $\bu = \big[ \bu_0; \dots; \bu_{K-1} \big]$. The finite-horizon discrete-time optimal control problem can be formulated as finding the optimal control sequence $\bu^*$ such that
\begin{subequations}
\begin{align}
& \bu^* =  \argmin_{\bu} J(\bx, \bu) = \sum_{k=0}^{K-1} \ell(\bx_k, \bu_k) + \phi(\bx_K) \\
&~~~~~~~~~ \mathrm{s.t.} \quad \bx_{k+1} = \bff(\bx_k, \bu_k),
\ \ \bx_0: \text{given},
\end{align}
\label{DDP Optimal Control Problem Formulation}%
\end{subequations}
with $J
: \Rb^{(K+1)p} \times \Rb^{K q} \rightarrow \Rb
$, $\ell
: \Rb^p \times \Rb^q \rightarrow \Rb
$ and $\phi
: \Rb^p \rightarrow \Rb
$ being the total, running and terminal costs, respectively.
Problem \eqref{DDP Optimal Control Problem Formulation} can be addressed with dynamic programming \cite{bellman1966dynamic} since, given $\bx_k$ at time $k$, the optimality of the future controls is independent of the past states and controls. The value function that describes the optimal ``cost-to-go'' starting from $\bx_k$ is provided by:
\begin{equation}
V_k(\bx_k):=\min_{\bu_k, \dots, \bu_{K-1}} \sum_{\kappa=k}^{K-1} \ell(\bx_{\kappa}, \bu_{\kappa}) + \phi(\bx_K).
\end{equation}
Hence, solving \eqref{DDP Optimal Control Problem Formulation} is equivalent with finding a control sequence $\bu$ such that $J(\bx, \bu) = V_0(\bx_0)$. Bellman's Principle of Optimality \cite{bellman1966dynamic} states that $V_k(\bx_k)$ can be found through the following backpropagation rule
\begin{subequations}
\begin{align}
V_k(\bx_k) &=  \min_{\bu_k} Q_k(\bx_k, \bu_k),
\label{Bellman rule}
\\
Q_k(\bx_k, \bu_k) &= \ell(\bx_k, \bu_k) + V_{k+1}(\bx_{k+1}),
\label{Bellman Q}
\end{align}
\label{Bellman}
\end{subequations}
with terminal condition $V_K(\bx_K) = \phi(\bx_K)$. 

DDP is a second-order method that solves problem \eqref{DDP Optimal Control Problem Formulation} locally through \eqref{Bellman rule}, by taking an approximation of $Q_k$. The algorithm operates in an iterative backward and forward pass fashion. During the \textit{backward} pass, both sides of \eqref{Bellman rule} are expanded around some nominal trajectories $\bar{\bx}$ and $\bar{\bu}$. In particular, by defining the deviations $\delta \bx_k = \bx_k - \bar{\bx}_k$, $\delta \bu_k = \bu_k - \bar{\bu}_k$, the quadratic expansion of $Q_k$ is given by
\begin{align}
& Q_k(\bx_k, \bu_k) \approx  
Q_{\bx,k}\T \delta \bx_k + Q_{\bu,k}\T \delta \bu_k + \frac{1}{2} \delta \bx_k\T Q_{\bx \bx,k} \delta \bx_{k}
\nonumber
\\ 
& ~~~~~~~~~~~~~~~~~ + \delta \bx_k Q_{\bx \bu,k} \delta \bu_{k} 
+ \frac{1}{2} \delta \bu_k \T {Q}_{\bu \bu,k}{\delta}{\bu}_k,
\end{align}
with
%
%
\begin{subequations}
\begin{align}
Q_{\bx \bx, k} &= \ell_{\bx \bx} + \bff_{\bx} \T {V}_{\bx \bx, k+1} \bff_{\bx} + V_{\bx,k+1} \cdot \bff_{\bx \bx},
\label{DDP: Qxx}
\\
Q_{\bx \bu,k} &= \ell_{\bx \bu} + \bff_{\bx}\T V_{\bx \bx,k+1} \bff_{\bu} +
V_{\bx,k+1} \cdot \bff_{\bx \bu},
\label{DDP: Qxu}
\\
Q_{\bu \bu,k} &= \ell_{\bu \bu} + \bff_{\bu}\T V_{\bx \bx,k+1} \bff_{\bu} +
V_{\bx,k+1} \cdot \bff_{\bu \bu},
\label{DDP: Quu}
\\
Q_{\bx,k} &= \ell_{\bx} + \bff_{\bx}\T V_{\bx,k+1},
\\
Q_{\bu,k} &= \ell_{\bu} + \bff_{\bu} \T V_{\bx,k+1},
\end{align}
\label{DDP: Q expressions}%
\end{subequations}
%
%
%
%
%
%
where the cost and dynamics derivatives are evaluated at $\bar{\bx}_k$ and $\bar{\bu}_k$. The last terms in \eqref{DDP: Qxx}-\eqref{DDP: Quu} are vector-tensor products that appear in DDP if the second-order approximation of the dynamics is considered. By only taking the first-order dynamics expansion, these terms are omitted and DDP coincides with iLQR \cite{todorov2005iLQR}. Using \eqref{DDP: Q expressions}, the minimization of the RHS of \eqref{Bellman rule} yields the following optimal control deviations
\begin{equation}
\delta \bu_k^* = \bk_k + \bK_k \delta \bx_k,
\label{DDP Optimal Control Deviations}
\end{equation}
with $\bk_k = - Q_{\bu \bu,k}^{-1} Q_{\bu,k}$ and $\bK_k = - Q_{\bu \bu,k}^{-1} Q_{\bu \bx,k}$ being the feed-forward and feedback gains, respectively. By also expanding $V_k$ quadratically, \eqref{Bellman rule} leads to the following backpropagation rules
\begin{subequations}
\begin{align}
V_k &= - \frac{1}{2} Q_{\bu,k}\T Q_{\bu \bu,k}^{-1} Q_{\bu,k}, 
\\
V_{\bx,k} &= Q_{\bx,k} - Q_{\bu \bx,k}\T Q_{\bu \bu,k}^{-1} Q_{\bu,k}, \\
V_{\bx \bx, k} &= Q_{\bx \bx,k} - Q_{\bu \bx,k}\T Q_{\bu \bu,k}^{-1} Q_{\bu \bx, k}, 
\end{align}
\end{subequations}
with $V_K = \phi(\bx_K)$, $V_{\bx, K} = \phi_{\bx}(\bx_K)$ and $V_{\bx \bx, K} = \phi_{\bx \bx }(\bx_K)$. 

After completing the backward pass, the new trajectories $\bx$ and $\bu$ are obtained during the \textit{forward} pass as follows
\begin{subequations}
\begin{align}
\bu_k & = \bar{\bu}_k + \bk_k + \bK_k (\bx_k - \bar{\bx}_k), \\
\bx_{k+1} & = \bff(\bx_k, \bu_k), \ \ \bx_0: \text{given}.
\end{align}
\end{subequations}
These new trajectories will be used as the nominal ones during the next backward pass, etc. This iterative procedure continues until a predefined termination criterion is satisfied. 

While the original DDP method was developed for unconstrained optimal control problems, several variations have been proposed for handling control and/or state constraints such as \cite{murray1979constrained, tassa2014controlddp, xie2017differential, pavlov2021interior, jallet2022constrained, howell2019altro,  aoyama2020constrainedDDPicra}. Out of them, methods that incorporate constraints through an AL on the cost have been shown to be particularly effective \cite{howell2019altro, aoyama2020constrainedDDPicra}. In this paper, we will use the recently suggested AL-DDP combination \cite{aoyama2020constrainedDDPicra} as a baseline for addressing constrained optimal control problems with DDP.

\subsection{Alternating Direction Method of Multipliers}
Subsequently, we provide a brief overview of ADMM. For more details, the reader is referred to \cite{boyd2011distributed}. The standard (two-block) version of ADMM considers an optimization problem of the following form
\begin{equation}
\min_{\bx, \bz} \ f(\bx) + g(\bz)  \quad
\mathrm{s.t.} \ 
\bA \bx + \bB \bz = \bc, 
\label{ADMM Standard Version}
\end{equation}
where $\bx \in \Rb^p$, $\bz \in \Rb^q$ are the primal variables, $f: \Rb^p\rightarrow \Rb$, $g: \Rb^q \rightarrow \Rb$, $\bA \in \Rb^{r \times p}$, $\bB \in \Rb^{r \times q}$ and $\bc \in \Rb^r$. For problem \eqref{ADMM Standard Version}, the AL is given by
\begin{align*}
\calL (\bx,\bz,\blambda)
& =
f(\bx) + g(\bz)
+ \blambda \T ( \bA \bx + \bB \bz - \bc ) \\
& + \frac{\rho}{2} \| \bA \bx + \bB \bz - \bc \|_2^2,
\end{align*}
where $\blambda \in \Rb^r$ is the dual variable for the constraint $\bA \bx + \bB \bz - \bc = 0$ and $\rho > 0$ is a penalty parameter. Classical ADMM consists of the following sequential updates
\begin{subequations}
\begin{align}
&\bx^{n+1} = \argmin_{\bx} \calL (\bx,\bz^n,\blambda^n) \\
&\bz^{n+1} = \argmin_{\bz} \calL (\bx^{n+1},\bz,\blambda^n) \\
&\blambda^{n+1} = \blambda^n + \rho ( \bA \bx^{n+1} + \bB \bz^{n+1} - \bc ),
\end{align}
\end{subequations}
where $n$ denotes the algorithm iteration. The method can be extended to a multi-block version with variables $\bx_1, \dots, \bx_N$, $N \geq 3$, that are updated in a sequential manner \cite{tao2011recovering, tang2019distributed}.

\IEEEpubidadjcol
\section{Multi-Agent Optimal Control Problem Formulation}
\label{sec:Multi-AgentFormulation}

Let us consider a team of $M$ agents given by the set $\calV = \{ 1, \dots, M \}$. We first introduce the notion of \textit{neighbors}.
\begin{definition}[Neighborhood Set] \label{def:neigh}
The ``neighborhood'' set $\calN_i \subseteq \calV$ of an agent $i \in \calV$ is the set that contains all agents $j \in \calV$ that are neighbors of $i$ (including $i$).
\end{definition}
\begin{definition}[Neighbor-of Set] \label{def:neigh of}
The ``neighbor-of'' set $\calP_i \subseteq \calV$ of an agent $i \in \calV$ is defined as $\calP_i = \{ j \in \calV : i \in \calN_j \}$. In other words, it is the set that contains all agents $j \in \calV$ that consider $i$ as a neighbor.
\end{definition}
\begin{assumption}
\label{time invariant assumption}
The sets $\calN_i, \ i \in \calV$, are time-invariant. It follows that $\calP_i, \ i \in \calV$, are also time-invariant.
\end{assumption}

In a multi-robot setting, we would typically have $j \in \calN_i$ if agent $j$ is within a close distance from $i$. Note that there is no assumption of mutual neighbors, i.e., it is not required that $\calN_i = \calP_i$.
Furthermore, we only require local communication between the agents through the following assumption.
\begin{assumption}
\label{communication assumption}
Each agent $i \in \calV$ is able to exchange information only with agents $j \in \calN_i \cup \calP_i$.
\end{assumption}

The dynamics of the $i$-th agent are provided by 
the following discrete-time nonlinear equations
\begin{equation}
\bx_{i,k+1} = \bff_i (\bx_{i,k}, \bu_{i,k}), \ \ \bx_{i,0}: \text{given}, 
\label{Multi-Agent DDP Dynamics}
\end{equation}
where $\bx_{i,k} \in \Rb^{p_i}$, $\bu_{i,k} \in \Rb^{q_i}$ are the state and control input of agent $i$ at time $k$. Let $\bx_i = 
[
\bx_{i,0}; \dots; \bx_{i,K}
]
$
and
$\bu_i = 
[
\bu_{i,0}; \dots; \bu_{i,K-1}
]
$
be the state and control trajectories of agent $i$ over a time horizon $K$. The global cost function that all agents aim to collectively minimize is
\begin{equation}
J = \sum_{i=1}^M J_i (\bx_i, \bu_i)
\end{equation}
where each local component $J_i: \Rb^{(K+1)p_i} \times \Rb^{K q_i} \rightarrow \Rb$ has the form
\begin{equation}
J_i  = \sum_{k=0}^{K-1} 
\Big[ \ell_i (\bx_{i,k}, \bu_{i,k}) 
\Big]
+ \phi_i(\bx_{i,K}), 
\label{Multi-Agent DDP Costs}
\end{equation}
with $\ell_i:\Rb^{p_i} \times \Rb^{q_i} \rightarrow \Rb$ and $\phi_i:\Rb^{p_i} \rightarrow \Rb$ being the running and terminal costs, respectively. All agents are subject to the following \textit{single-agent} control and state constraints
%
\begin{align}
\bb_{i,k} (\bu_{i,k}) & \leq 0, \ k \in \llbracket 0, K-1 \rrbracket,
\label{Multi-Agent DDP Single-agent control constraints}
\\
\bg_{i,k}(\bx_{i,k}) & \leq 0, \ k \in \llbracket 0, K \rrbracket.
\label{Multi-Agent DDP Single-agent state constraints}
\end{align}
In multi-robot problems, the former usually correspond to actuation limitations, while the latter often represent position, velocity, obstacle avoidance constraints etc.
Finally, the following \textit{inter-agent} state constraints must also be satisfied between any agent $i \in \calV$ and its  neighbors,
\begin{equation}
\bh_{ij,k}(\bx_{i,k},\bx_{j,k}) \leq 0, \ k \in \llbracket 0, K \rrbracket, \ j \in \calN_i \backslash \{i\}.
\label{Multi-Agent DDP Inter-agent constraints}
\end{equation}
Such constraints could enforce collision avoidance or connectivity maintenance between neighbors. Note that the use of the latter supports introducing Assumption \ref{time invariant assumption} since neighboring agents would always stay within a close distance. Consequently, we proceed with formulating the multi-agent optimal control problem.
\begin{problem}[Multi-Agent Optimal Control Problem]
\label{multi-agent optimal control problem 1}
Find the optimal control sequences $\bu_i^*, \ \forall i \in \calV$, such that
%
\begin{align*}
& \{ \bu_i^* \}_{i \in \calV} = \argmin \sum_{i=1}^M J_i (\bx_i, \bu_i)
\nonumber
\\
\mathrm{s.t.} \quad & \bx_{i,k+1} = \bff_i (\bx_{i,k}, \bu_{i,k}), \ \bx_{i,0}: \text{given},
\nonumber
\\
& \bb_{i,k} (\bu_{i,k}) \leq 0, 
\nonumber
\ \bg_{i,k}(\bx_{i,k}) \leq 0, \ 
\nonumber
\\ 
&
\bh_{ij,k}(\bx_{i,k},\bx_{j,k}) \leq 0,
\ j \in \calN_i \backslash \{i\}, \ i \in \calV.
\label{Multi-Agent DDP Centralized Problem}
\end{align*}
\end{problem}

\section{Nested Distributed DDP}
\label{sec:Nested_DDP}

\begin{figure}[t]
\centering
\begin{tikzpicture}
    \node[anchor=south west,inner sep=0] at (0,0){\includegraphics[width=0.48\textwidth, trim={0cm 3cm 0cm 2.5cm},clip]{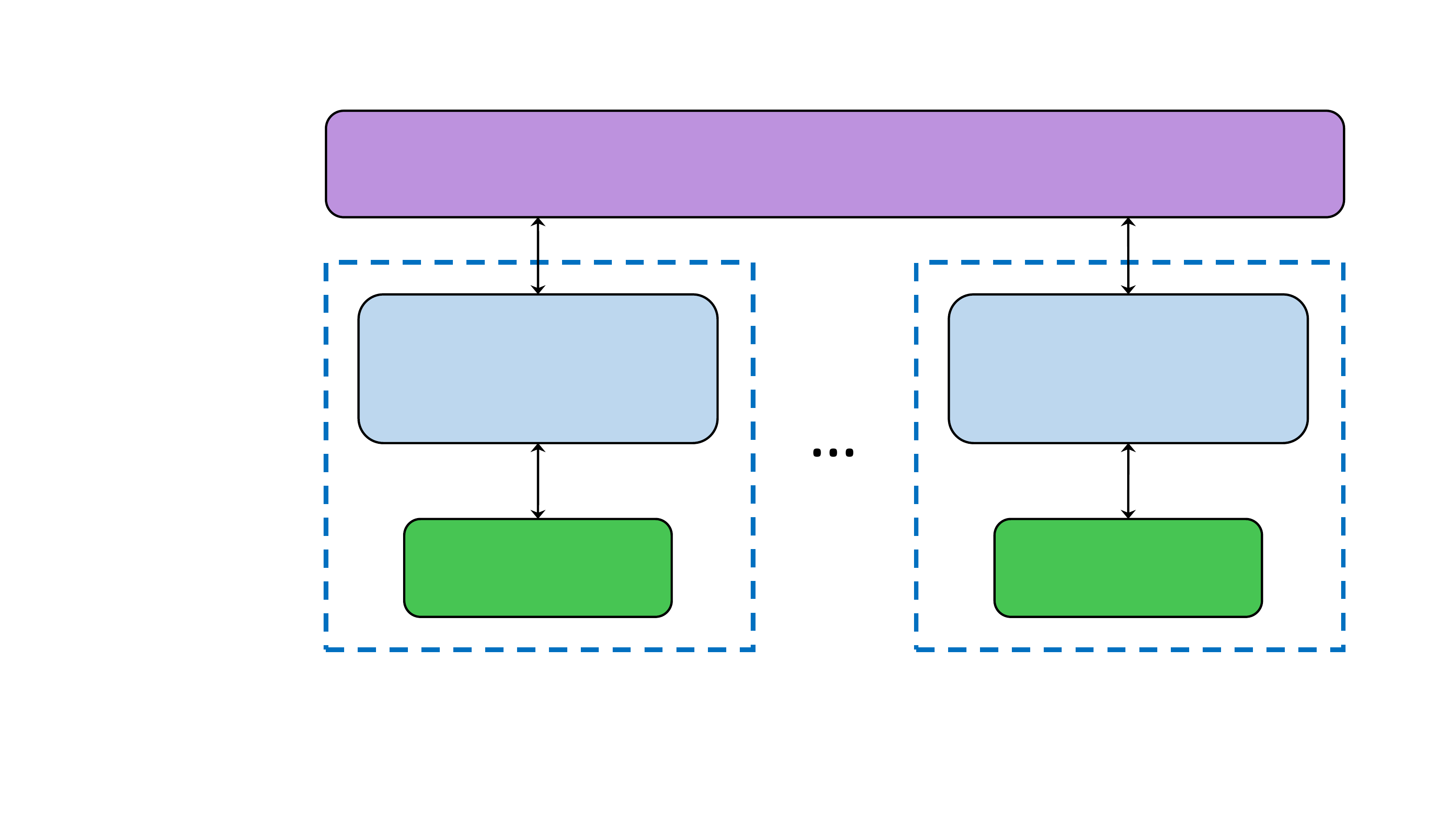}
    };
    \node[align=center] (c) at (4.99,3.15) {\textbf{ADMM for Consensus}};
    \node[align=center] (c) at (3.22,1.94) {\small{\textbf{AL for Local}} \\ \small{\textbf{Constraints}}};
    \node[align=center] (c) at (3.22,0.74) {\small{\textbf{DDP}}};
    \node[align=center] (c) at (6.80,1.94) {\small{\textbf{AL for Local}} \\ \small{\textbf{Constraints}}};
    \node[align=center] (c) at (6.80,0.74) {\small{\textbf{DDP}}};
    \node[align=center, text=NavyBlue] (c) at (3.27,-0.14) {Agent $i=1$ };
    \node[align=center, text=NavyBlue] (c) at (6.88,-0.14) {Agent $i=M$ };
    \node[align=center] (c) at (1.1,3.15) {Level $1$ };
    \node[align=center] (c) at (1.1,1.94) {Level $2$ };
    \node[align=center] (c) at (1.1,0.74) {Level $3$ };
\end{tikzpicture}
\caption{The ND-DDP (``nested'') architecture. Level 1: ADMM for consensus between different agents; Level 2: AL for local constraints; Level 3: DDP.}
\label{fig:ND-DDP}
\end{figure}
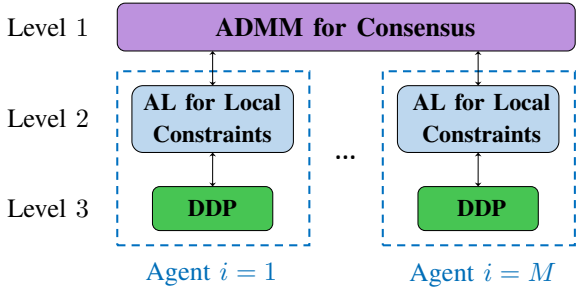

The first proposed architecture (Fig.  \ref{fig:ND-DDP}) uses an ADMM layer for enforcing a consensus between the decisions of all agents (Level 1). Next, each agent's neighborhood constraints are captured through a local AL layer (Level 2). Finally, the resulting problems are solved by each agent through DDP (Level 3). Due to the ``nested'' structure of this framework, we refer to it as Nested Distributed DDP (ND-DDP).

\subsection{Problem Transformation}

Prior to the method derivation, it is necessary that Problem \ref{multi-agent optimal control problem 1} is transformed to an equivalent form that is suitable for distributed optimization. The coupling that prohibits directly solving it in a decentralized manner is the one induced by the inter-agent constraints \eqref{Multi-Agent DDP Inter-agent constraints}. To address this, we propose that each agent contains copy variables regarding the states and controls of their neighbors. In other words, we define the variables $\bx_{j,k}^i \in \Rb^{p_j}, \ \bu_{j,k}^i \in \Rb^{q_j}, \ j \in \calN_i, \ i \in \calV$, where each $\bx_{j,k}^i$ (or $\bu_{j,k}^i$) refers to what would be safe for agent $j$ from the \textit{point of view} of agent $i$. Of course, the variables $\bx_{i,k}^i$ and $\bu_{i,k}^i$ coincide with $\bx_{i,k}$ and $\bu_{i,k}$, respectively.
Thus, the following augmented state and control variables can be defined for each agent:
\begin{align}
\bx_{i,k}^{\text{a}} 
= [\{ \bx_{j,k}^i \}_{j \in \calN_i} ]
\in \Rb^{\tilde{p}_i}, \ \tilde{p}_i = \sum_{j \in \calN_i} p_j
, \\ 
\bu_{i,k}^{\text{a}} 
= [\{ \bu_{j,k}^i \}_{j \in \calN_i} ]
\in \Rb^{\tilde{q}_i}, \ \tilde{q}_i = \sum_{j \in \calN_i} q_j.
\end{align}
The trajectories $\bx_j^i, \bu_j^i, \bx_i^{\text{a}}, \bu_i^{\text{a}}$ are defined as $\bx_i, \bu_i$ earlier. Thanks to introducing the copy variables, the inter-agent constraints \eqref{Multi-Agent DDP Inter-agent constraints} can be rewritten from the perspective of each agent $i$ as $\bh_{ij,k}(\bx_{i,k},\bx_{j,k}^i) \leq 0$, or more compactly as
\begin{equation}
\bh_{i,k}^{\text{a}}(\bx_{i,k}^{\text{a}}) \leq 0, \ i \in \calV,
\end{equation}
where $\bh_{i,k}^{\text{a}}(\bx_{i,k}^{\text{a}}) = \big[ \{\bh_{ij,k}(\bx_{i,k},\bx_{j,k}^i)\}_{j \in \calN_i \backslash \{i\}} \big]$.

Nevertheless, since there might be multiple variables referring to the same agent, it becomes necessary to enforce a consensus between such variables. To achieve this, the following global state and control variables are also introduced
\begin{align}
\bz_k & = [ \{ \bz_{i,k} \}_{i \in \calV}] \in \Rb^p, \ p = \sum_{i \in \calV} p_i, \ k \in \llbracket 0, K \rrbracket, 
\label{ND-DDP Global state}
\\ 
\bw_k & = [ \{ \bw_{i,k} \}_{i \in \calV}] \in \Rb^q, \ q = \sum_{i \in \calV} q_i, \ k \in \llbracket 0, K-1 \rrbracket,
\end{align}
with the trajectories $\bz$, $\bw$, $\bz_i$ and $\bw_i$ defined accordingly.
Thus, we can impose the following consensus constraints
\begin{equation}
\bx_j^i = \bz_j, \ \bu_j^i = \bw_j, \ j \in \calN_i, \ i \in \calV,
\label{ND-DDP consensus constraints}
\end{equation}
which can be written in a more compact form as
\begin{equation}
\bx_i^{\text{a}} = \bz_i^{\text{a}}, \ 
\bu_i^{\text{a}} = \bw_i^{\text{a}}, \ i \in \calV,
\label{ND-DDP consensus constraints compact}
\end{equation}
with $\bz_{i,k}^{\text{a}} = \big[ \{ \bz_{j,k} \}_{j \in \calN_i} \big] \in \Rb^{\tilde{p}_i}$ and $\bw_{i,k}^{\text{a}} = \big[ \{ \bw_{j,k} \}_{j \in \calN_i} \big] \in \Rb^{\tilde{q}_i}$. Let us also introduce the augmented dynamics $\bx_{i,k+1}^{\text{a}} = \bff_i^{\text{a}} (\bx_{i,k}^{\text{a}}, \bu_{i,k}^{\text{a}})$
and constraints $\bb_{i,k}^{\text{a}}(\bu_{i,k}^{\text{a}}) \leq 0$, $\bg_{i,k}^{\text{a}}(\bx_{i,k}^{\text{a}}) \leq 0$, accordingly. 
Finally, we suggest the new local cost functions:
\begin{equation}
J_i^{\text{a}} (\bx_i^{\text{a}}, \bu_i^{\text{a}}) = \sum_{j \in \calN_i} \frac{1}{|\calP_j|} J_j (\bx_j^i, \bu_j^i),
\label{ND-DDP augmented cost}
\end{equation}
for every agent $i \in \calV$. Therefore, Problem \ref{multi-agent optimal control problem 1} can be transformed to the following equivalent form.
\begin{problem}[Multi-Agent Optimal Control Problem II]
\label{multi-agent optimal control problem 2}
Find the optimal control sequences $\bu_i^{\text{a}*}, \ \forall i \in \calV$, such that
%
\begin{align*}
& \{ \bu_i^{\text{a}*} \}_{i \in \calV} = \argmin \sum_{i=1}^M J_i^{\text{a}} (\bx_i^{\text{a}}, \bu_i^{\text{a}})
\nonumber
\\
\mathrm{s.t.} \quad & \bx_{i,k+1}^{\text{a}} = \bff_i^{\text{a}} (\bx_{i,k}^{\text{a}}, \bu_{i,k}^{\text{a}}), \ \bx_{i,0}^{\text{a}}: \text{given},
\nonumber
\\
& \bb_{i,k}^{\text{a}}(\bu_{i,k}^{\text{a}}) \leq 0, 
\ \bg_{i,k}^{\text{a}}(\bx_{i,k}^{\text{a}}) \leq 0, \ 
\bh_{i,k}^{\text{a}}(\bx_{i,k}^{\text{a}}) \leq 0,
\nonumber
\\
& \bx_i^{\text{a}} = \bz_i^{\text{a}}, 
\ \bu_i^{\text{a}} = \bw_i^{\text{a}},
\ i \in \calV.
\end{align*}
\end{problem}

Note that in \eqref{ND-DDP augmented cost}, each $J_j$ is multiplied with a factor $1/|\calP_j|$ so that the cost functions in Problems \ref{multi-agent optimal control problem 1} and \ref{multi-agent optimal control problem 2} are equivalent.
%

\IEEEpubidadjcol
\subsection{Method Derivation}

Subsequently, the derivation of the ND-DDP method is presented. Let us start with rewriting Problem \ref{multi-agent optimal control problem 2} by replacing part of the constraints with their indicator functions as follows
\begin{align}
& \min \sum_{i=1}^M J_i^{\text{a}} (\bx_i^{\text{a}}, \bu_i^{\text{a}}) 
+ \calI_{\bff_i^{\text{a}}} (\bx_i^{\text{a}}, \bu_i^{\text{a}}) 
+ \calI_{\bb_i^{\text{a}}} (\bu_i^{\text{a}}) 
\nonumber
\\ 
& ~~~~~~~~~~~ + \calI_{\bg_i^{\text{a}}} (\bx_i^{\text{a}}) 
+ \calI_{\bh_i^{\text{a}}} (\bx_i^{\text{a}})
\nonumber
\\[0.1cm]
& ~~~~~~~~ \mathrm{s.t.} \quad  
\bx_i^{\text{a}} = \bz_i^{\text{a}}, 
\ \bu_i^{\text{a}} = \bw_i^{\text{a}},
\ i \in \calV.
\label{ND-DDP Problem Ind}
\end{align}
The AL for problem \eqref{ND-DDP Problem Ind} can be formulated as
\begin{align}
\calL = 
& \sum_{i=1}^M J_i^{\text{a}} (\bx_i^{\text{a}}, \bu_i^{\text{a}}) 
+ \calI_{\bff_i^{\text{a}}} (\bx_i^{\text{a}}, \bu_i^{\text{a}}) 
+ \calI_{\bb_i^{\text{a}}} (\bu_i^{\text{a}})
\nonumber
\\ 
& ~~~~~ + \calI_{\bg_i^{\text{a}}} (\bx_i^{\text{a}})
+ \blambda_i \T (\bx_i^{\text{a}} - \bz_i^{\text{a}}) 
+ \by_i \T (\bu_i^{\text{a}} - \bw_i^{\text{a}})
\nonumber
\\
& 
~~~~~ + \frac{\rho}{2} \| \bx_i^{\text{a}} - \bz_i^{\text{a}} \|_2^2
+ \frac{\mu}{2} \| \bu_i^{\text{a}} - \bw_i^{\text{a}} \|_2^2,
\label{NDDDP AL}
\end{align}
where $\blambda_i, \by_i$ are the corresponding dual variables of the consensus constraints and $\rho, \mu > 0$ are penalty parameters. Next, the ADMM update rules are derived. In the following, ADMM iterations are denoted with $n$.

\textbf{Optimization Step 1: Local AL-DDP Updates}. In the first block of updates, the AL \eqref{NDDDP AL} is minimized w.r.t. the local variables $\bx_i^{\text{a}}, \bu_i^{\text{a}}$, i.e.,
\begin{equation*}
\{ \bx_i^{\text{a}}, \bu_i^{\text{a}} \}^{n+1} = \argmin \calL(\bx_i^{\text{a}}, \bu_i^{\text{a}}, \bz^n, \bw^n, \blambda_i^n, \by_i^n), 
\end{equation*}
for all agents $i \in \calV$. This results to the following $M$ subproblems, that can be solved in parallel from all agents
\begin{align}
\{ \bx_i^{\text{a}}, \bu_i^{\text{a}} & \}^{n+1} = \argmin \sum_{k=0}^{K-1} 
\Big[ \hat{\ell}_i (\bx_{i,k}^{\text{a}}, \bu_{i,k}^{\text{a}}) 
\Big]
+ \hat{\phi}_i (\bx_{i,K}^{\text{a}}),
\nonumber
\\
\mathrm{s.t.} \quad & \bx_{i,k+1}^{\text{a}} = \bff_{i,k}^{\text{a}} (\bx_{i,k}^{\text{a}}, \bu_{i,k}^{\text{a}}), \ \bx_{i,0}^{\text{a}}: \text{given},
\label{ND-DDP primal update}
\\
& \bb_{i,k}^{\text{a}}(\bu_{i,k}^{\text{a}}) \leq 0, 
\ \bg_{i,k}^{\text{a}}(\bx_{i,k}^{\text{a}}) \leq 0, \ 
\bh_{i,k}^{\text{a}}(\bx_{i,k}^{\text{a}}) \leq 0,
\nonumber
\end{align}
with the running and terminal costs being
\begin{align*}
& \hat{\ell}_i (\bx_{i,k}^{\text{a}}, \bu_{i,k}^{\text{a}}) = 
\sum_{j \in \calN_i} \frac{1}{|\calP_j|} \ell_j (\bx_{j,k}^i, \bu_{j,k}^i)
\nonumber
\\
& ~~~~~ + \frac{\rho}{2} \left\| \bx_{i,k}^{\text{a}} - \bz_{i,k}^{\text{a}} + \frac{\blambda_{i,k}}{\rho} \right\|_2^2
+ \frac{\mu}{2} \left\| \bu_{i,k}^{\text{a}} - \bw_{i,k}^{\text{a}} + \frac{\by_{i,k}}{\mu} \right\|_2^2,
\\[0.1cm]
& \hat{\phi}_i (\bx_{i,K}^{\text{a}}) = 
\sum_{j \in \calN_i} \frac{1}{|\calP_j|} \phi_j (\bx_{j,K}^i)
+ \frac{\rho}{2} \left\| \bx_{i,K}^{\text{a}} - \bz_{i,K}^{\text{a}} + \frac{\blambda_{i,k}}{\rho} \right\|_2^2.
\end{align*}
Each of these local subproblems is solved by incorporating the neighborhood constraints into the cost through an AL approach \cite[Section IV.A]{aoyama2020constrainedDDPicra} as follows
%
\begin{align}
\{ \bx_i^{\text{a}}, \bu_i^{\text{a}} & \}^{n+1} = \argmin \sum_{k=0}^{K-1} 
\Big[ \hat{\ell}_i (\bx_{i,k}^{\text{a}}, \bu_{i,k}^{\text{a}}) 
\Big]
+ \hat{\phi}_i (\bx_{i,K}^{\text{a}}) 
\nonumber
\\ 
& ~~~~~~~~~~ + \sum_{k=0}^{K} C_{i,k}(\bx_{i,k}^{\text{a}}, \bu_{i,k}^{\text{a}})
\label{ND-DDP AL-DDP}
\\
& \mathrm{s.t.} \quad \bx_{i,k+1}^{\text{a}} = \bff_{i,k}^{\text{a}} (\bx_{i,k}^{\text{a}}, \bu_{i,k}^{\text{a}}), \ \bx_{i,0}^{\text{a}}: \text{given},
\nonumber
\end{align}
where
\begin{align*}
\bs_{i,k}^{\text{a}} (\bx_{i,k}^{\text{a}}, \bu_{i,k}^{\text{a}}) &= \big[ \bb_{i,k}^{\text{a}}(\bu_{i,k}^{\text{a}}); \ \bg_{i,k}^{\text{a}}(\bx_{i,k}^{\text{a}}); \ \bh_{i,k}^{\text{a}}(\bx_{i,k}^{\text{a}}) \big],
\\
C_{i,k}(\bx_{i,k}^{\text{a}}, \bu_{i,k}^{\text{a}}) &= P \big( \bw_{i,k}, \beta_{i,k}, \bs_{i,k}^{\text{a}} (\bx_{i,k}^{\text{a}}, \bu_{i,k}^{\text{a}}) \big),
\end{align*} 
and $\bw_{i,k}$, $\beta_{i,k}$ are the Lagrange multipliers and penalty parameters for the constraint $\bs_{i,k}^{\text{a}} (\bx_{i,k}^{\text{a}}, \bu_{i,k}^{\text{a}}) \leq 0$, respectively. During each local AL iteration $l = 1,\dots,L$, problem \eqref{ND-DDP AL-DDP} is solved with DDP.  Details on the form of $P(\cdot)$ and on how $\bw_{i,k}$, $\beta_{i,k}$ are updated, can be found in \cite{aoyama2020constrainedDDPicra}. The backward pass rules for DDP will obtain the following form
\begin{align*}
Q_{\bx_i^{\text{a}} \bx_i^{\text{a}}, k} &= \bigg( \sum_{j \in \calN_i} \frac{1}{|\calP_j|} \ell_j \bigg)_{\bx_i^{\text{a}} \bx_i^{\text{a}}} + \rho \bI 
+ C_{\bx_i^{\text{a}} \bx_i^{\text{a}}, k}
\\ &~~~~~~+
\bff_{\bx_i^{\text{a}}}^{\text{a}} {}\T {V}_{\bx_i^{\text{a}} \bx_i^{\text{a}}, k+1} \bff_{\bx_i^{\text{a}}}^{\text{a}},
\end{align*}
\begin{align*}
Q_{\bx_i^{\text{a}} \bu_i^{\text{a}}, k} &= \bigg( \sum_{j \in \calN_i} \frac{1}{|\calP_j|} \ell_j \bigg)_{\bx_i^{\text{a}} \bu_i^{\text{a}}} 
+ \bff_{\bx_i^{\text{a}}}^{\text{a}} {}\T {V}_{\bx_i^{\text{a}} \bx_i^{\text{a}}, k+1} \bff_{\bu_i^{\text{a}}}^{\text{a}}, 
\\
Q_{\bu_i^{\text{a}} \bu_i^{\text{a}}, k} &= \bigg( \sum_{j \in \calN_i} \frac{1}{|\calP_j|} \ell_j \bigg)_{\bu_i^{\text{a}} \bu_i^{\text{a}}} + \mu \bI + C_{\bu_i^{\text{a}} \bu_i^{\text{a}}, k}
\\ &~~~~~~+
\bff_{\bu_i^{\text{a}}}^{\text{a}} {}\T {V}_{\bx_i^{\text{a}} \bx_i^{\text{a}}, k+1} \bff_{\bu_i^{\text{a}}}^{\text{a}},
\\
Q_{\bx_i^{\text{a}}, k} &= \bigg( \sum_{j \in \calN_i} \frac{1}{|\calP_j|} \ell_j \bigg)_{\bx_i^{\text{a}}} + \rho(\bx_{i,k}^{\text{a}} - \bz_{i,k}^{\text{a}}) + \blambda_{i,k}
\\ &~~~~~~ + C_{\bx_i^{\text{a}}, k} +
\bff^{\text{a}}_{\bx_i^{\text{a}}} {}\T V_{\bx_i^{\text{a}}, k+1},
\\
Q_{\bu_i^{\text{a}}, k} &= \bigg( \sum_{j \in \calN_i} \frac{1}{|\calP_j|} \ell_j \bigg)_{\bu_i^{\text{a}}} + \mu(\bu_{i,k}^{\text{a}} - \bw_{i,k}^{\text{a}}) + \by_{i,k} 
\\ &~~~~~~ + C_{\bu_i^{\text{a}}, k} +
\bff^{\text{a}}_{\bu_i^{\text{a}}} {}\T V_{\bx_i^{\text{a}}, k+1},
\end{align*}
where, for simplicity, only the first-order dynamics expansion is considered. The terminal conditions for the value functions and their derivatives will be given by
\begin{align*}
V_{i,K} & = \sum_{j \in \calN_i} \frac{1}{|\calP_j|} \phi_j (\bx_{j,K}^i)
+ \frac{\rho}{2} \left\| \bx_{i,K}^{\text{a}} - \bz_{i,K}^{\text{a}} + \frac{\blambda_{i,K}}{\rho} \right\|_2^2
\\ &~~~~~~ + C_{i,K}, 
\\
V_{\bx_i^{\text{a}}, K} &= \bigg( \sum_{j \in \calN_i} \frac{1}{|\calP_j|} \phi_j (\bx_{j,K}^i) \bigg)_{\bx_i^{\text{a}}} + \rho(\bx_{i,K}^{\text{a}} - \bz_{i,K}^{\text{a}}) \\ 
&~~~~~~+ \blambda_{i,K}
+ C_{\bx_i^{\text{a}}, K},
\\
V_{\bx_i^{\text{a}} \bx_i^{\text{a}}, K} &= \bigg( \sum_{j \in \calN_i} \frac{1}{|\calP_j|} \phi_j (\bx_{j,K}^i) \bigg)_{\bx_i^{\text{a}} \bx_i^{\text{a}}} + \rho \bI + C_{\bx_i^{\text{a}} \bx_i^{\text{a}}, K}.
\end{align*}
%
\begin{remark}
The new local problems \eqref{ND-DDP AL-DDP} include four significant modifications compared to their unconstrained single-agent counterparts. First, each agent optimizes for its augmented state and control which allows for handling inter-agent constraints - from its own perspective. Second, all agents partially take into account the objectives of their neighbors. Third, all neighborhood constraints are incorporated through the local AL for each agent. Finally, the new costs include extra terms that encourage a consensus between local and global variables, and therefore, a consensus between all agents. 
\end{remark}

\textbf{Optimization Step 2: Global Updates}. In the second block, the global variables $\bz, \bw$ are updated, i.e., 
\begin{equation*}
\{ \bz, \bw \}^{n+1} = \argmin \calL(\bx_i^{\text{a},n+1}, \bu_i^{\text{a},n+1}, \bz, \bw, \blambda_i^n, \by_i^n), 
\end{equation*}
\noindent
which leads to
\begin{align*}
\{ \bz, \bw \}^{n+1} & = \argmin \sum_{i=1}^M \sum_{j \in \calN_i}
\frac{\rho}{2} \| \bx_j^i - \bz_j \|_2^2
+ \frac{\mu}{2} \| \bu_j^i - \bw_j \|_2^2
\\
&  
~~~~~~~~~~~~~~~~~~~~~~~ - \blambda_j^i {}\T \bz_j  
- \by_j^i {}\T \bw_j.
\end{align*}
This minimization can be decomposed for all $\bz_i, \bw_i, \ i \in \calV$, leading to the following update rules
\begin{subequations}
\begin{align}
\bz_i^{n+1} & = \frac{1}{|\calP_i|} \sum_{j \in \calP_i} \bx_i^{j,n+1} + \frac{1}{\rho} \blambda_i^{j,n},
\label{ND-DDP Global state update}
\\
\bw_i^{n+1} & = \frac{1}{|\calP_i|} \sum_{j \in \calP_i} \bu_i^{j,n+1} + \frac{1}{\mu} \by_i^{j,n}.
\label{ND-DDP Global control update}
\end{align}
\label{ND-DDP Global update}%
\end{subequations}
For $n>0$, the parts involving the dual variables in \eqref{ND-DDP Global update}, will become zero with a similar argument as in \cite[Section 7.2]{boyd2011distributed}. Therefore, the global updates  are essentially taking an average of all variables (actual or copy ones) that refer to agent $i$. Finally, the dual variables are updated as follows
\begin{subequations}
\begin{align}
\blambda_i^{n+1} & = \blambda_i^n + \rho(\bx_i^{\text{a},n+1} - \bz_i^{\text{a},n+1}), 
\label{ND-DDP Dual state update}
\\
\by_i^{n+1} & = \by_i^n + \mu(\bu_i^{\text{a},n+1} - \bw_i^{\text{a},n+1}).
\label{ND-DDP Dual control update}
\end{align}
\label{ND-DDP Dual update}
\end{subequations}
\vspace{-0.5cm}
\begin{algorithm}[t]
\caption{Nested Distributed DDP (ND-DDP)}\label{ND-DDP Algorithm}
\begin{algorithmic}[1] 
\State $\bx_i', \bu_i' \gets$ Solve single-agent problems in parallel $ \forall \ i \in \calV$.
\State \textit{Each agent $i \in \calV$ receives $\bx_j', \bu_j'$ from all $j \in \calN_i \backslash \{i\}$.} 
\State \textbf{Initialize:} 
$\bx_i^{\text{a}} \leftarrow [ \{ \bx_j' \}_{j \in \calN_i} ]$,
$\bu_i^{\text{a}} \leftarrow [ \{ \bu_j' \}_{j \in \calN_i} ]$,
$\bz_i^{\text{a}} \leftarrow \bx_i^{\text{a}}$, 
$\bw_i^{\text{a}} \leftarrow \bu_i^{\text{a}}$,
$\blambda_i \leftarrow 0$, $\by_i \leftarrow 0$ .
\While{$n \leq N$}
\While{$l \leq L$} (in parallel $ \forall \ i \in \calV$)
\State $\bx_i^{\text{a}}, \bu_i^{\text{a}} \leftarrow$ Solve \eqref{ND-DDP AL-DDP} in parallel $ \forall \ i \in \calV$.
\State $\bw_{i,k}, \beta_{i,k} \leftarrow$ Update in parallel $ \forall \ i \in \calV, \forall \ k \in$ 
\State $\llbracket 0, K-1 \rrbracket$.
\If{$\| \bs_{i,k}^{\text{a}} \|_2 \leq \epsilon_{\text{AL}}$}
\State \textbf{break}
\EndIf
\EndWhile
\State \textit{Each agent $i \in \calV$ receives $\bx_i^j, \bu_i^j$ from all $j \in \calP_i \backslash \{i\}$.} 
\State $\bz_i, \bw_i \leftarrow$ Update with \eqref{ND-DDP Global state update}, \eqref{ND-DDP Global control update} in parallel $ \forall \ i \in \calV$.
\State \textit{Each agent $i \in \calV$ receives $\bz_j, \bw_j$ from all $j \in \calN_i \backslash \{i\}$.} 
\State $\blambda_i, \by_i \leftarrow$ Update with \eqref{ND-DDP Dual state update}, \eqref{ND-DDP Dual control update} in parallel $ \forall \ i \in \calV$.
\EndWhile
\end{algorithmic}
\end{algorithm}

\subsection{Algorithm}

The ND-DDP algorithm is presented in Alg. \ref{ND-DDP Algorithm}. To warmstart the variables (Line 1), each agent $i \in \calV$ first obtains $\bx_i', \bu_i'$ through solving once with DDP its unconstrained single-agent problem
%
%
which only takes into account the dynamics constraints \eqref{Multi-Agent DDP Dynamics}. Next, every agent $j \in \calN_i \backslash \{i\}$ sends $\bx_j', \bu_j'$ to agent $i$ (Line 2). These are used by each agent $i$ to initialize $\bx_i^{\text{a}}$, $\bu_i^{\text{a}}$, $\bz_i^{\text{a}}$ and $\bw_i^{\text{a}}$ (Line 3). 
Subsequently, the iterative ADMM algorithm starts. In Lines 5-12, $\bx_i^{\text{a}}$ and $\bu_i^{\text{a}}$ are obtained by solving \eqref{ND-DDP primal update} with AL-DDP in parallel for all agents. This involves $L$ loops, wherein $\bx_i^{\text{a}}$, $\bu_i^{\text{a}}$ (Line 6) and $\bw_{i,k}, \beta_{i,k}$ (Line 7) are updated iteratively. The loop terminates early if the residuals of the local neighborhood constraints get below some threshold, i.e., $\| \bs_{i,k}^{\text{a}} \|_2 \leq \epsilon_{\text{AL}}$. After that, each agent $j \in \calP_i \backslash \{i\}$ sends $\bx_i^j, \bu_i^j$ to agent $i$ (Line 13), so that the latter is able to perform the global updates (Line 14) using \eqref{ND-DDP Global update}. Next, every agent $j \in \calN_i \backslash \{i\}$ sends $\bz_j, \bw_j$ to agent $i$, so that the latter can construct $\bz_i^{\text{a}}$, $\bw_i^{\text{a}}$ (Line 15). Thus, every agent $i$ can now perform its dual updates (Line 16) with \eqref{ND-DDP Dual update}. This iterative procedure (Lines 4-17) continues until $n$ reaches to the maximum number of ADMM iterations $N$.
\begin{remark}
All computations (Lines 1,5-12,14,16) can be performed in parallel by every agent $i$. Furthermore, all necessary communication steps (Lines 2,13,15) only take place in a neighborhood level. These two attributes characterize the \textit{fully decentralized} nature of ND-DDP from both \textit{computational} and \textit{communication} aspects.
\end{remark}
\begin{remark}
It is possible to use a termination criterion that would check whether the ADMM primal and dual residuals get below some threshold \cite[Section 3.3]{boyd2011distributed}. Nevertheless, this would require collecting information from all agents at every ADMM iteration, thus violating the fully distributed form of the algorithm. For this reason, we empirically set a maximum number of iterations $N$ that leads to sufficient convergence.
\end{remark}
\begin{remark}[Computational Complexity of ND-DDP]
\label{ND-DDP comp complexity}
The most computationally demanding part of ND-DDP are the local AL-DDP updates (Step 1). Let us denote the DDP iterations with $D$. The computational bottleneck of the DDP algorithm itself is the matrix inversion $Q_{\bu \bu,i,k}^{-1}$ required for \eqref{DDP Optimal Control Deviations} which is performed $K$ times per DDP iteration. Therefore, the ND-DDP algorithm has the following computational complexity
\begin{equation}
O\big( N \cdot  L \cdot D \cdot K \cdot
\tilde{q}_i^3 \big).  
\end{equation}
Note that applying AL-DDP directly on Problem \ref{multi-agent optimal control problem 1} in a centralized fashion would lead to the following computational complexity 
\begin{equation}
O \big( L \cdot D \cdot K \cdot 
( M q_i )^3 \big),
\end{equation}
since the matrix inversion in \eqref{DDP Optimal Control Deviations} would have the dimensionality of the concatenated control vector of all agents. Thus, it is clear that ND-DDP can achieve dramatic computational gains against centralized AL-DDP as the number of agents $M$ increases.
\end{remark}
\begin{remark}
Any other constrained DDP method such as \cite{xie2017differential, pavlov2021interior, jallet2022constrained, howell2019altro,  aoyama2020constrainedDDPicra} could also be used instead of the AL-DDP combination (Lines 5-12) for solving problems \eqref{ND-DDP primal update} within ND-DDP.
\end{remark}

\begin{figure}[t]
\centering
\begin{tikzpicture}
    \node[anchor=south west,inner sep=0] at (0,0){\includegraphics[width=0.48\textwidth, trim={0cm 2cm 0cm 1.5cm},clip]{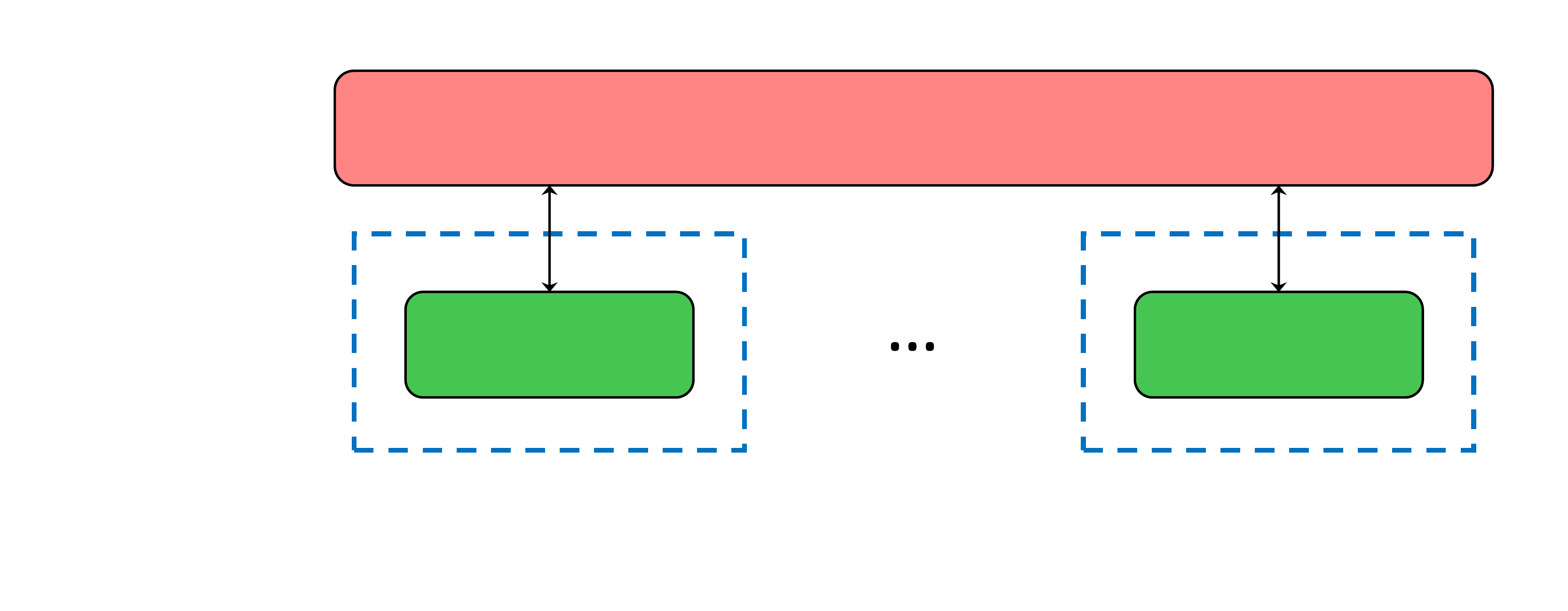}
    };
    \node[align=center] (c) at (5.07,2.18) {\small{\textbf{ADMM for Consensus \& Local Constraints}}};
    \node[align=center] (c) at (3.05,0.97) {\small{\textbf{DDP}}};
    \node[align=center] (c) at (7.12,0.97) {\small{\textbf{DDP}}};
    \node[align=center, text=NavyBlue] (c) at (3.1,0.0) {Agent $i=1$ };
    \node[align=center, text=NavyBlue] (c) at (7.16,0.0) {Agent $i=M$ };
    \node[align=center] (c) at (0.95,2.2) {Level $1$ };
    \node[align=center] (c) at (0.95,1.0) {Level $2$ };
\end{tikzpicture}
\caption{The MD-DDP (``merged'') architecture. Level 1: ADMM for consensus and local constraints; Level 2: DDP.}
\label{fig:MD-DDP}
\end{figure}
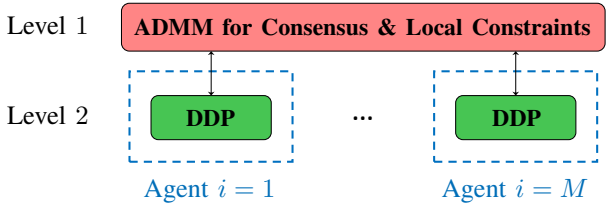

\section{Merged Distributed DDP}
\label{sec:Merged_DistributedDDP}

In the second proposed architecture, we wish to take advantage of the fact that ADMM can also be used for handling the neighborhood constraints locally rather than only the consensus ones. This would lead into ``merging'' the consensus and local constraints levels of ND-DDP into a single level (Fig. \ref{fig:MD-DDP}). For this reason, we name this framework as Merged Distributed DDP (MD-DDP).

\subsection{Problem Transformation}

To handle the local state and control constraints with ADMM, it is necessary to first introduce the ``safe'' copy variables $\tilde{\bx}_{i,k} \in \Rb^{p_i},\  \tilde{\bu}_{i,k} \in \Rb^{q_i}, \ i \in \calV$. These variables must satisfy the local single-agent constraints, i.e., $\bb_{i,k}(\tilde{\bu}_{i,k}) \leq 0$ and $\bg_{i,k}(\tilde{\bx}_{i,k}) \leq 0$, while of course being in consensus with the actual states and controls 
\begin{equation}
\tilde{\bx}_{i,k} = \bx_{i,k}, \
\tilde{\bu}_{i,k} = \bu_{i,k}, \ i \in \calV.
\end{equation}
Moreover, each agent will also contain the following safe copy variables for its neighbors, $\tilde{\bx}_{j,k}^i \in \Rb^{p_j}, \ j \in \calN_i \backslash \{i \}, \ i \in \calV$. Thus, in a similar manner as in Section \ref{sec:Nested_DDP}, we can introduce the augmented ``safe'' states as follows
\begin{equation}
\tilde{\bx}_{i,k}^{\text{a}} = [\tilde{\bx}_{i,k}; \{ \tilde{\bx}_{j,k}^i \}_{j \in \calN_i \backslash \{i\}} ] \in \Rb^{\tilde{p}_i}, \ i \in \calV,
\end{equation}
for which we impose that $ \bh_{i,k}^{\text{a}}(\tilde{\bx}_{i,k}^{\text{a}}) \leq 0$. The trajectories $\tilde{\bx}_i, \tilde{\bx}_j^i, \tilde{\bx}_i^{\text{a}}$ are defined accordingly.
Finally, we define a global state $\bz_k$ as in \eqref{ND-DDP Global state} and enforce the consensus constraints
\begin{equation}
\tilde{\bx}_i^{\text{a}} = \bz_i^{\text{a}}, \ i \in \calV.
\label{MD-DDP consensus constraints compact}
\end{equation}
As a result, we can formulate the following problem which is equivalent to Problem \ref{multi-agent optimal control problem 1}.
\begin{problem}[Multi-Agent Optimal Control Problem III]
\label{multi-agent optimal control problem 3}
Find the optimal control sequences $\bu_i^*, \ \forall i \in \calV$, such that
%
\begin{align*}
& \{ \bu_i^* \}_{i \in \calV} = \argmin \sum_{i=1}^N J_i (\bx_i, \bu_i)
\nonumber
\\
\mathrm{s.t.} \quad & \bx_{i,k+1} = \bff_i (\bx_{i,k}, \bu_{i,k}), \ \bx_{i,0}: \text{given},
\nonumber
\\
& \bb_{i,k}(\tilde{\bu}_{i,k}) \leq 0, 
\nonumber
\ \bg_{i,k}(\tilde{\bx}_{i,k}) \leq 0, \ 
\bh_{i,k}^{\text{a}}(\tilde{\bx}_{i,k}^{\text{a}}) \leq 0,
\nonumber
\\
& \bu_i = \tilde{\bu}_i, \
\bx_i = \tilde{\bx}_i, \  \tilde{\bx}_i^{\text{a}} = \bz_i^{\text{a}},
\ i \in \calV.
\end{align*}
\end{problem}
%
%
%

\subsection{Method Derivation}

The derivation of the MD-DDP updates follows. First, let us restate Problem \ref{multi-agent optimal control problem 3} as
\begin{align}
& \min \sum_{i=1}^M J_i (\bx_i, \bu_i)
+ \calI_{\bff_i} (\bx_i, \bu_i) 
+ \calI_{\bb_i} (\tilde{\bu}_i) 
\nonumber
\\ 
& ~~~~~~~~~~~ 
+ \calI_{\bg_i} (\tilde{\bx}_i)
+ \calI_{\bh_i^{\text{a}}} (\tilde{\bx}_i^{\text{a}})
\nonumber
\\[0.1cm]
& ~~ \mathrm{s.t.} \quad  
\bu_i = \tilde{\bu}_i, \
\bx_i = \tilde{\bx}_i, \  
\tilde{\bx}_i^{\text{a}} = \bz_i^{\text{a}}, 
\ i \in \calV.
\label{MD-DDP Problem Ind}
\end{align}
For this problem, the AL can be written as
\begin{align}
\calL = 
& \sum_{i=1}^M J_i (\bx_i, \bu_i) 
+ \calI_{\bff_i} (\bx_i, \bu_i) 
+ \calI_{\bb_i} (\tilde{\bu}_i) 
+ \calI_{\bg_i} (\tilde{\bx}_i)
\nonumber
\\ 
& 
+ \calI_{\bh_i^{\text{a}}} (\tilde{\bx}_i^{\text{a}})
+ \bxi_i \T (\bu_i - \tilde{\bu}_i) 
+ \blambda_i \T (\bx_i - \tilde{\bx}_i)
\nonumber
\\
&
+ \by_i \T (\tilde{\bx}_i^{\text{a}} - \bz_i^{\text{a}}) 
+ \frac{\tau}{2} \| \bu_i - \tilde{\bu}_i \|_2^2
+ \frac{\rho}{2} \| \bx_i - \tilde{\bx}_i \|_2^2
\nonumber
\\
& + \frac{\mu}{2} \| \tilde{\bx}_i^{\text{a}} - \bz_i^{\text{a}} \|_2^2.
\label{MDDDP AL}
\end{align}

\textbf{Optimization Step 1: Local DDP Updates}. In the first block of updates, we minimize the AL w.r.t. the local variables $\bx_i, \bu_i$, i.e.,
\begin{equation*}
\{ \bx_i, \bu_i \}^{n+1} = \argmin \calL(\bx_i, \bu_i, \tilde{\bx}_i^{\text{a},n}, \tilde{\bu}_i^n, \bz^n, \bxi_i^n, \blambda_i^n, \by_i^n), 
\end{equation*}
for all agents $i \in \calV$. This results to the following $M$ subproblems
\begin{align}
& \{ \bx_i, \bu_i \}^{n+1} = \argmin \sum_{k=0}^{K-1} 
\Big[ \hat{\ell}_i (\bx_{i,k}, \bu_{i,k}) 
\Big]
+ \hat{\phi}_i (\bx_{i,K}),
\nonumber
\\
& ~~~~~~~~ \mathrm{s.t.} \quad  \bx_{i,k+1} = \bff_i (\bx_{i,k}, \bu_{i,k}), \ \bx_{i,0}: \text{given}, 
\label{MD-DDP primal update}
\end{align}
with 
%
%
\begin{align*}
\hat{\ell}_i (\bx_{i,k}, \bu_{i,k}) & = 
\ell_i (\bx_{i,k}, \bu_{i,k}) 
+ \frac{\rho}{2} \left\| \bx_{i,k} - \tilde{\bx}_{i,k} + \frac{\blambda_{i,k}}{\rho} \right\|_2^2
\nonumber
\\
& \quad 
+ \frac{\tau}{2} \left\| \bu_{i,k} - \tilde{\bu}_{i,k} + \frac{\bxi_{i,k}}{\tau} \right\|_2^2,
\nonumber
\\[0.1cm] 
\hat{\phi}_i (\bx_{i,K}) & = 
\phi_i (\bx_{i,K}) + \frac{\rho}{2} \left\| \bx_{i,K} - \tilde{\bx}_{i,K} + \frac{\blambda_{i,K}}{\rho} \right\|_2^2.
\nonumber
\end{align*}
These problems are solved in parallel by every agent $i \in \calV$, with DDP. The backward pass rules will obtain the following form 
\begin{align*}
Q_{\bx_i \bx_i, k} &= \ell_{\bx_i \bx_i} + \rho \bI +
\bff_{\bx_i} \T {V}_{\bx_i \bx_i, k+1} \bff_{\bx_i},
\\
Q_{\bx_i \bu_i,k} &= \ell_{\bx_i \bu_i} + \bff_{\bx_i}\T V_{\bx_i \bx_i,k+1} \bff_{\bu_i}, 
\\
Q_{\bu_i \bu_i,k} &= \ell_{\bu_i \bu_i} + \tau \bI + \bff_{\bu_i}\T V_{\bx_i \bx_i,k+1} \bff_{\bu_i}, 
\\
Q_{\bx_i,k} &= \ell_{\bx_i} + \rho(\bx_{i,k} - \tilde{\bx}_{i,k}) + \blambda_{i,k} + \bff_{\bx_i}\T V_{\bx_i,k+1},
\\
Q_{\bu_i,k} &= \ell_{\bu_i} + \tau(\bu_{i,k} - \tilde{\bu}_{i,k}) + \bxi_{i,k} + \bff_{\bu_i} \T V_{\bx_i,k+1},
\end{align*}
where only the first-order dynamics approximation is considered for brevity. The terminal conditions for the value functions and their derivatives will be
\begin{align*}
V_{i,K} & = \phi_i(\bx_{i,K}) + \frac{\rho}{2} \left\| \bx_{i,K} - \tilde{\bx}_{i,K} + \frac{\blambda_{i,K}}{\rho} \right\|_2^2, 
\\
V_{\bx_i, K} &= \phi_{\bx_i}(\bx_{i,K}) + \rho(\bx_{i,K} - \tilde{\bx}_{i,K}) + \blambda_{i,K},
\\
V_{\bx_i \bx_i, K} &= \phi_{\bx_i \bx_i }(\bx_{i,K}) + \rho \bI.
\end{align*}

\textbf{Optimization Step 2: Local Safe Updates}. The second block yields the following update rules 
\begin{equation*}
\{ \tilde{\bx}_i^{\text{a}}, \tilde{\bu}_i \}^{n+1} = \argmin \calL(\bx_i^{n+1}, \bu_i^{n+1}, \tilde{\bx}_i^{\text{a}}, \tilde{\bu}_i, \bz^n, \bxi_i^n, \blambda_i^n, \by_i^n), 
\end{equation*}
which leads to the following $M$ (for every agent $i = 1,\dots,M)$ state subproblems
\begin{align}
& \tilde{\bx}_i^{\text{a}, n+1} = \argmin \frac{\rho}{2} \left\| \bx_i - \tilde{\bx}_i + \frac{\blambda_i}{\rho} \right\|_2^2
+ \frac{\mu}{2} \left\| \tilde{\bx}_i^{\text{a}} - \bz_i^{\text{a}} + \frac{\by_i}{\mu} \right\|_2^2
\nonumber
\\[0.1cm]
& ~~~ ~\mathrm{s.t.} \quad \bg_{i,k}(\tilde{\bx}_{i,k}) \leq 0, \ 
\tilde{\bh}_{i,k} (\tilde{\bx}_{i,k}^{\text{a}}) \leq 0, \ k \in \llbracket 0, K \rrbracket,
\label{MDDDP safe update}
\end{align}
and $M$ control subproblems
\begin{align}
& \tilde{\bu}_i^{n+1} = \argmin \frac{\tau}{2} \left\| \bu_i - \tilde{\bu}_i + \frac{\bxi_i}{\tau} \right\|_2^2 
\nonumber
\\[0.1cm]
& \mathrm{s.t.} \quad \bb_{i,k}(\tilde{\bu}_{i,k}) \leq 0, \ k \in \llbracket 0, K-1 \rrbracket.
\label{MDDDP safe control update}
\end{align}
Note that each agent's state subproblem \eqref{MDDDP safe update} can be further decomposed into $K+1$ smaller subproblems (for each time instant $k = 0, \dots, K$), i.e.,
\begin{align}
\tilde{\bx}_{i,k}^{\text{a}, n+1} = \argmin & \frac{\rho}{2} \left\| \bx_{i,k} - \tilde{\bx}_{i,k} + \frac{\blambda_{i,k}}{\rho} \right\|_2^2
\nonumber
\\
& + \frac{\mu}{2} \left\| \tilde{\bx}_{i,k}^{\text{a}} - \bz_{i,k}^{\text{a}} + \frac{\by_{i,k}}{\mu} \right\|_2^2
\nonumber
\\[0.1cm]
\mathrm{s.t.} \quad \bg_{i,k}(\tilde{\bx}_{i,k}) & \leq 0, \ 
\bh_{i,k}^{\text{a}}(\tilde{\bx}_{i,k}^{\text{a}}) \leq 0,
\label{MDDDP safe update 2}
\end{align}
that can be solved in parallel as well. The same holds for the control subproblems, which can be decomposed into the following $K$ subproblems (for $k = 0, \dots, K-1$)
\begin{align}
& \tilde{\bu}_{i,k}^{n+1} = \argmin \frac{\tau}{2} \left\| \bu_{i,k} - \tilde{\bu}_{i,k} + \frac{\bxi_{i,k}}{\tau} \right\|_2^2
\nonumber
\\[0.1cm]
& ~~~~~~~~~~~~ \mathrm{s.t.} \quad \bb_{i,k}(\tilde{\bu}_{i,k}) \leq 0.
\label{MDDDP safe control update 2}
\end{align}
Moreover, note that \eqref{MDDDP safe update 2} and \eqref{MDDDP safe control update 2} are actually projection steps that admit trivial solutions in the case of box constraints. For example, if we have $\bu_{i,\text{min}} \leq \tilde{\bu}_{i,k} \leq \bu_{i,\text{max}}$, then \eqref{MDDDP safe control update 2} yields 
\begin{equation}
\tilde{\bu}_{i,k}^{n+1} = \Pi_{[\bu_{i,\text{min}}, \bu_{i,\text{max}}]}
\left( \bu_{i,k} + \frac{\bxi_{i,k}}{\tau} \right),
\end{equation}
which can be computed by clamping all elements of $\bu_{i,k} + \bxi_{i,k}/\tau$, so that they are within $[\bu_{i,\text{min}}, \bu_{i,\text{max}}]$.
\begin{remark}
\label{MDDDP remark high par}
It follows that Step 2 of MD-DDP is a \textit{highly parallelizable} step, since it can be decomposed w.r.t. i) all agents, ii) states and controls and iii) all time instants. Therefore, it results into $M (2K+1)$ small-dimensional subproblems that can be solved in parallel.
\end{remark}
\begin{algorithm}[t]
\caption{Merged Distributed DDP (MD-DDP)}\label{MD-DDP Algorithm}
\begin{algorithmic}[1] 
\State $\bx_i', \bu_i' \gets$ Solve single-agent problems in parallel $ \forall \ i \in \calV$.
\State \textit{Each agent $i \in \calV$ receives $\bx_j'$ from all $j \in \calN_i \backslash \{i\}$.} 
\State \textbf{Initialize:} 
$\bx_i \leftarrow \bx_i'$, 
$\bu_i \leftarrow \bu_i'$, 
$\tilde{\bx}_i^{\text{a}} \leftarrow [ \{ \bx_j' \}_{j \in \calN_i} ]$,
$\tilde{\bu}_i \leftarrow \bu_i'$, 
$\bz_i^{\text{a}} \leftarrow \tilde{\bx}_i^{\text{a}}$, 
$\bxi_i \leftarrow 0$, 
$\blambda_i \leftarrow 0$, 
$\by_i \leftarrow 0$. 
\While{$n \leq N$}
\State $\bx_i, \bu_i \leftarrow$ Solve \eqref{MD-DDP primal update} in parallel $ \forall \ i \in \calV$.
\State $\tilde{\bx}_i^{\text{a}} \leftarrow$ Solve \eqref{MDDDP safe update 2} in parallel $ \forall \ i \in \calV$, $\forall \ k \in \llbracket 0, K \rrbracket$.
\State $\tilde{\bu}_i \leftarrow$ Solve \eqref{MDDDP safe control update 2} in parallel $ \forall \ i \in \calV$, $\forall \ k \in \llbracket 0, K-1 \rrbracket$.
\State \textit{Each agent $i \in \calV$ receives $\tilde{\bx}_i^j$ from all $j \in \calP_i \backslash \{i\}$.} 
\State $\bz_i \leftarrow$ Update with \eqref{MD-DDP Global state update} in parallel $ \forall \ i \in \calV$.
\State \textit{Each agent $i \in \calV$ receives $\bz_j$ from all $j \in \calN_i \backslash \{i\}$.} 
\State $\bxi_i, \blambda_i, \by_i \leftarrow$ Update \eqref{MD-DDP Dual safe control update}-\eqref{MD-DDP Dual consensus state update} in parallel $ \forall \ i \in \calV$.
\EndWhile
\end{algorithmic}
\end{algorithm}

\textbf{Optimization Step 3: Global Updates}. Similarly with Step 2 of ND-DDP, the global updates will be given by
\begin{equation}
\bz_{i,k}^{n+1} = \frac{1}{|\calP_i|} \sum_{j \in \calP_i} \tilde{\bx}_i^{j,n+1} + \frac{1}{\mu} \by_i^{j,n}.
\label{MD-DDP Global state update}
\end{equation}
Finally, the dual updates are provided by
\begin{subequations}
\begin{align}
\bxi_i^{n+1} & = \bxi_i^n + \tau(\bu_i^{n+1} - \tilde{\bu}_i^{n+1}), 
\label{MD-DDP Dual safe control update}
\\
\blambda_i^{n+1} & = \blambda_i^n + \rho(\bx_i^{n+1} - \tilde{\bx}_i^{n+1}), 
\label{MD-DDP Dual safe state update}
\\
\by_i^{n+1} & = \by_i^n + \mu(\tilde{\bx}_i^{\text{a},n+1} - \bz_i^{\text{a},n+1}).
\label{MD-DDP Dual consensus state update}
\end{align}
\end{subequations}
\vspace{-0.85cm}

\subsection{Algorithm}
The MD-DDP algorithm is illustrated in Alg. \ref{MD-DDP Algorithm}. As in ND-DDP, the single-agent unconstrained problems are initially solved once (Line 1), so that $\bx_i$, $\bu_i$, $\tilde{\bx}_i^{\text{a}}$, $\tilde{\bu}_i$, $\bz_i^{\text{a}}$ are warmstarted (Line 3). During every ADMM iteration, all agents first solve problems \eqref{MD-DDP primal update} in parallel with DDP, to obtain $\bx_i, \bu_i$ (Line 5). Subsequently, problems \eqref{MDDDP safe update 2},\eqref{MDDDP safe control update 2} are solved in parallel for all agents and all time instants (Lines 6,7). Afterwards, every agent $i \in \calV$ receives $\tilde{\bx}_i^j$ from all agents $j \in \calP_i \backslash \{i\}$ (Line 8), so that the global updates \eqref{MD-DDP Global state update} are executed (Line 9). Finally, each agent $i$ gets $\bz_j$ from all $j \in \calN_i \backslash \{i\}$ (Line 10), so that the variables $\bz_i^{\text{a}}$ are constructed and the dual updates can be performed (Line 11). Lines 5-11 repeat until $n$ reaches to $N$.
\begin{remark}
As with ND-DDP, the MD-DDP algorithm is \textit{fully decentralized} since all computations can be performed in parallel among all agents (Lines 1,5-7,9,11), while the required communication steps (Lines 2,8,10) take place in a neighborhood level.
\end{remark}
\begin{remark}[Computational Complexity of MD-DDP] 
\label{MD-DDP comp complexity}
Using a similar justification as in Remark \ref{ND-DDP comp complexity}, the computational complexity of MD-DDP is given by
\begin{equation}
O\big( N \cdot D \cdot K \cdot q_i^3 \big).    
\end{equation}
Note that this is a major computational improvement even compared to ND-DDP, since the AL loop has been eliminated and the DDP subproblems only have a single-agent dimensionality, i.e., the matrix inversion in \eqref{DDP Optimal Control Deviations} has a $O(q_i^3)$ cost. 
\end{remark}

\section{Architectures Improvements and Details}
\label{sec:improvements}
\IEEEpubidadjcol
In this section, we propose two improvements for enhancing the performance of ND-DDP and MD-DDP. First, we show how Nesterov acceleration can be incorporated through the ADMM updates to accelerate the convergence of the algorithms. Second, we propose a decentralized \textit{``agent-specific''} scheme for adapting the ADMM penalty parameters that is suitable for multi-agent optimal control. For the sake of brevity, both are being presented only in the context of MD-DDP. Further details on the algorithms are also provided.

\subsection{Nesterov Acceleration}
A Nesterov accelerated version of MD-DDP is presented here. For more details on ADMM with Nesterov acceleration, the reader is referred to \cite{goldstein2014fast, tang2019distributed}. Let us introduce the Nesterov duplicate variables $\bar{\bx}_i^{\text{a}}$, $\bar{\bu}_i$, $\bar{\bz}$, $\bar{\bxi}_i$, $\bar{\blambda}_i$ and $\bar{\by}_i$. The updates of the modified MD-DDP algorithm are carried out as detailed below. In Step 1 (Local DDP updates), the variables $\bx_i, \bu_i$ are updated through 
\begin{equation*}
\{ \bx_i, \bu_i \}^{n+1} = \argmin \calL(\bx_i, \bu_i, \bar{\bx}_i^{\text{a},n}, \bar{\bu}_i^n, \bar{\bz}^n, \bar{\bxi}_i^n, \bar{\blambda}_i^n, \bar{\by}_i^n)
\end{equation*}
which leads to problems \eqref{MD-DDP primal update}, but with $\bar{\bx}_i$, $\bar{\bu}_i$, $\bar{\blambda}_i$, $\bar{\bxi}_i$ in place of $\tilde{\bx}_i$, $\tilde{\bu}_i$, $\blambda_i$, $\bxi_i$, respectively. Step 2 (Local Safe updates) yields the following updates for $\tilde{\bx}_i^{\text{a}}$, $\tilde{\bu}_i$
\begin{equation*}
\{ \tilde{\bx}_i^{\text{a}}, \tilde{\bu}_i \}^{n+1} = \argmin \calL(\bx_i^{n+1}, \bu_i^{n+1}, \tilde{\bx}_i^{\text{a}}, \tilde{\bu}_i, \bar{\bz}^n, \bar{\bxi}_i^n, \bar{\blambda}_i^n, \bar{\by}_i^n)
\end{equation*}
which results to \eqref{MDDDP safe update 2}, \eqref{MDDDP safe control update 2} using $\bar{\bz}_{i,k}^{\text{a}}$, $\bar{\bxi}_{i,k}$, $\bar{\blambda}_{i,k}$ and $\bar{\by}_{i,k}$ instead of $\bz_{i,k}^{\text{a}}$, $\bxi_{i,k}$, $\blambda_{i,k}$ and $\by_{i,k}$, respectively. Subsequently, Step 3 (Global updates) will have the following form
\begin{equation}
\bar{\bz}_i^{n+1} = \frac{1}{|\calP_i|} \sum_{j \in \calP_i} \tilde{\bx}_i^{j,n+1} + \frac{1}{\mu} \bar{\by}_i^{j,n}.
\end{equation}
The dual updates are provided by
\begin{subequations}
\begin{align}
\bxi_i^{n+1} & = \bar{\bxi}_i^n + \tau(\bu_i^{n+1} - \tilde{\bu}_i^{n+1}),
\label{MD-DDP Nesterov Dual safe control update}
\\
\blambda_i^{n+1} & = \bar{\blambda}_i^n + \rho(\bx_i^{n+1} - \tilde{\bx}_i^{n+1}),
\label{MD-DDP Nesterov Dual safe state update}
\\
\by_i^{n+1} & = \bar{\by}_i^n + \mu(\tilde{\bx}_i^{\text{a},n+1} - \bz_i^{\text{a},n+1}).
\label{MD-DDP Nesterov Dual consensus state update}
\end{align}
\end{subequations}
Finally, the Nesterov duplicate variables are updated as follows
\vspace{-0.4cm}
\begin{subequations}
\begin{align}
\bar{\bx}_i^{\text{a},n+1} & = \tilde{\bx}_i^{\text{a},n+1} + \gamma_n (\tilde{\bx}_i^{\text{a},n+1} - \tilde{\bx}_i^{\text{a},n}),
\\
\bar{\bu}_i^{n+1} & = \tilde{\bu}_i^{n+1} + \gamma_n (\tilde{\bu}_i^{n+1} - \tilde{\bu}_i^n),
\\
\bar{\bz}_i^{n+1} & = \bz_i^{n+1} + \gamma_n (\bz_i^{n+1} - \bz_i^n),
\\
\bar{\bxi}_i^{n+1} & = \bxi_i^{n+1} + \gamma_n (\bxi_i^{n+1} - \bxi_i^n),
\\
\bar{\blambda}_i^{n+1} & = \blambda_i^{n+1} + \gamma_n (\blambda_i^{n+1} - \blambda_i^n),
\\
\bar{\by}_i^{n+1} & = \by_i^{n+1} + \gamma_n (\by_i^{n+1} - \by_i^n),
\end{align}
\end{subequations}
where $\gamma_n = \eta \displaystyle{\frac{\alpha_n - 1}{\alpha_{n+1}}}$, $\{\alpha_n\}_{n=0,1,\dots}$ is the Nesterov sequence
\begin{equation*}
\alpha_{n+1} = \frac{1 + \sqrt{1 + 4 \alpha_n^2}}{2}, \quad \alpha_1 = 1,
\end{equation*}
and $\eta \in [0,1)$ is a tuning parameter. For $\eta = 0$, the algorithm collapses to vanilla MD-DDP.

\subsection{Decentralized Penalty Parameters Adaptation}
\label{sec: arch improvements ppa}
Proper selection of the penalty parameters in ADMM algorithms is significant since it affects how fast they will converge to a satisfying solution, i.e., the primal and dual residuals will reach below some sufficient thresholds \cite{boyd2011distributed}. In general, these parameters are treated as tuning parameters, although several adaptation schemes have been proposed in the literature such as \cite{rockafellar1976monotone, he2000alternating, song2016fast, xu2017adaptiveB}. Most of them rely on steering the total primal and dual residuals ratio to some desired value \cite{rockafellar1976monotone, he2000alternating}, since high or low values of the parameters prioritize the reduction of the primal or dual residuals, respectively. Of course, computing the total residuals of the global problem would require a central node, sacrificing the fully decentralized form of ND-DDP and MD-DDP. In \cite{song2016fast}, the residual balancing idea is extended by using different penalty parameters per node/agent. 

Our observations have shown that schemes of this nature can be particularly useful in optimal control for multi-agent systems, while also admitting an intuitive interpretation. In fact, we take this idea one step further and propose (in the context of MD-DDP) that each agent contains the following diagonal \textit{penalty parameter matrices}
\begin{equation}
\bT_i \in \Rb^{q_i \times q_i}, \ \bP_i \in \Rb^{p_i \times p_i}, \ \bM_i \in \Rb^{\tilde{p}_i \times \tilde{p}_i}, \ i \in \calV,
\end{equation}
with every diagonal element corresponding to a particular control/state component. The intuition here is that not only each agent should be penalized differently based on whether its local constraints are satisfied or not, but moreover, each control/state component should also be penalized separately. This is crucial in the case where the original costs \eqref{Multi-Agent DDP Costs} assign significantly different weights to these components, so the penalty parameters should be attuned to such discrepancies. 

The resulting modified updates for MD-DDP are provided in Appendix \ref{sec: PPA-MD-DDP}. 
%
At each ADMM iteration, the penalty parameter matrices are updated as follows
\begin{subequations}
\begin{align}
\bT_i^{n+1} &= a_{1,i}^{n+1} \bT_i^0, \\  
\bP_i^{n+1} &= a_{2,i}^{n+1} \bP_i^0, \\ 
\bM_i^{n+1} &= a_{3,i}^{n+1} \bM_i^0,
\end{align}
\label{parameter adaptation}%
\end{subequations}
where $\bT_i^0$, $\bP_i^0$, $\bM_i^0$ are their initially assigned values.
The update rules for the parameters $a_{b,i}^{n+1}, \ b=1,2,3$, depend on the corresponding primal and dual residuals ratios, with detailed expressions provided in Appendix \ref{sec: PPA details}. 

Of particular interest is the quite common case, where the costs \eqref{Multi-Agent DDP Costs} are quadratic, i.e., 
\begin{align}
J_i & = \sum_{k=0}^{K-1} 
\Big[ (\bx_{i,k} - \bx_i^\text{g}) \T \bQ_i (\bx_{i,k} - \bx_i^\text{g}) +  \bu_{i,k}\T \bR_i \bu_{i,k} 
\Big]
\nonumber
\\
& ~~~~~~~~~~~ + (\bx_{i,K} - \bx_i^\text{g}) \T \bQ_i^{\text{f}} (\bx_{i,K} - \bx_i^\text{g}), 
\label{quadratic agent cost}
\end{align}
with $\bQ_i, \ \bR_i$ being diagonal matrices and  $\bx_i^\text{g}$ being the target state of the $i$-th agent. In this case, a well-suited initialization is $\bT_i^0 = c_1 \cdot \bR_i$, $\bP_i^0 = c_2 \cdot \bQ_i$, $\bM_i^0 = c_3 \cdot \bdiag(\{\bQ_i\}_{j \in \calN_i})$, where $c_1, c_2, c_3 > 0$ with reasonable values, say $0.1$ to $10$. Choosing values $c_1, c_2, c_3 > 1$ would assign more importance to consensus in the MD-DDP updates (see Appendix \ref{sec: PPA-MD-DDP}), while the opposite would assign more importance to the original costs. The intuition behind this choice is that each state/control component of each agent should be penalized by a symmetric amount as in the original costs. Subsequently, depending on which constraints are violated or not, the updates \eqref{parameter adaptation} will lead the penalty parameters to appropriate values, while still respecting the original relative weighting that was assigned to each state/control component by the cost matrices $\bQ_i, \ \bR_i$. It is straightforward to apply the same ideas in ND-DDP as well, by replacing the parameters $\rho$ and $\mu$ in Section \ref{sec:Nested_DDP}, with matrices $\bP_i \in \Rb^{\tilde{p}_i \times \tilde{p}_i}$ and $\bM_i \in \Rb^{\tilde{q}_i \times \tilde{q}_i}$ for every agent $i$.

\subsection{Convergence}
DDP is known to admit a quadratic convergence rate to a locally optimal solution \cite{yakowitz1984computational, liao1991convergence}. 
In addition, convergence guarantees for ADMM are well-established in convex optimization problems \cite{boyd2011distributed}. While convergence results on problems with nonconvex constraints are limited in the literature and require significant modifications that reduce computational efficiency \cite{sun2019two, tang2021fast, zhang2020improved}, extensive computational experience has shown that ADMM works well in practice for nonconvex problems \cite{fortin2000augmented, glowinski1989augmented, jiang2014alternating, magnusson2015distributed, erseghe2014distributed, shen2014augmented, xu2016empirical}. In fact, after thorough testing, no divergence cases have been observed for both vanilla ND-DDP and MD-DDP. For their Nesterov accelerated versions, a sufficient tuning of parameter $\eta$ is necessary to ensure that the algorithms will not diverge \cite{tang2019distributed}. 
%
\subsection{DDP Details}
The performance of DDP can be further improved with two main modifications, which we adopt in the proposed frameworks. First, during the backward pass of DDP, we perform a sufficient regularization on the $\bQ_{\bu\bu}$ matrix such that it is ensured that it is a positive definite matrix using the adaptation rule proposed in \cite{liao1991convergence}. Second, at every DDP forward pass, we use the following modified version of \eqref{DDP Optimal Control Deviations},
\begin{equation}
\delta \bu_k^* = \alpha \bk_k + \bK_k \delta \bx_k,
\label{DDP Optimal Control Deviations Linesearch}
\end{equation}
where a line search is performed to find the optimal $\alpha$ that leads to the highest cost reduction \cite{jacobson1970differential}. In all cases, DDP is warmstarted with the previous solution of each agent.

\section{Simulation Results}
\label{sec:Simulation_Results}
In this section, the efficacy of the proposed methods is verified through extensive simulation results. We start by demonstrating their performance in various challenging multi-vehicle problems and gradually increase to large-scale scenarios with thousands of vehicles. Subsequently, the capability of the algorithms to handle more complex dynamics in 3D space is illustrated by successfully testing them in multi-UAV control problems with up to hundreds of drones. An ablative analysis which shows how the improvements proposed in Section \ref{sec:improvements} can enhance performance is provided afterwards. Finally, the superior scalability of ND-DDP and MD-DDP against popular multi-agent optimal control alternatives such as centralized/decentralized SQP and centralized DDP is highlighted for large-scale problems. In all results, only the first-order expansions of the dynamics are used. For all tasks, the results of both MD-DDP and ND-DDP are included in the supplementary video\footnote{\url{https://youtu.be/tluvENcWldw}}. In the main paper, due to space limitations, only snapshots using MD-DDP are presented.

\setlength{\textfloatsep}{0pt}

\begin{figure*}[!t]
\centering
\hfil
\subfloat{
\begin{tikzpicture}
    \node[anchor=south west,inner sep=0] at (0,0){    \includegraphics[width=0.28\textwidth, trim={0cm 0cm 0cm 0cm},clip]{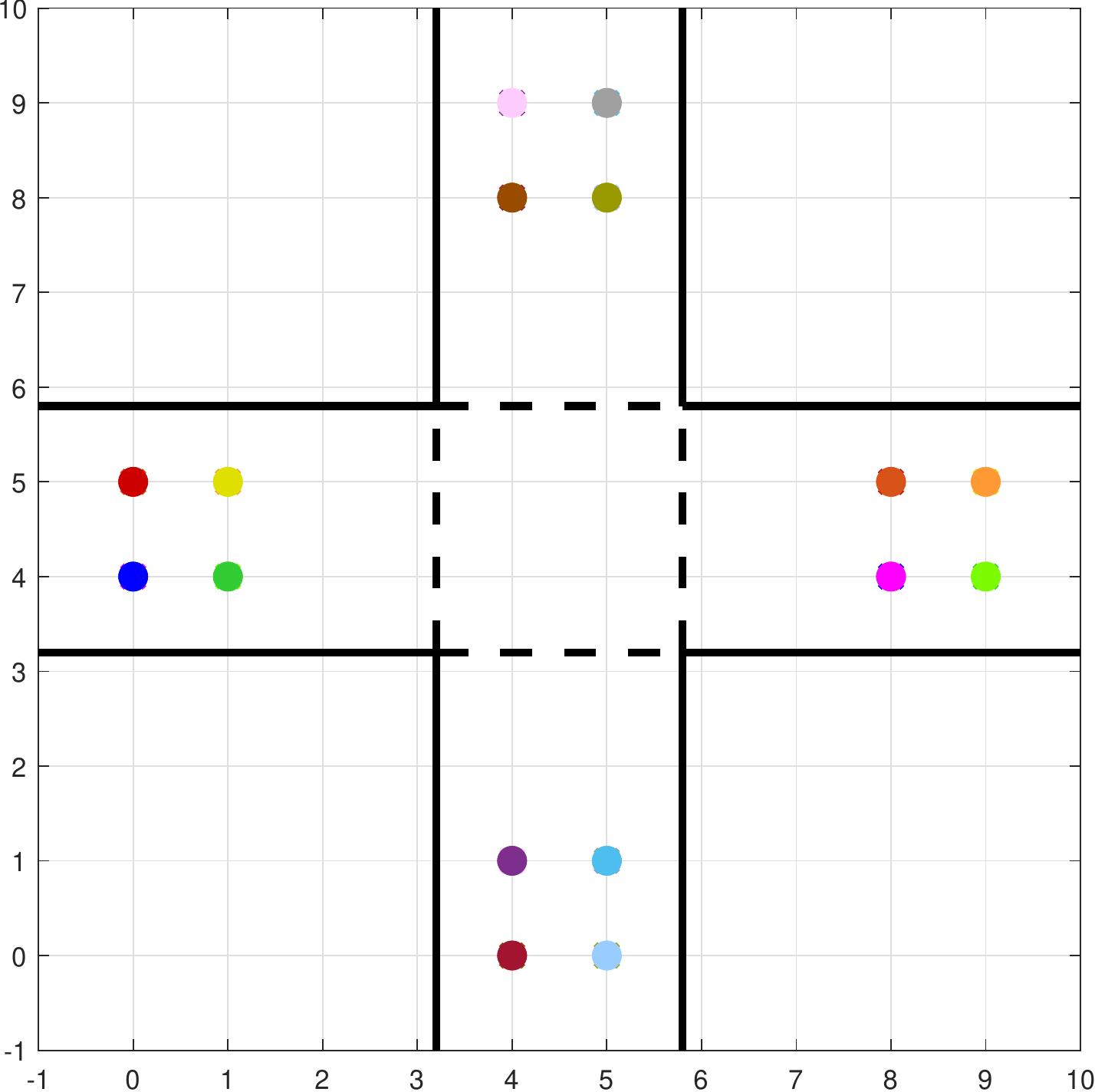}};
    \node[align=center, text=NavyBlue] (c) at (4.48, 0.45) {$k = 0$};
\end{tikzpicture}
\label{fig_inter_1}}
\hfil
\subfloat{
\begin{tikzpicture}
    \node[anchor=south west,inner sep=0] at (0,0){    \includegraphics[width=0.28\textwidth, trim={0cm 0cm 0cm 0cm},clip]{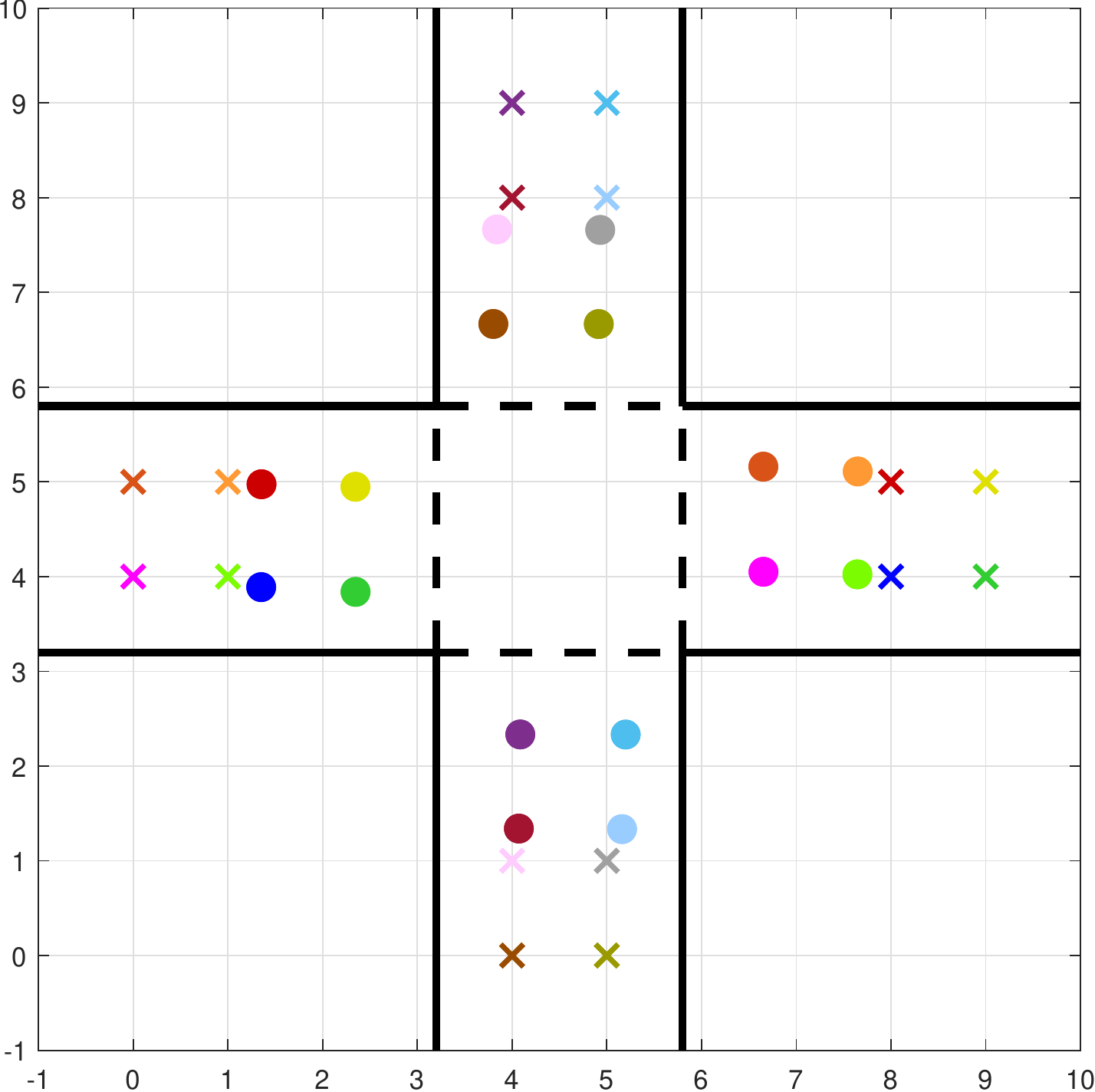}};
    \node[align=center, text=NavyBlue] (c) at (4.4, 0.45) {$k = 15$};
\end{tikzpicture}
\label{fig_inter_2}}
\hfil
\subfloat{
\begin{tikzpicture}
    \node[anchor=south west,inner sep=0] at (0,0){    \includegraphics[width=0.28\textwidth, trim={0cm 0cm 0cm 0cm},clip]{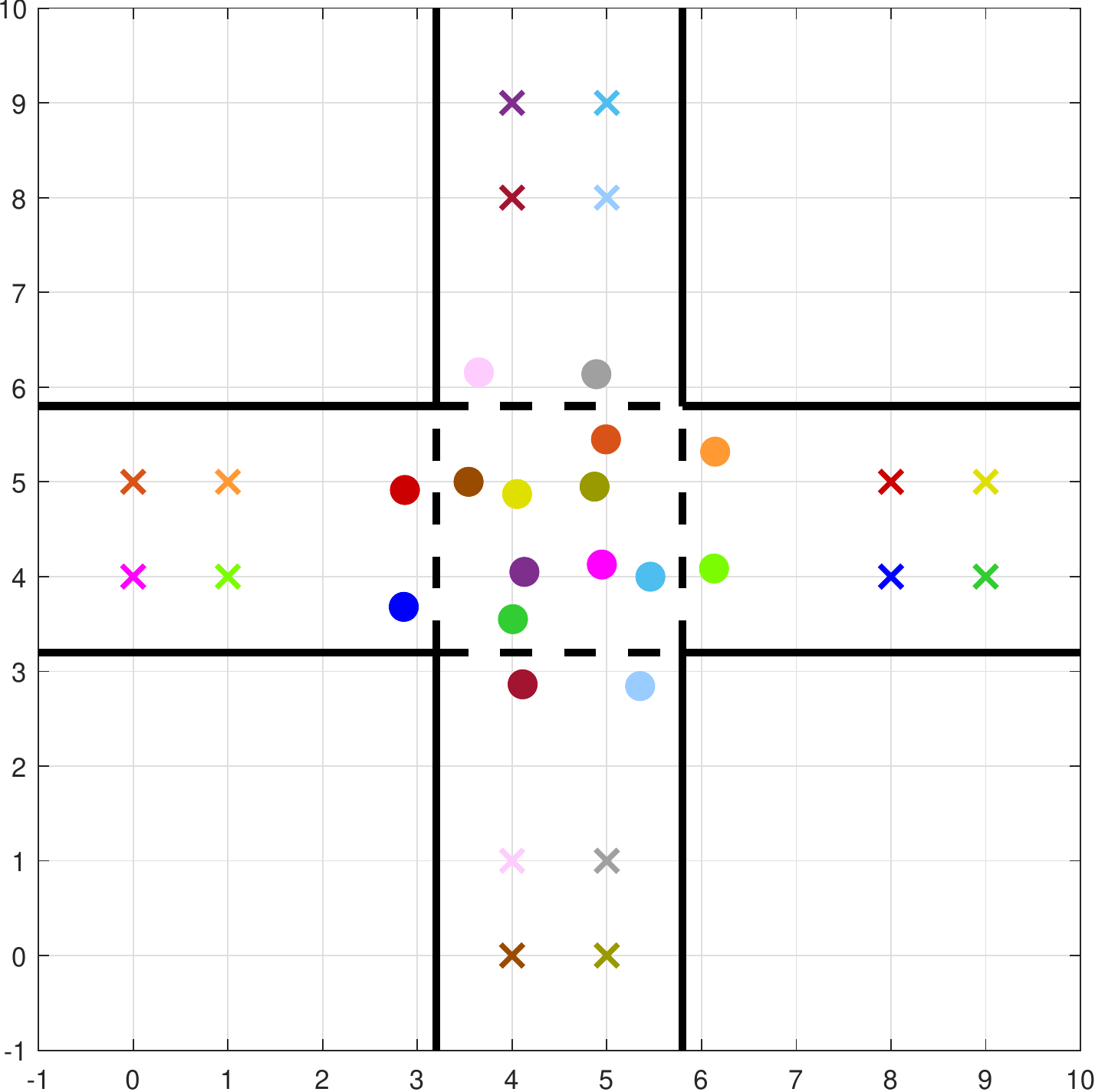}};
    \node[align=center, text=NavyBlue] (c) at (4.4, 0.45) {$k = 30$};
\end{tikzpicture}
\label{fig_inter_3}}
\hfil
\\
\hfil
\subfloat{
\begin{tikzpicture}
    \node[anchor=south west,inner sep=0] at (0,0){    \includegraphics[width=0.28\textwidth, trim={0cm 0cm 0cm 0cm},clip]{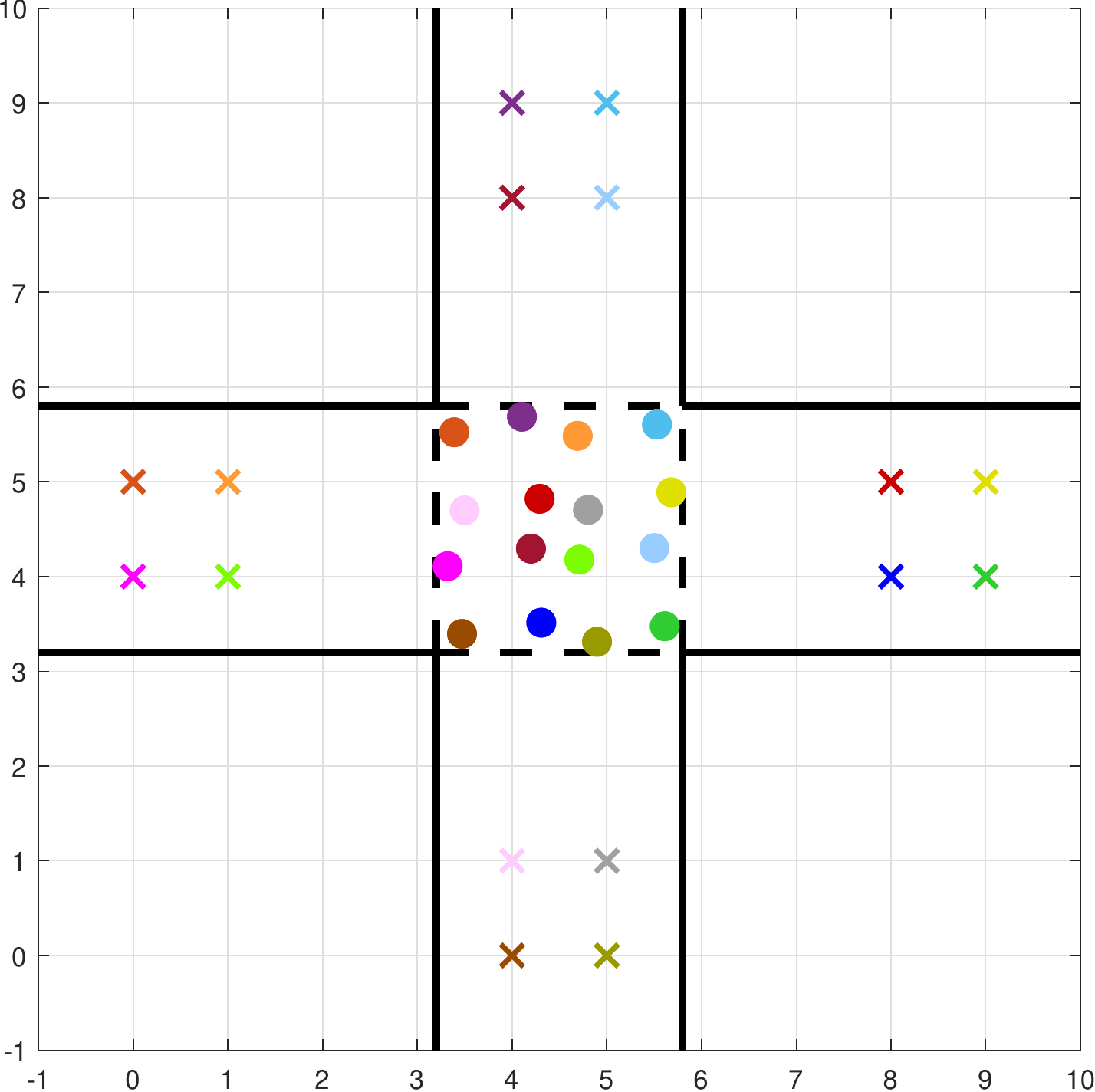}};
    \node[align=center, text=NavyBlue] (c) at (4.4, 0.45) {$k = 40$};
\end{tikzpicture}
\label{fig_inter_4}}
\hfil
\subfloat{
\begin{tikzpicture}
    \node[anchor=south west,inner sep=0] at (0,0){    \includegraphics[width=0.28\textwidth, trim={0cm 0cm 0cm 0cm},clip]{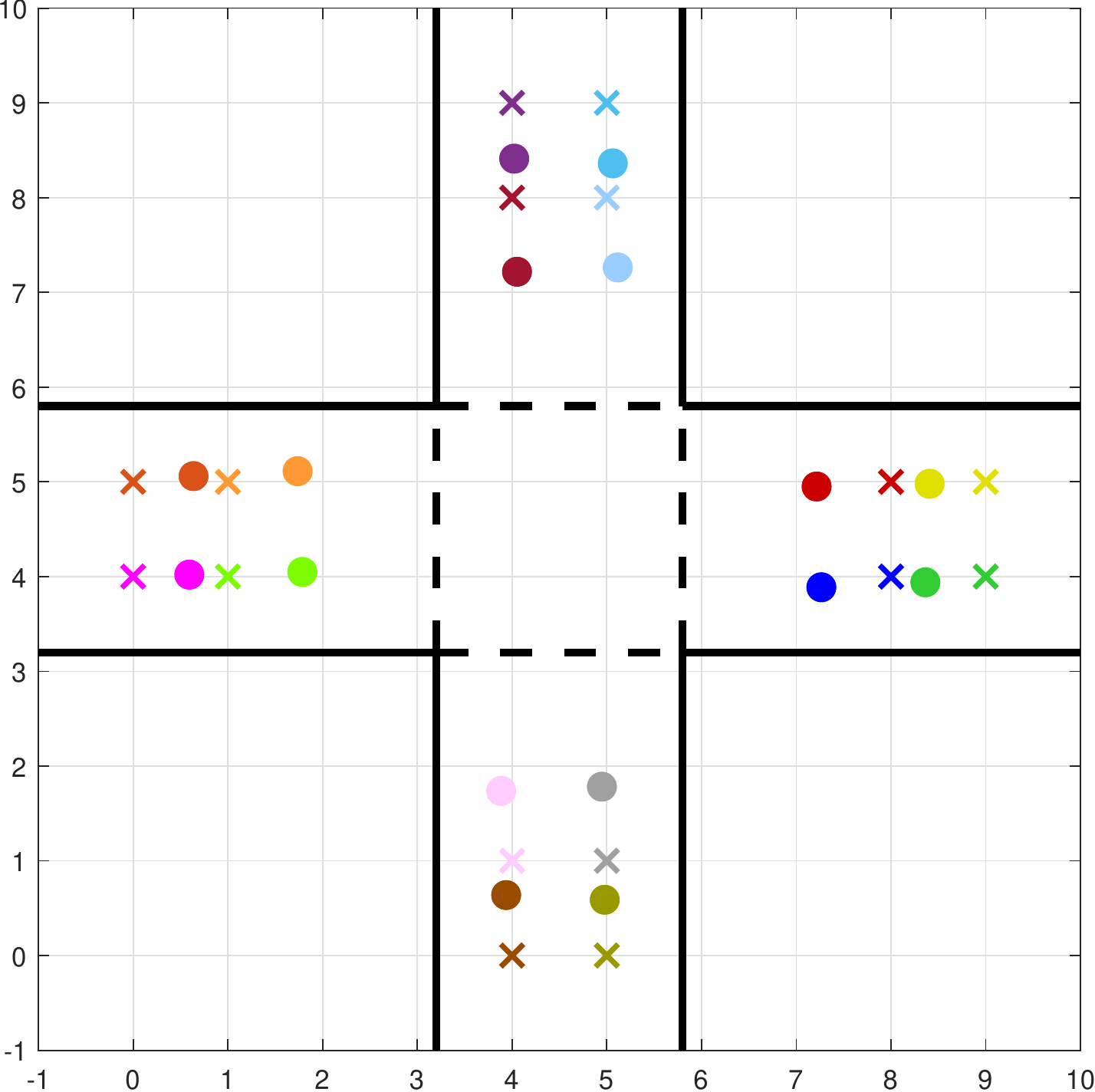}};
    \node[align=center, text=NavyBlue] (c) at (4.4, 0.45) {$k = 75$};
\end{tikzpicture}
\label{fig_inter_5}}
\hfil
\subfloat{
\begin{tikzpicture}
    \node[anchor=south west,inner sep=0] at (0,0){    \includegraphics[width=0.28\textwidth, trim={0cm 0cm 0cm 0cm},clip]{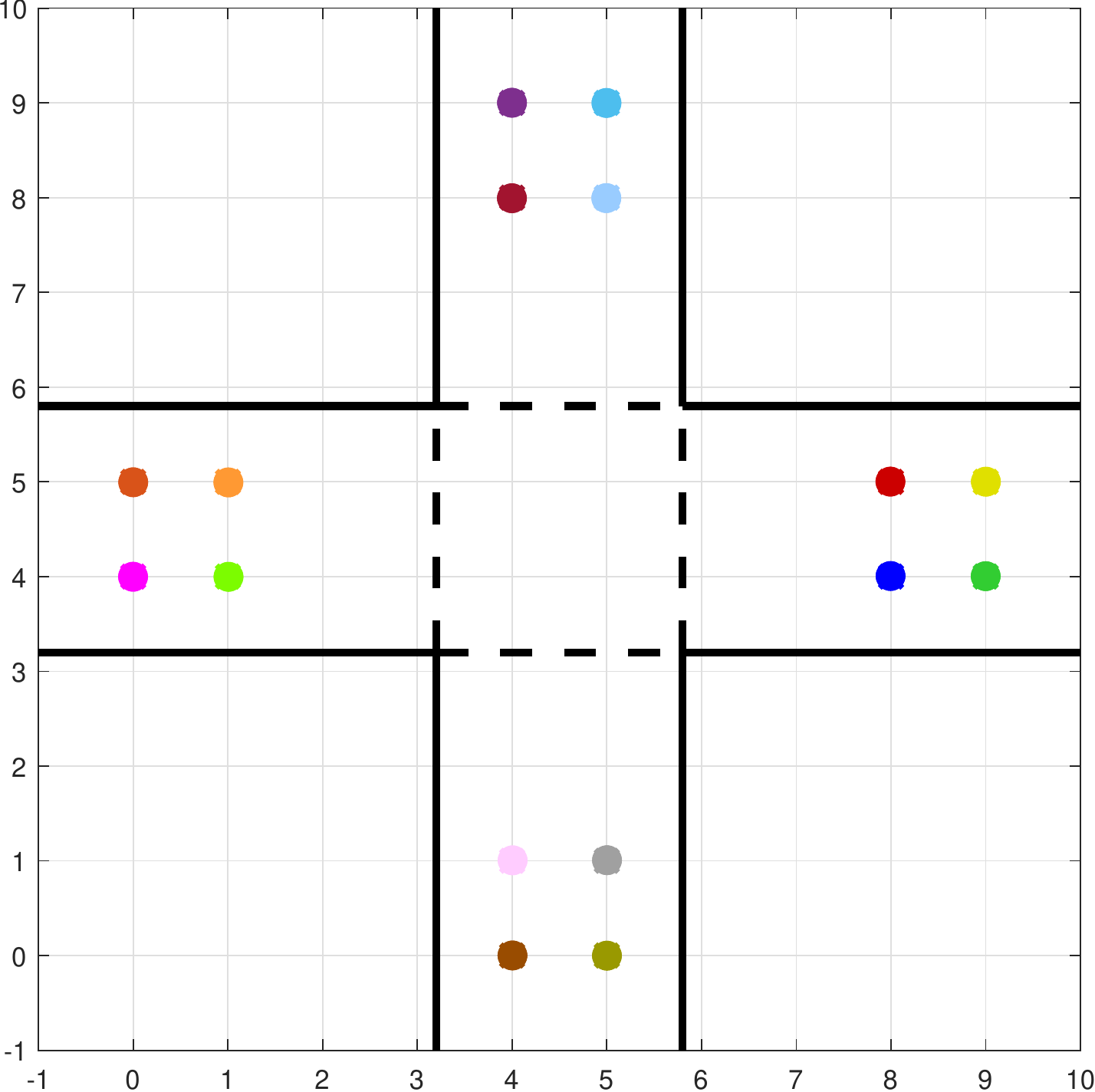}};
    \node[align=center, text=NavyBlue] (c) at (4.32, 0.45) {$k = 150$};
\end{tikzpicture}
\label{fig_inter_6}}
\hfil
\caption{Multi-vehicle intersection task with MD-DDP including $16$ cars: Snapshots at different time instants. Each `x' marker indicates the target of the car with the same color. The black lines indicate the lane bounds of each team of cars.}
\label{fig_inter}
\end{figure*}
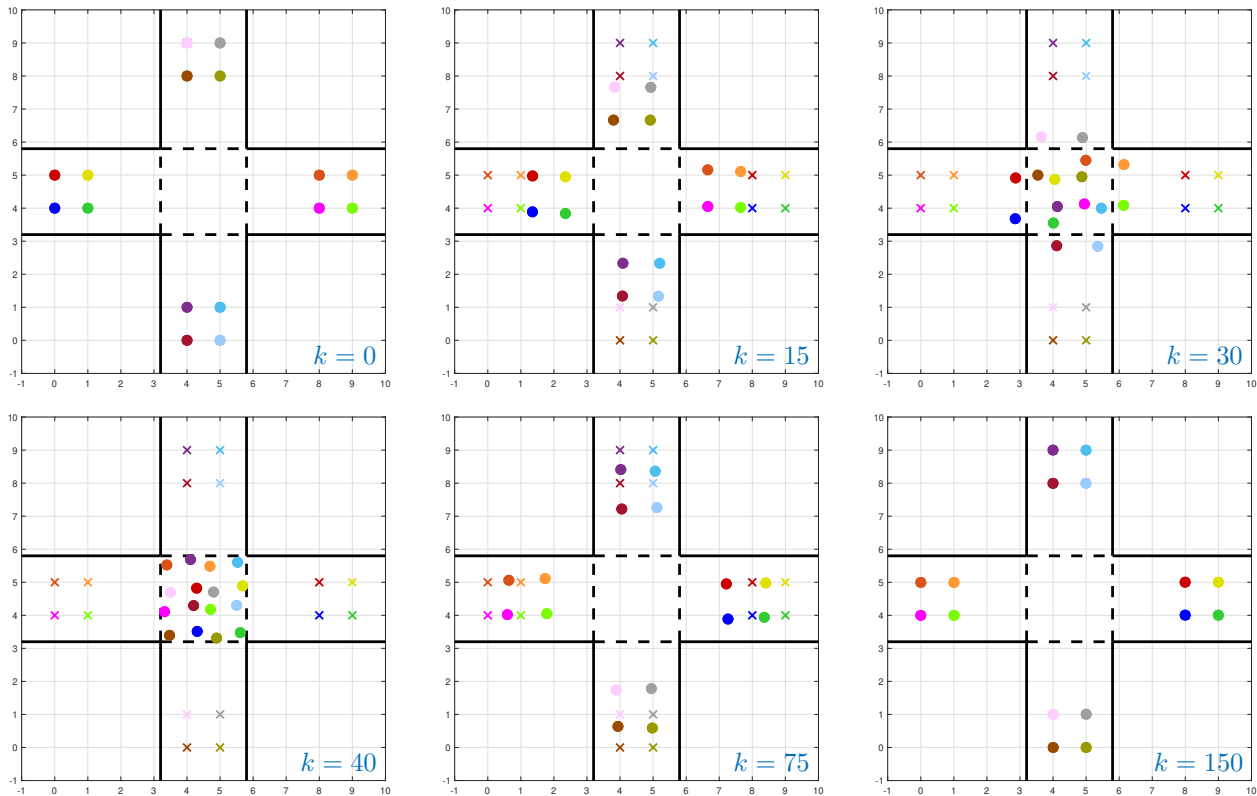

\subsection{Multi-Vehicle Control}
\label{sec: sim car}
We consider a problem setup where each agent $i$ has Dubin's car dynamics with state $\bx_{i,k} = \begin{bmatrix}
\mathrm{x}_{i,k} & \mathrm{y}_{i,k} & \theta_{i,k} & v_{i,k}
\end{bmatrix}\T \in \Rb^4
$ 
and control input 
$\bu_{i,k} = \begin{bmatrix}
a_{i,k} & \omega_{i,k}
\end{bmatrix}\T \in \Rb^2
$, 
where $(\mathrm{x}_{i,k}, \mathrm{y}_{i,k})$ is the 2D position, $\theta_{i,k}$ is the angle, $v_{i,k}, \ \omega_{i,k}$ are the linear and angular velocities and $a_{i,k}$ is the linear acceleration of vehicle $i$ at time instant $k$. For all tasks, we use time step $dt = 0.02$s and time horizon $K = 150$, unless specified otherwise. Each agent's cost is of the form \eqref{quadratic agent cost} with $\bQ_i = \diag (30,30,0,6)$, $\bR_i = \diag(0.5, 0.5)$ and $\bQ_i^{\text{f}} = \diag(100, 100, 0, 100)$. All cars are subject to the following box control constraints 
\begin{equation}
- a_{\text{max}} \leq a_{i,k} \leq a_{\text{max}}, \quad 
- \omega_{\text{max}} \leq \omega_{i,k} \leq \omega_{\text{max}},
\end{equation}
%
with 
$a_{\text{max}} = 10 \text{m}/\text{s}^2$ and $\omega_{\text{max}} = 30^{\circ}/\text{s}$. The following box state constraints must also be satisfied
%
\begin{align}
\mathrm{x}_{\text{min}} \leq &\mathrm{x}_{i,k} \leq \mathrm{x}_{\text{max}}, \quad
\mathrm{y}_{\text{min}} \leq \mathrm{y}_{i,k} \leq \mathrm{y}_{\text{max}}, 
\label{car state box y}
\\
& ~~ - v_{\text{max}} \leq v_{i,k} \leq v_{\text{max}},
\label{car state box v}
\end{align}
where $\mathrm{x}_{\text{min/max}}$, $\mathrm{y}_{\text{min/max}}$ are the field bounds for each task and $v_{\text{max}} = 10 \text{m}/\text{s}$. Furthermore, all vehicles are subject to the obstacle avoidance constraints
\begin{equation}
\| \bp_{i,k} - \bp_{o} \|_2 
\geq r_o + d_o, 
\ o \in \calO, \ i \in \calV,
\label{obs avoidance constraints}
\end{equation}
where $\bp_{i,k} = 
\begin{bmatrix}
\mathrm{x}_{i,k}, \mathrm{y}_{i,k}
\end{bmatrix}\T$,
while $\bp_o$, $r_o$ and $d_o$ are the position, radius and desired distance from obstacle $o \in \calO$, and $\calO$ is the set of all obstacles. For all obstacles, we set $d_o = 0.3 \text{m}$.
Lastly, the following inter-agent collision avoidance and connectivity maintenance constraints are also enforced between each vehicle $i$ and its neighbors
\begin{align}
\| \bp_{i,k} - \bp_{j,k} \|_2  & \geq d_{\text{col}}, 
\ j \in \calN_i \backslash \{i\}, \ i \in \calV,
\label{coll avoidance constraints}
\\
\| \bp_{i,k} - \bp_{j,k} \|_2  & \leq d_{\text{con}}, 
\ j \in \calN_i \backslash \{i\}, \ i \in \calV,
\label{conn maint constraints}
\end{align}
with $d_{\text{col}} = 0.3 \text{m}$ and $d_{\text{con}} = 2.0 \text{m}$. While constraints \eqref{coll avoidance constraints} ensure that the cars will not collide with each other, the ones in \eqref{conn maint constraints} force neighboring cars to stay within a close distance. Note that \eqref{obs avoidance constraints}-\eqref{conn maint constraints} can be handled directly through the local ALs in ND-DDP. For MD-DDP, we opt to linearize them \cite[Section III.A]{augugliaro2012generation} at every ADMM iteration $n$, around the trajectories $\bx_i^n$ obtained from Step 1. Hence, problems \eqref{MDDDP safe update 2} will be small-dimensional QPs that can be solved very fast - especially considering Remark \ref{MDDDP remark high par}. An additional communication step will be needed in MD-DDP between Steps 1 and 2 where every agent $i$ must send $\bx_i^n$ to agents $j \in \calP_i \backslash \{i\}$, so that subsequently, each agent $i$ linearizes the constraints \eqref{coll avoidance constraints} and \eqref{conn maint constraints} around the trajectories $\bx_j^n, \ j \in \calN_i$. In Sections \ref{sec: sim car} and \ref{sec: sim drone}, we use the vanilla versions of ND-DDP and MD-DDP with constant agent-specific penalty parameter matrices - but without parameter adaptation or Nesterov acceleration. For ND-DDP, we set $\bM_i = 2 \cdot \bdiag(\{\bR_i\}_{j \in \calN_i})$, $\bP_i = 8 \cdot \bdiag(\{\bQ_i\}_{j \in \calN_i})$ and for MD-DDP, we assign $\bT_i = 2 \cdot \bR_i$, $\bP_i = 8 \cdot \bQ_i$, $\bM_i = 8 \cdot \bdiag(\{\bQ_i\}_{j \in \calN_i})$ for every vehicle. The maximum number of ADMM iterations is set to $N=50$ for ND-DDP and $N=200$ for MD-DDP. 

\setlength{\textfloatsep}{0pt}

\begin{figure*}[!t]
\centering
\hfil
\subfloat{
\begin{tikzpicture}
    \node[anchor=south west,inner sep=0] at (0,0){    \includegraphics[width=0.28\textwidth, trim={0cm 0cm 0cm 0cm},clip]{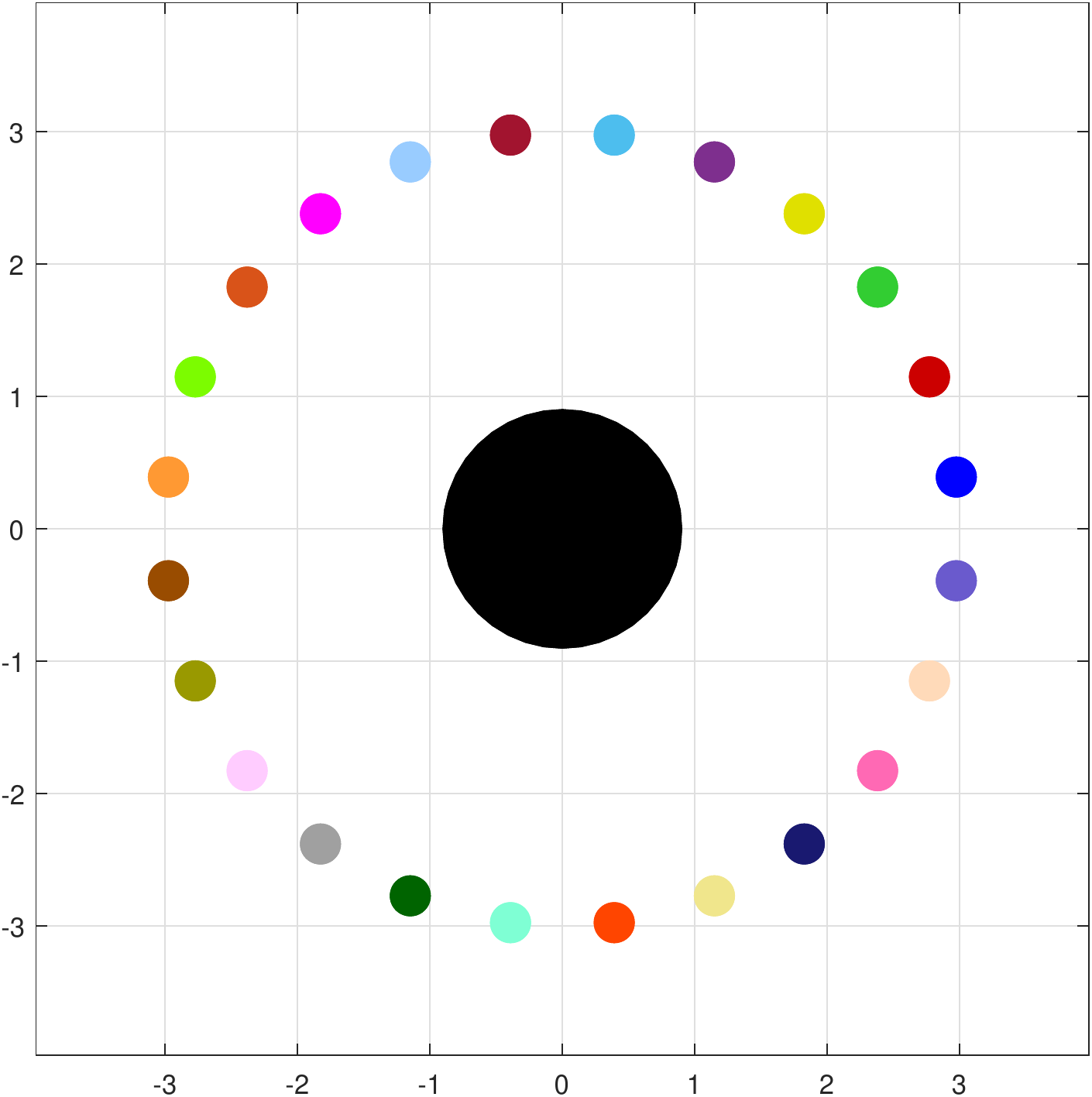}};
    \node[align=center, text=NavyBlue] (c) at (4.48, 0.45) {$k = 0$};
\end{tikzpicture}
\label{fig_swapp_1}}
\hfil
\subfloat{
\begin{tikzpicture}
    \node[anchor=south west,inner sep=0] at (0,0){    \includegraphics[width=0.28\textwidth, trim={0cm 0cm 0cm 0cm},clip]{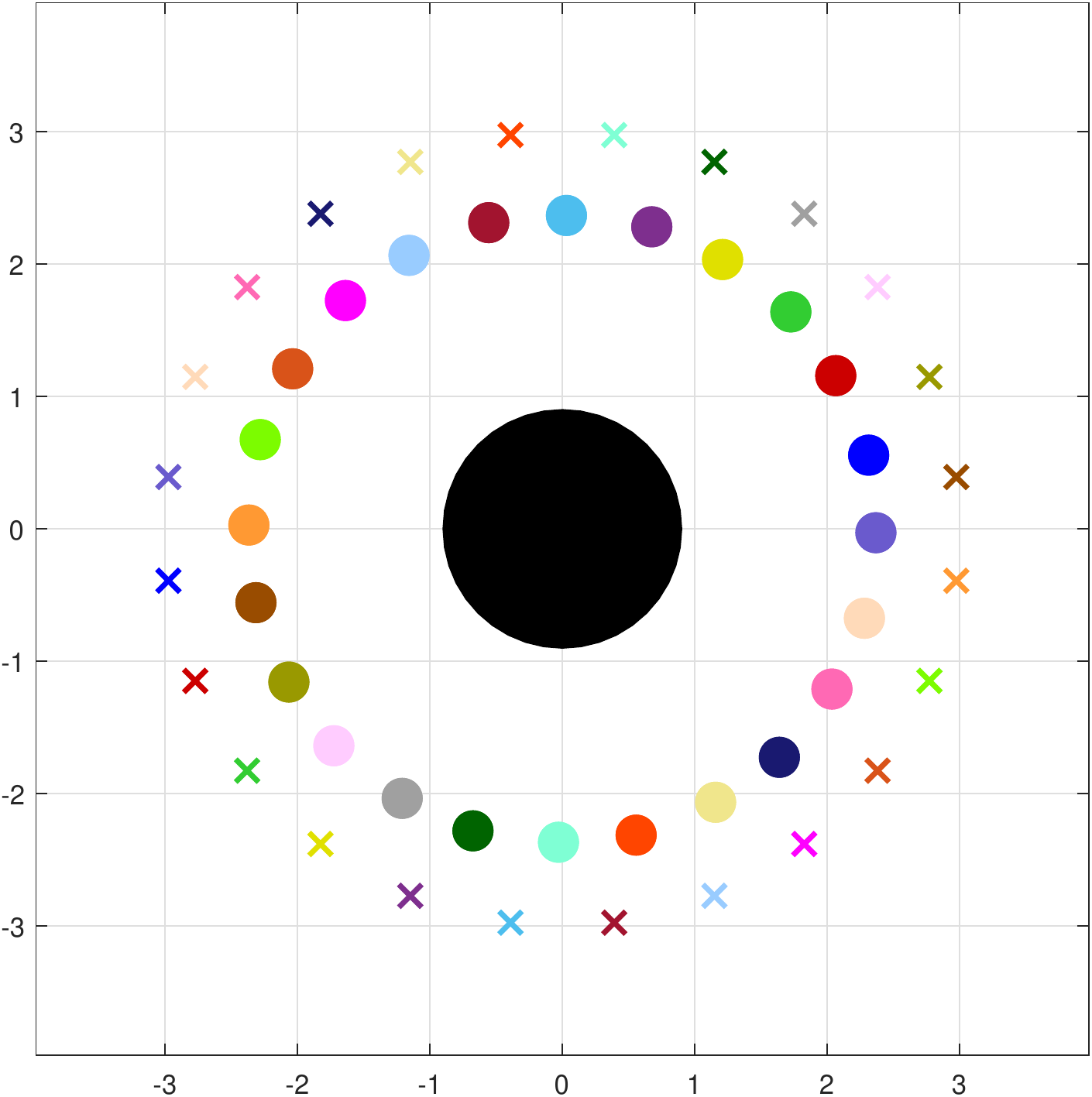}};
    \node[align=center, text=NavyBlue] (c) at (4.4, 0.45) {$k = 20$};
\end{tikzpicture}
\label{fig_swapp_2}}
\hfil
\subfloat{
\begin{tikzpicture}
    \node[anchor=south west,inner sep=0] at (0,0){    \includegraphics[width=0.28\textwidth, trim={0cm 0cm 0cm 0cm},clip]{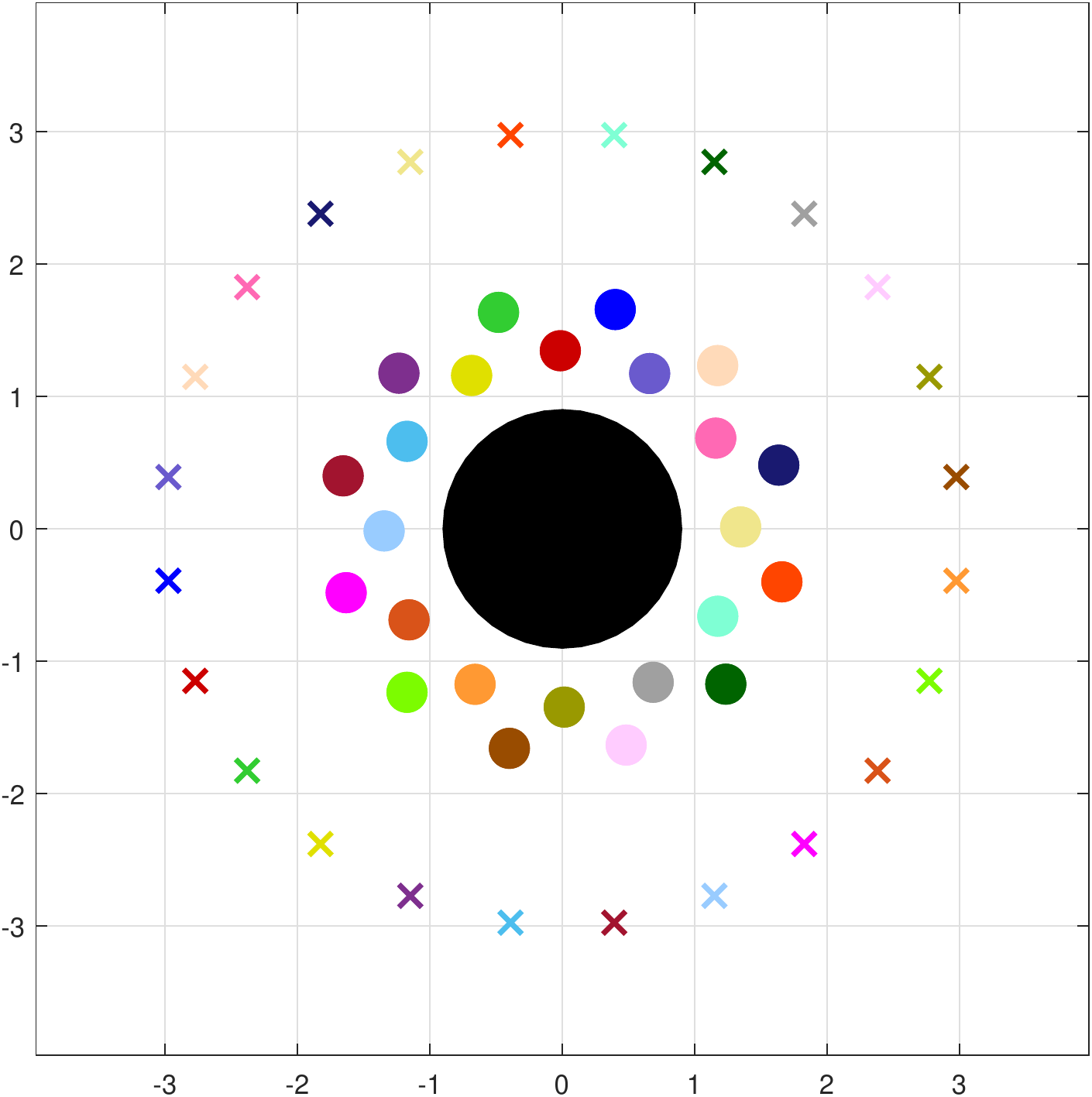}};
    \node[align=center, text=NavyBlue] (c) at (4.4, 0.45) {$k = 40$};
\end{tikzpicture}
\label{fig_swapp_3}}
\hfil
\\
\hfil
\subfloat{
\begin{tikzpicture}
    \node[anchor=south west,inner sep=0] at (0,0){    \includegraphics[width=0.28\textwidth, trim={0cm 0cm 0cm 0cm},clip]{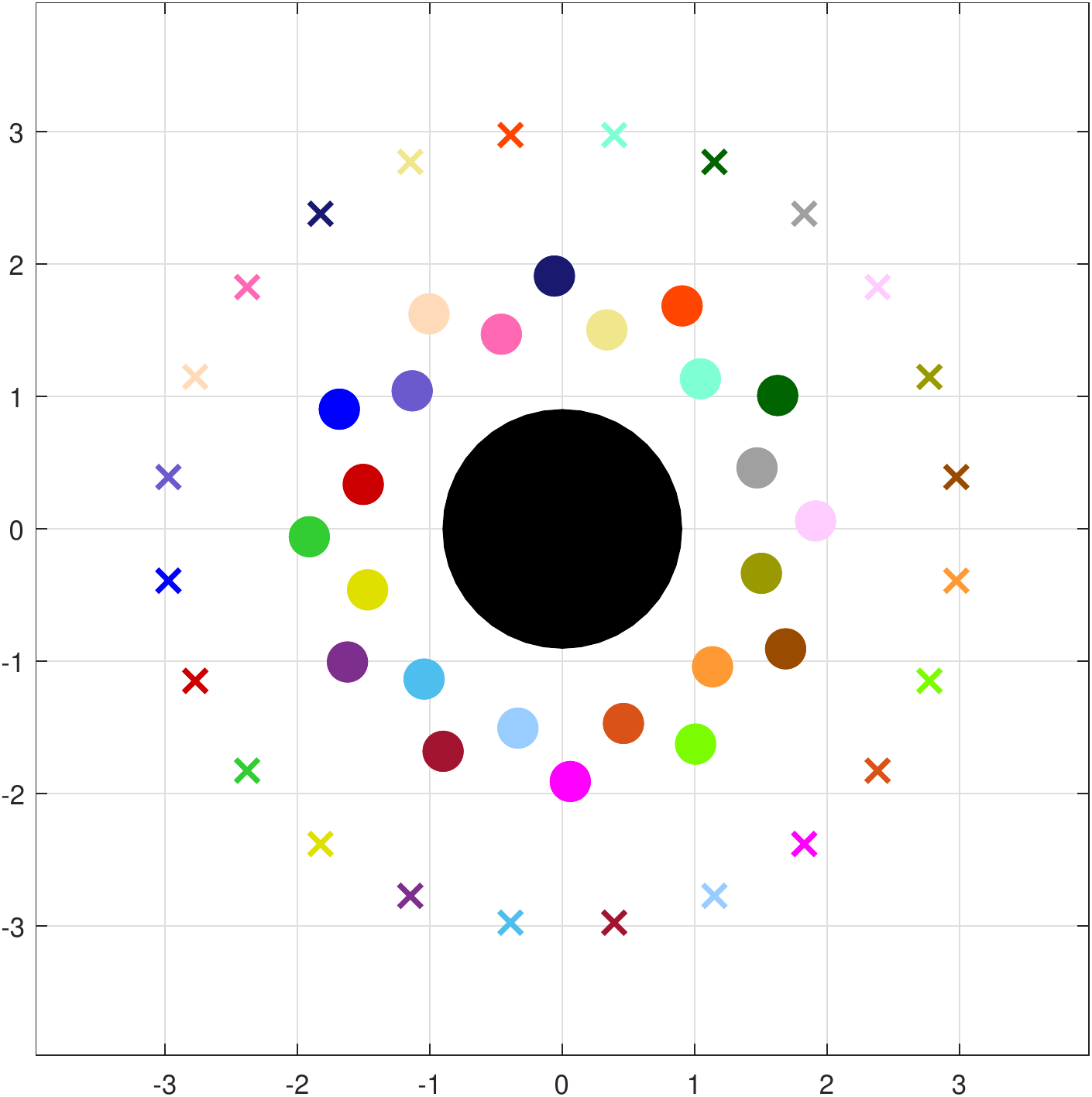}};
    \node[align=center, text=NavyBlue] (c) at (4.4, 0.45) {$k = 60$};
\end{tikzpicture}
\label{fig_swapp_4}}
\hfil
\subfloat{
\begin{tikzpicture}
    \node[anchor=south west,inner sep=0] at (0,0){    \includegraphics[width=0.28\textwidth, trim={0cm 0cm 0cm 0cm},clip]{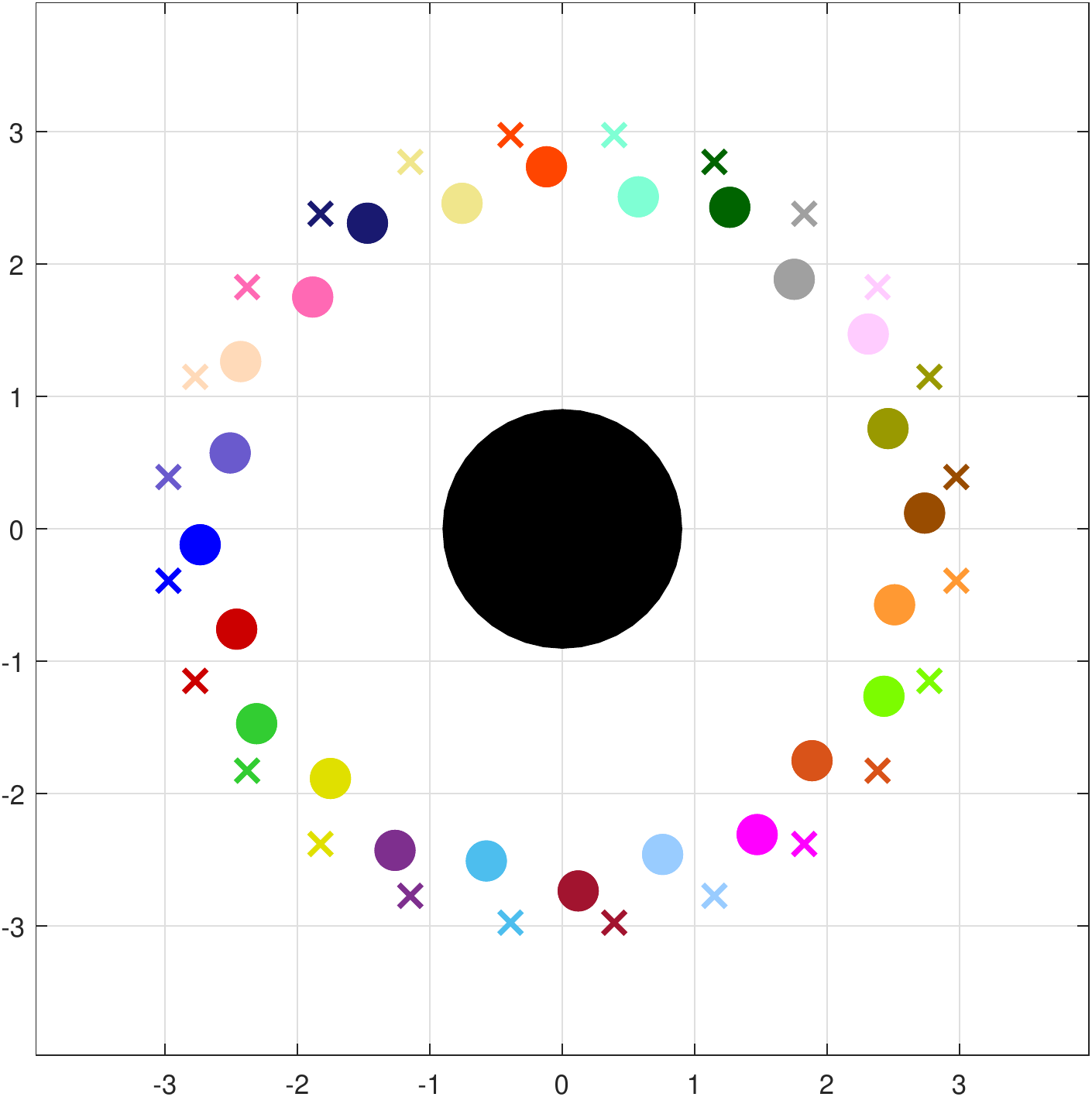}};
    \node[align=center, text=NavyBlue] (c) at (4.4, 0.45) {$k = 90$};
\end{tikzpicture}
\label{fig_swapp_5}}
\hfil
\subfloat{
\begin{tikzpicture}
    \node[anchor=south west,inner sep=0] at (0,0){    \includegraphics[width=0.28\textwidth, trim={0cm 0cm 0cm 0cm},clip]{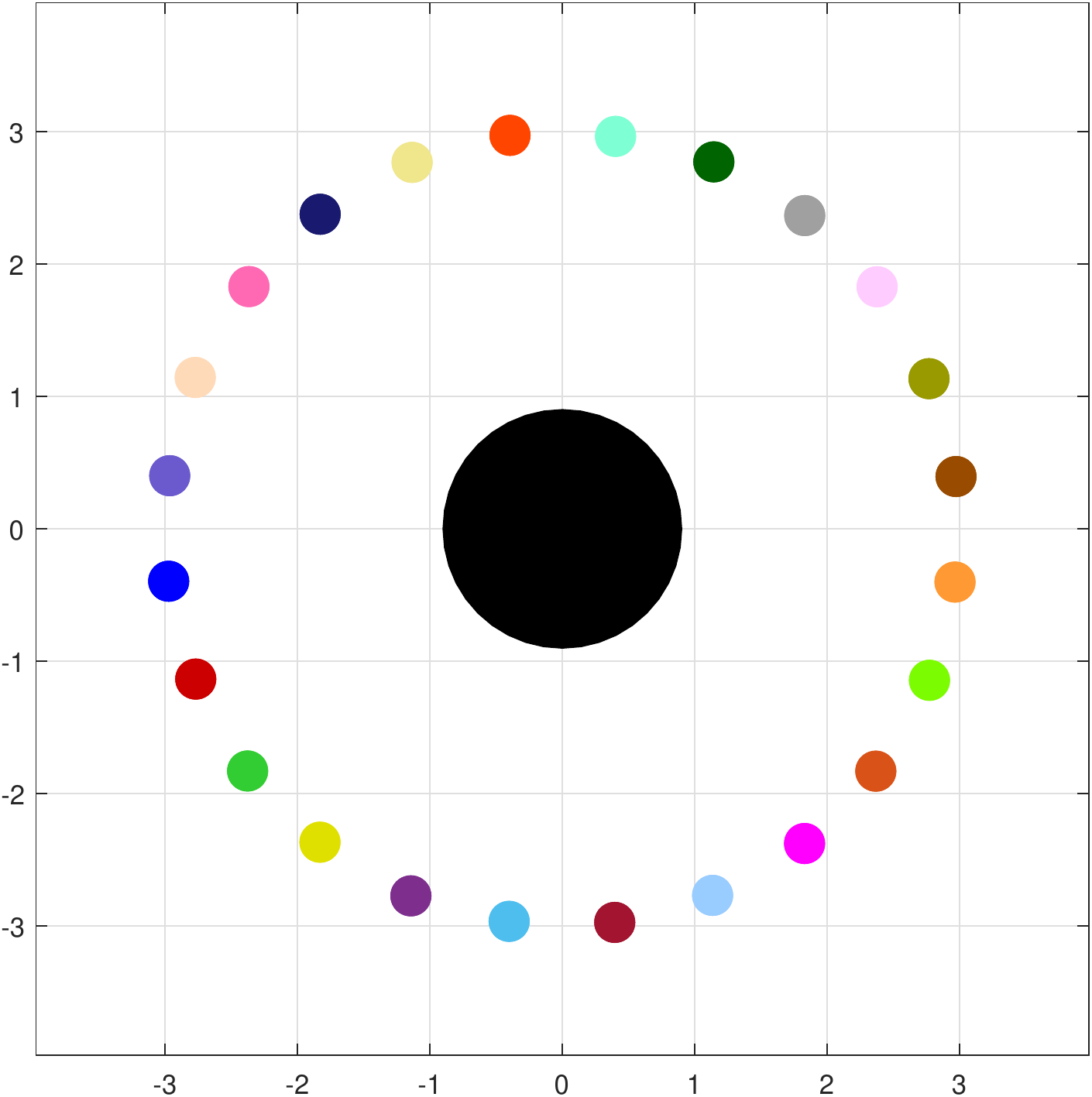}};
    \node[align=center, text=NavyBlue] (c) at (4.32, 0.45) {$k = 150$};
\end{tikzpicture}
\label{fig_swapp_6}}
\caption{Multi-vehicle ``circle swapping'' task with MD-DDP including $24$ cars: Snapshots at different time instants. Obstacles are displayed with black circles.}
\label{fig_swapp}
\end{figure*}

In the first task (Fig. \ref{fig_inter}), four groups of 4 cars are initialized with velocities $v_{i,0} = 3 \text{m/s}$, while moving towards an intersection with a high chance of colliding. Except for avoiding collisions with all other cars ($\calN_i = \calV$) and reaching their targets, the vehicles are also required to remain within their lanes. Since all cars are neighbors here, constraints \eqref{conn maint constraints} can be omitted. Figure \ref{fig_inter} shows six snapshots of the vehicles motion using the MD-DDP optimal solutions. All cars are able to reach their goals while maintaining their safety.
Note that in cases where the neighborhood size is close to the total number of agents, the computational benefit of ND-DDP against centralized AL-DDP decreases.
On the contrary, MD-DDP is immune to such choices, since the local DDP problems in Step 1 always maintain a single-agent dimensionality, irrespective of the neighborhood size (see Remark \ref{MD-DDP comp complexity}).

In the next task (Fig. \ref{fig_swapp}), 24 cars are initialized in a circle formation. Their goal is to swap positions with the diametrically opposite ones while avoiding each other and an obstacle in the center. Every car has a neighborhood size of $|\calN_i|=5$. The optimal car trajectories acquired with MD-DDP are shown in Figure \ref{fig_swapp}. All cars successfully complete the task while staying safe. A quite interesting policy emerges once the vehicles reach to consensus, as they ``agree'' to rotate in two circles of different radii around the obstacle, so that they manage to avoid collisions.

\begin{figure*}[!t]
\centering
\hfil
\subfloat{
\begin{tikzpicture}
    \node[anchor=south west,inner sep=0] at (0,0){    \includegraphics[width=0.48\textwidth, trim={0.0cm 0.0cm 0cm 0cm},clip]{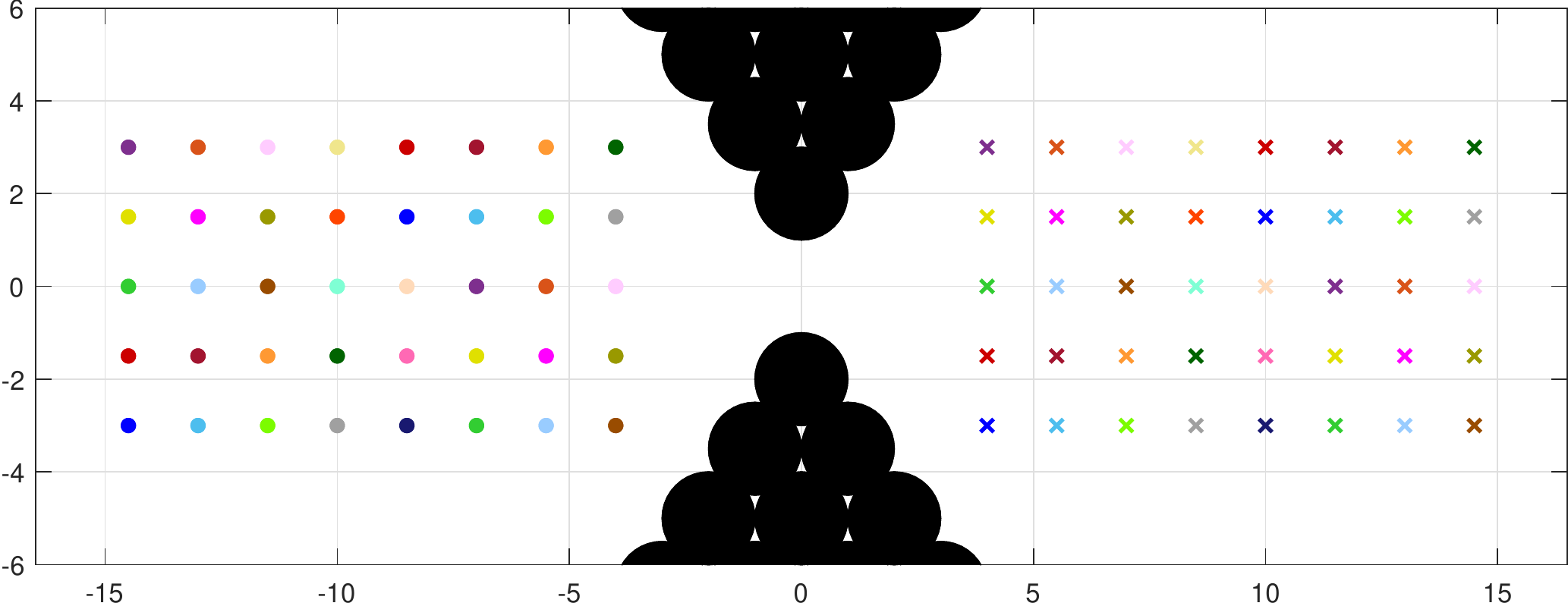}};
    \node[align=center, text=NavyBlue] (c) at (8.09, 0.5) {$k = 0$};
\end{tikzpicture}
\label{fig_bottle_1}}
\hfil
\subfloat{
\begin{tikzpicture}
    \node[anchor=south west,inner sep=0] at (0,0){    \includegraphics[width=0.48\textwidth, trim={0.0cm 0.0cm 0cm 0cm},clip]{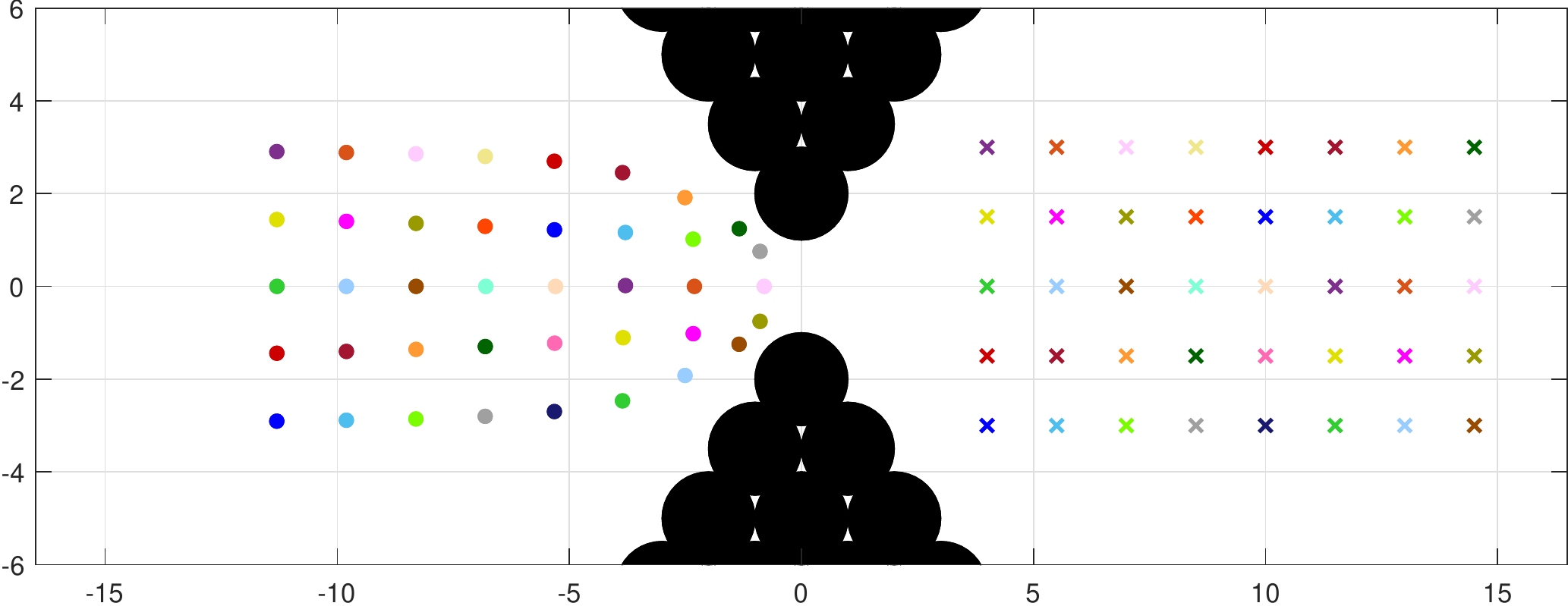}};
    \node[align=center, text=NavyBlue] (c) at (8.02, 0.5) {$k = 30$};
\end{tikzpicture}
\label{fig_bottle_2}}
\hfil
\\
\hfil
\subfloat{
\begin{tikzpicture}
    \node[anchor=south west,inner sep=0] at (0,0){    \includegraphics[width=0.48\textwidth, trim={0.0cm 0.0cm 0cm 0cm},clip]{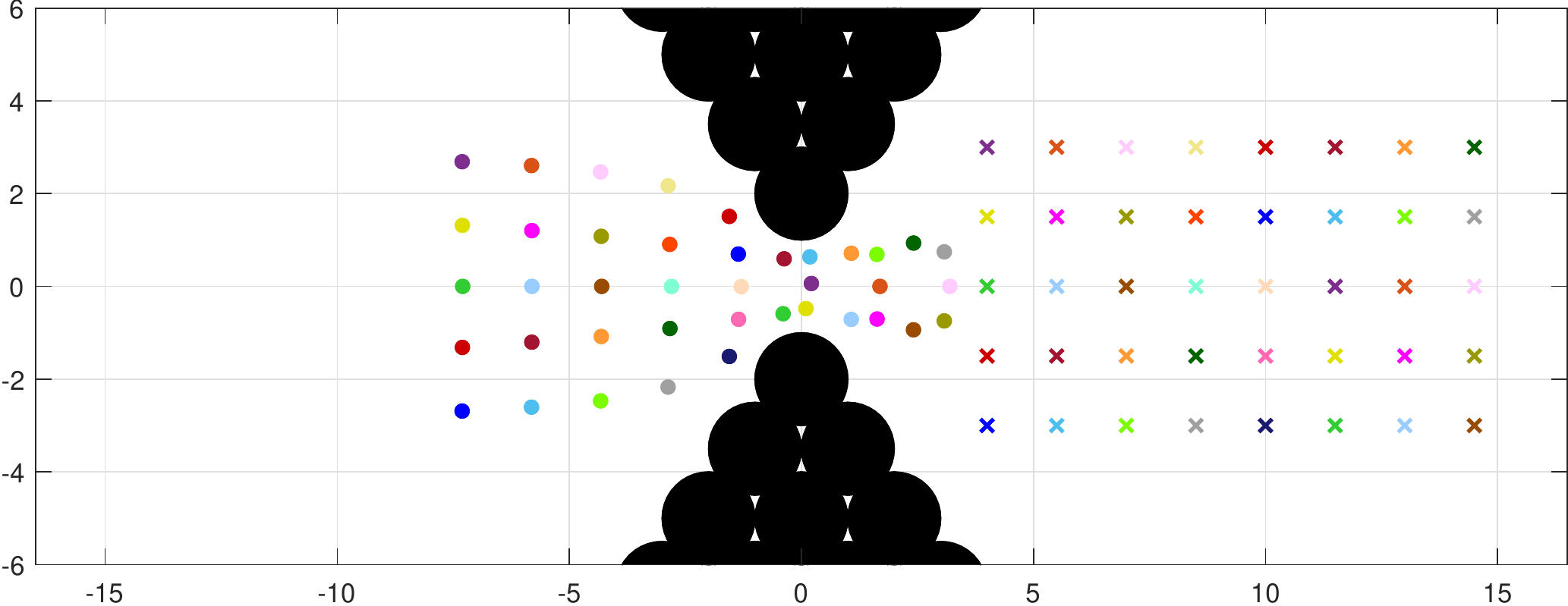}};
    \node[align=center, text=NavyBlue] (c) at (8.02, 0.5) {$k = 50$};
\end{tikzpicture}
\label{fig_bottle_3}}
\hfil
\subfloat{
\begin{tikzpicture}
    \node[anchor=south west,inner sep=0] at (0,0){    \includegraphics[width=0.48\textwidth, trim={0.0cm 0.0cm 0cm 0cm},clip]{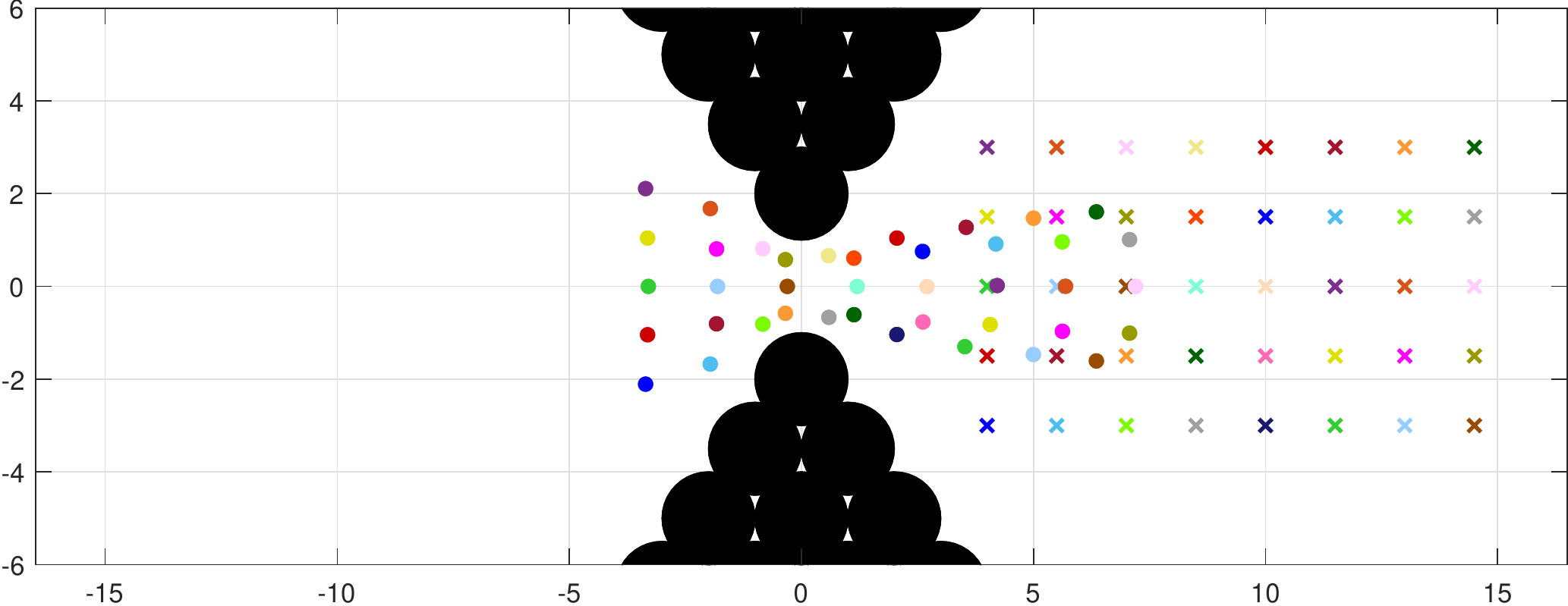}};
    \node[align=center, text=NavyBlue] (c) at (8.02, 0.5) {$k = 70$};
\end{tikzpicture}
\label{fig_bottle_4}}
\hfil
\\
\hfil
\subfloat{
\begin{tikzpicture}
    \node[anchor=south west,inner sep=0] at (0,0){    \includegraphics[width=0.48\textwidth, trim={0.0cm 0.0cm 0cm 0cm},clip]{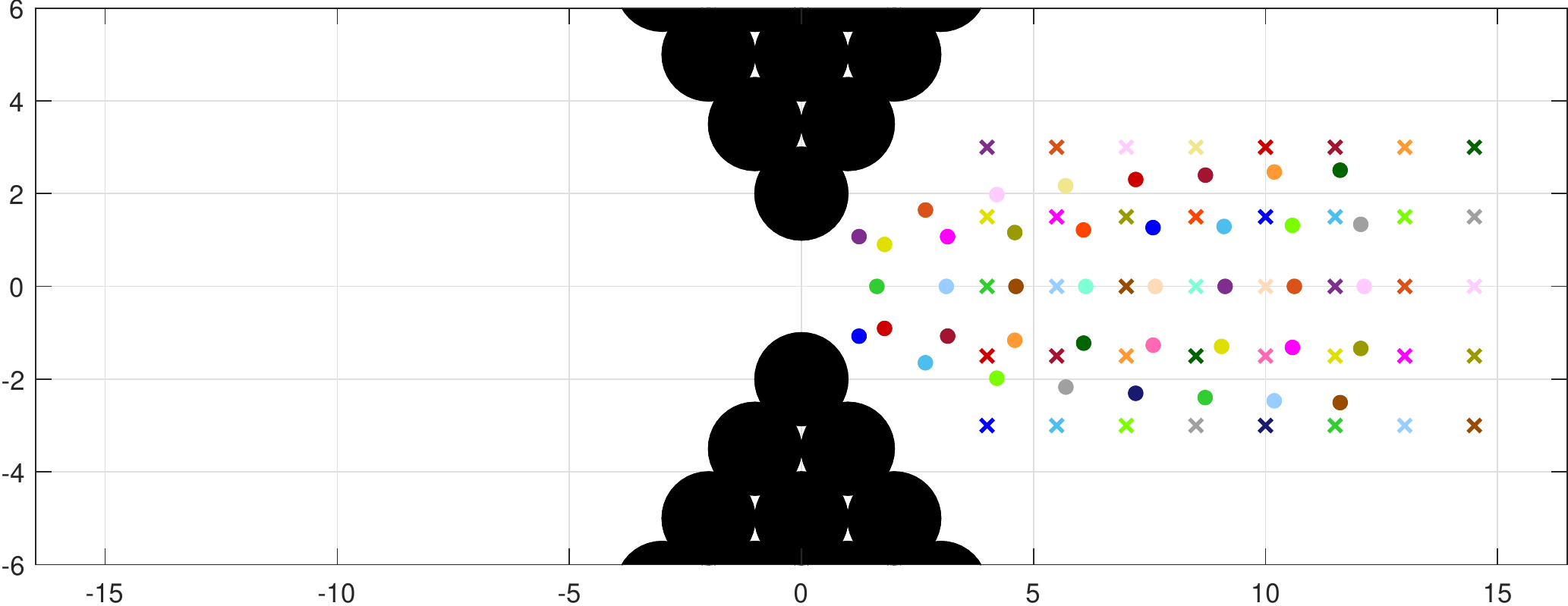}};
    \node[align=center, text=NavyBlue] (c) at (7.95, 0.5) {$k = 100$};
\end{tikzpicture}
\label{fig_bottle_5}}
\hfil
\subfloat{
\begin{tikzpicture}
    \node[anchor=south west,inner sep=0] at (0,0){    \includegraphics[width=0.48\textwidth, trim={0.0cm 0.0cm 0cm 0cm},clip]{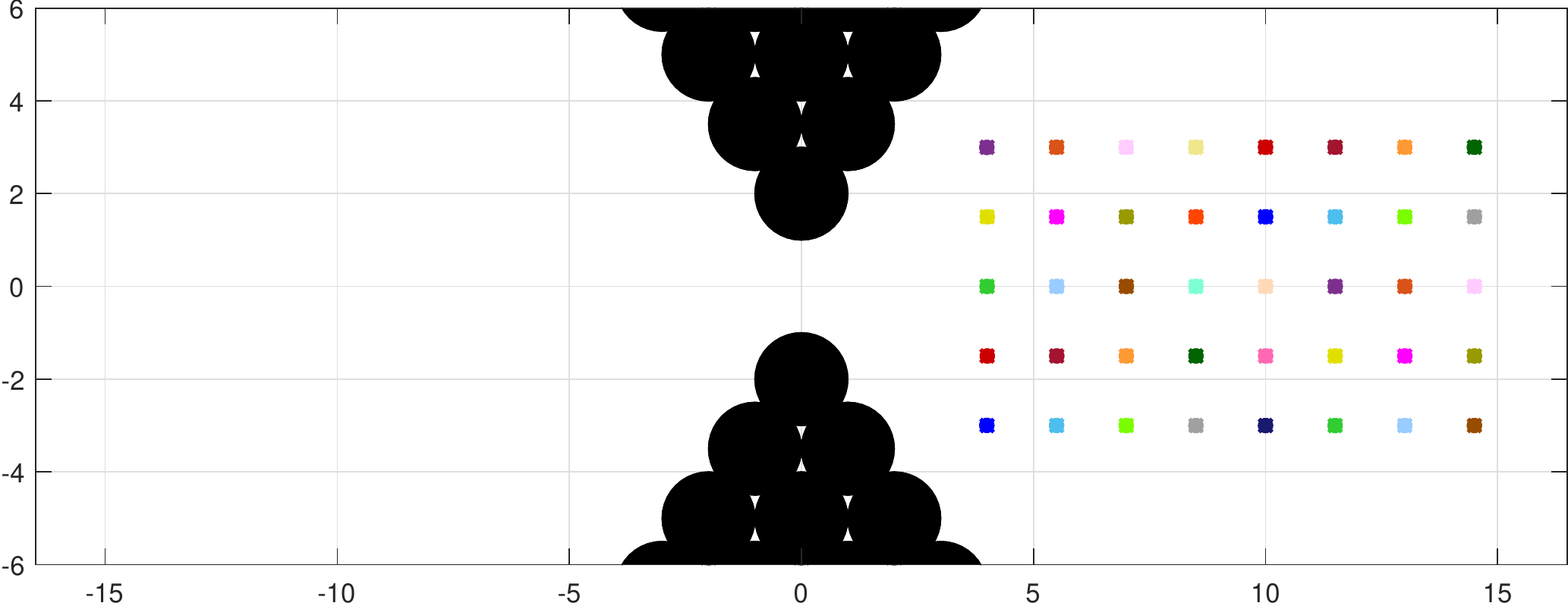}};
    \node[align=center, text=NavyBlue] (c) at (7.95, 0.5) {$k = 200$};
\end{tikzpicture}
\label{fig_bottle_6}}
\hfil
\caption{Multi-vehicle ``bottleneck'' task with MD-DDP including $40$ cars: Snapshots at different time instants.}
\label{fig_bottle}
\end{figure*}

The capabilities of the methods are further demonstrated in a problem where a team of 40 cars has to pass through a narrow ``bottleneck'' while avoiding collisions in order to reach their targets. Figure \ref{fig_bottle} illustrates how the cars are able to safely complete the task using MD-DDP. A time horizon $K=200$ and a neighborhood size of $|\calN_i|=9$ are used here.

\begin{figure*}[!t]
\centering
\hfil
\subfloat{
\begin{tikzpicture}
    \node[anchor=south west,inner sep=0] at (0,0){    \includegraphics[width=0.486\textwidth, trim={2.75cm 2.55cm 2.52cm 2.5cm},clip]{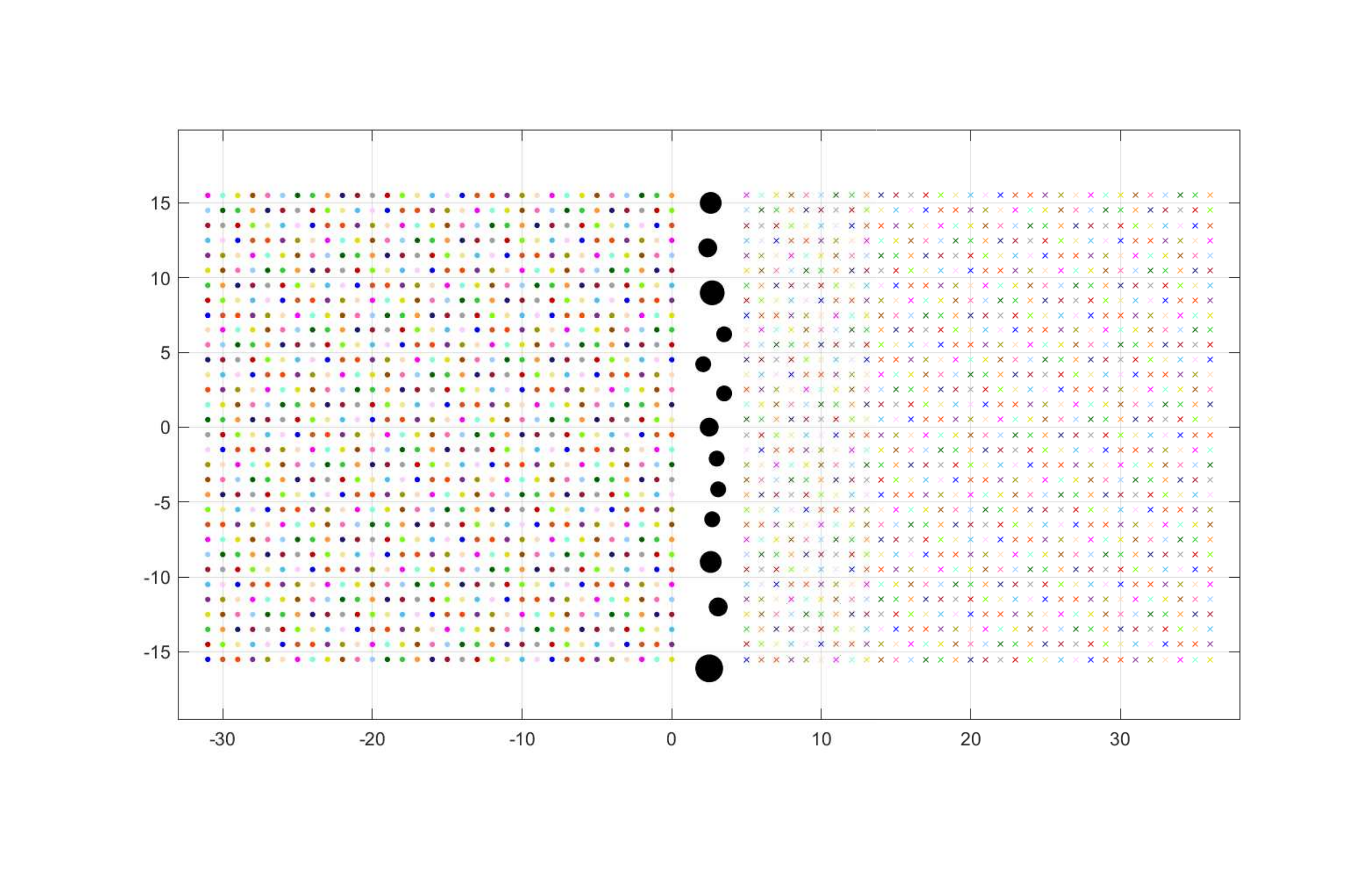}};
    \node[align=center, text=NavyBlue] (c) at (8.2, 0.44) {$k = 0$};
\end{tikzpicture}
\label{fig_formation_1}}
\hfil
\subfloat{
\begin{tikzpicture}
    \node[anchor=south west,inner sep=0] at (0,0){    \includegraphics[width=0.486\textwidth, trim={2.75cm 2.55cm 2.52cm 2.5cm},clip]{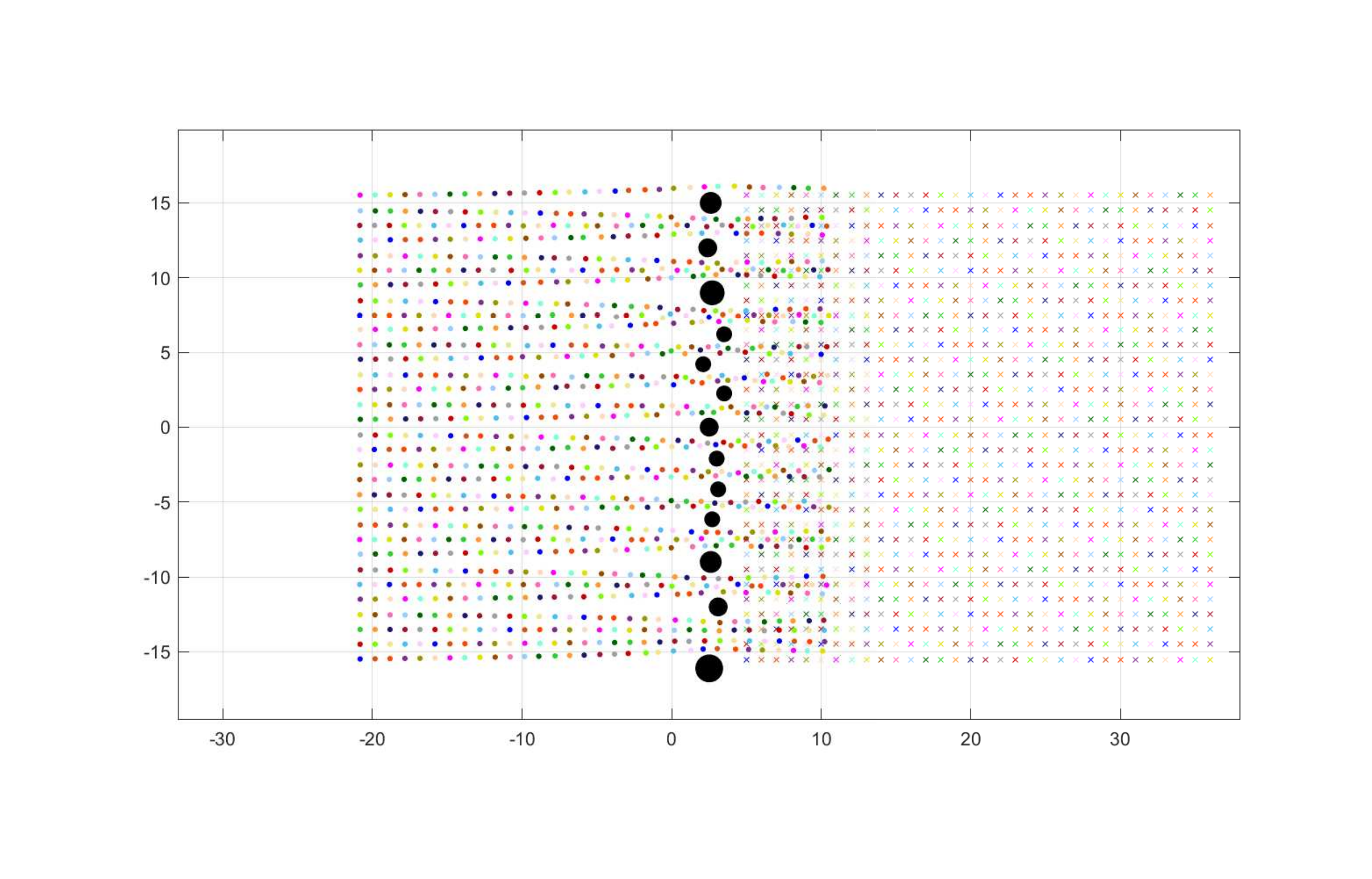}};
    \node[align=center, text=NavyBlue] (c) at (8.13, 0.44) {$k = 60$};
\end{tikzpicture}
\label{fig_formation_2}}
\hfil
\\
\hfil
\subfloat{
\begin{tikzpicture}
    \node[anchor=south west,inner sep=0] at (0,0){    \includegraphics[width=0.486\textwidth, trim={2.75cm 2.55cm 2.52cm 2.5cm},clip]{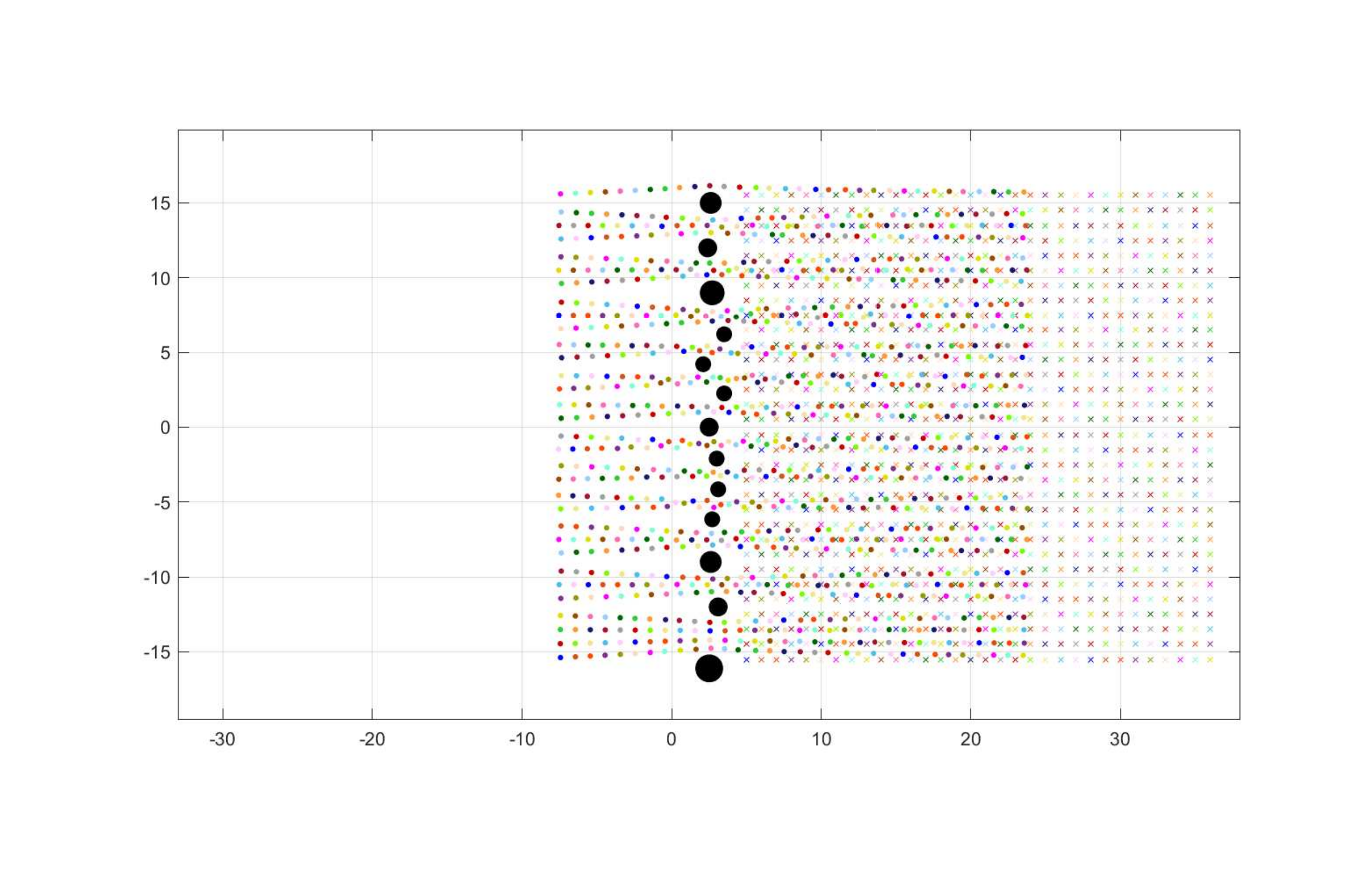}};
    \node[align=center, text=NavyBlue] (c) at (8.06, 0.44) {$k = 120$};
\end{tikzpicture}
\label{fig_formation_3}}
\hfil
\subfloat{
\begin{tikzpicture}
    \node[anchor=south west,inner sep=0] at (0,0){    \includegraphics[width=0.486\textwidth, trim={2.75cm 2.55cm 2.52cm 2.5cm},clip]{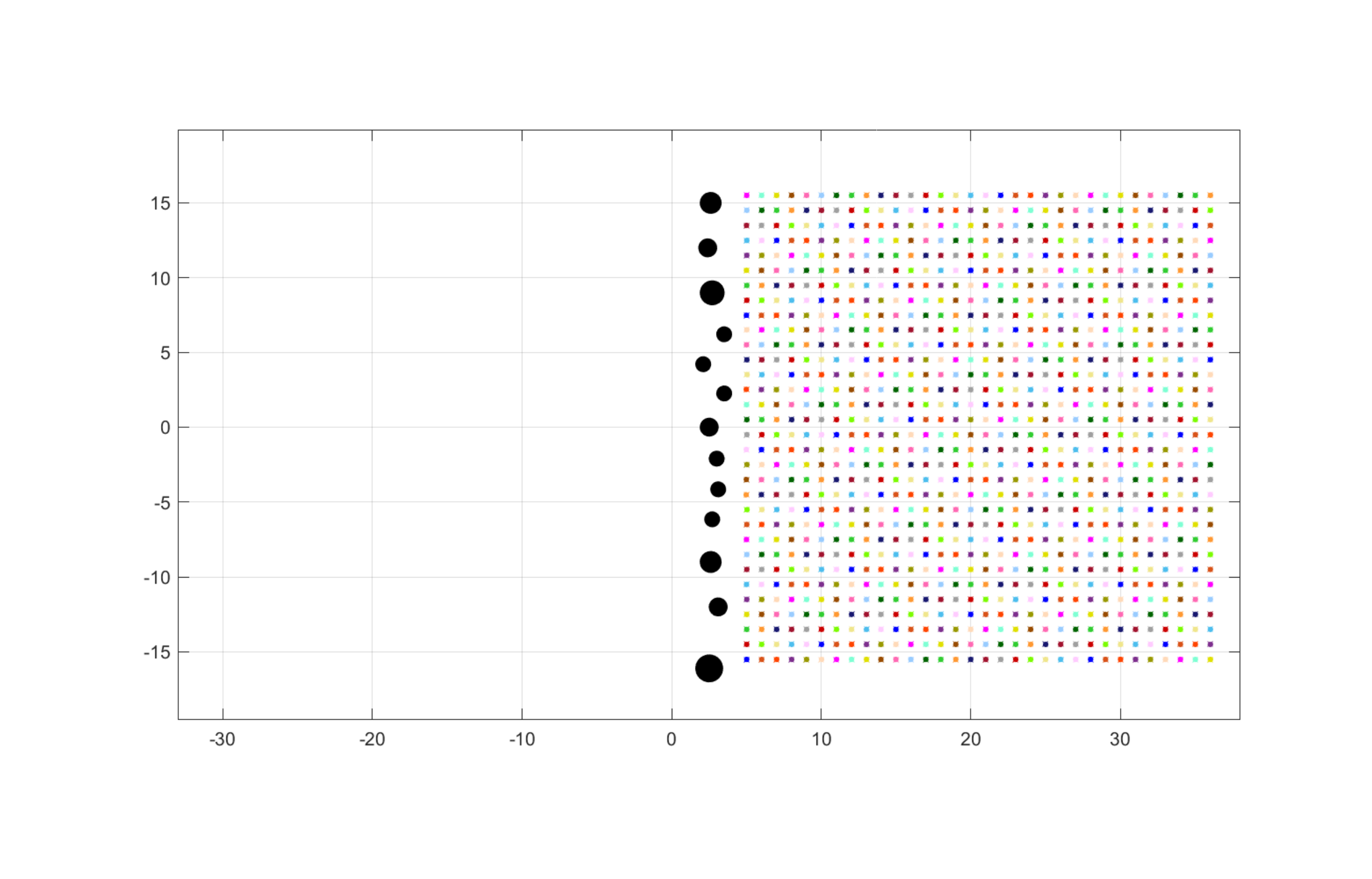}};
    \node[align=center, text=NavyBlue] (c) at (8.06, 0.44) {$k = 400$};
\end{tikzpicture}
\label{fig_formation_4}}
\hfil
\caption{Multi-vehicle formation task with MD-DDP including $1,024$ cars: Snapshots at different time instants.}
\label{fig_formation}
\end{figure*}
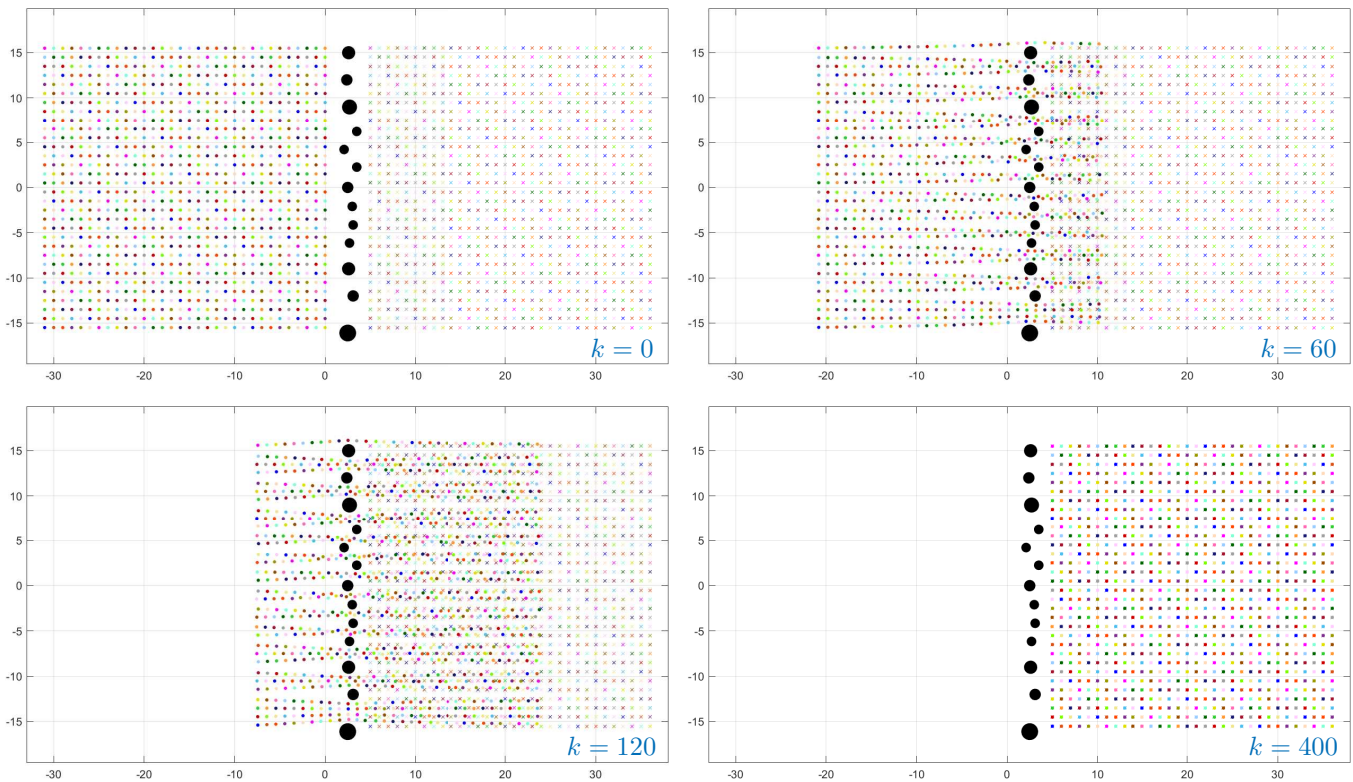

\newpage
Finally, to exhibit the scalability of both frameworks to large-scale problems, we consider a task where a team of vehicles has to move from one square grid to another while again avoiding collisions with each other and multiple obstacles. Figure \ref{fig_formation} shows a scenario where 1,024 cars successfully complete the task with MD-DDP. The supplementary video includes results with 256 cars using ND-DDP and with up to 4,096 cars using MD-DDP. 
The 256 vehicles case includes $1{,}024$ states and $512$ control inputs. With time horizon $K=200$, this is a problem with $307{,}200$ optimization variables in total. 
The 4,096 vehicles problem, that MD-DDP is able to solve, includes $16{,}384$ states and $8{,}192$ controls, which using a time horizon $K=800$, results to a $19{,}660{,}800$-dimensional optimization problem. To our best knowledge, ND-DDP and MD-DDP are the first decentralized DDP-based methods to exhibit such a scalability for large-scale multi-robot control.

\setlength{\textfloatsep}{0pt}

\setlength{\textfloatsep}{0pt}

\subsection{Multi-UAV Control}
\label{sec: sim drone}

Next, the proposed algorithms are tested in multi-UAV problems that involve more complex dynamics and a 3D environment. Each drone $i$ has a state $\bx_{i,k} = \big[
\bp_{i,k} ; \ \bv_{i,k} ; \ \beeta_{i,k} ; \ \bnu_{i,k}
\big] \in \Rb^{12}
$ and control input 
$\bu_{i,k} = \big[
F_{i,k}^{(1)} \ \ F_{i,k}^{(2)} \ \ F_{i,k}^{(3)} \ \ F_{i,k}^{(4)}
\big]\T \in \Rb^4
$, where $\bp_{i,k} \in \Rb^3$ is the center of mass absolute linear position, $\bv_{i,k} \in \Rb^3$ are the linear velocities, $\beeta_{i,k} \in \Rb^3$ are the Euler angles, $\bnu_{i,k} \in \Rb^3$ are the angular velocities and $F_{i,k}^{(c)}, \ c \in \llbracket 1, 4 \rrbracket$ is the thrust force of the $c$-th rotor. The drones dynamics and parameters can be found in \cite{luukkonen2011modelling}. The discrete dynamics are obtained through Euler discretization with time step $dt = 0.02\text{s}$. Each drone has a quadratic cost \eqref{quadratic agent cost} with $\bQ_i = \diag (150,150,150, 50, \dots, 50)$, $\bR_i = \diag(1, \dots, 1)$ and $\bQ_i^{\text{f}} = \bQ_i$. The following thrust forces limits must be satisfied
\begin{equation}
0 \leq F_{i,k}^{(c)} \leq F_{\text{max}}, \ c \in \llbracket 1, 4 \rrbracket,
\end{equation}
with 
$F_{\text{max}} = 30\text{N}$. Position constraints similar to \eqref{car state box y}, are imposed in 3D so that the drones remain inbound. Collision avoidance \eqref{coll avoidance constraints} and connectivity maintenance constraints \eqref{conn maint constraints} must also satisfied with $d_{\text{col}} = 0.5 \text{m}$ and $d_{\text{con}} = 2.0 \text{m}$. The constraints are handled by each method as in Section \ref{sec: sim car} and the drones are always initialized at hovering state. The same amounts of iterations $N$ as in Section \ref{sec: sim car} are used.

\begin{figure*}[!t]
\centering
\hfil
\subfloat{
\begin{tikzpicture}
    \node[anchor=south west,inner sep=0] at (0,0){    \includegraphics[width=0.315\textwidth, trim={1.9cm 2.2cm 2.0cm 2.4cm},clip]{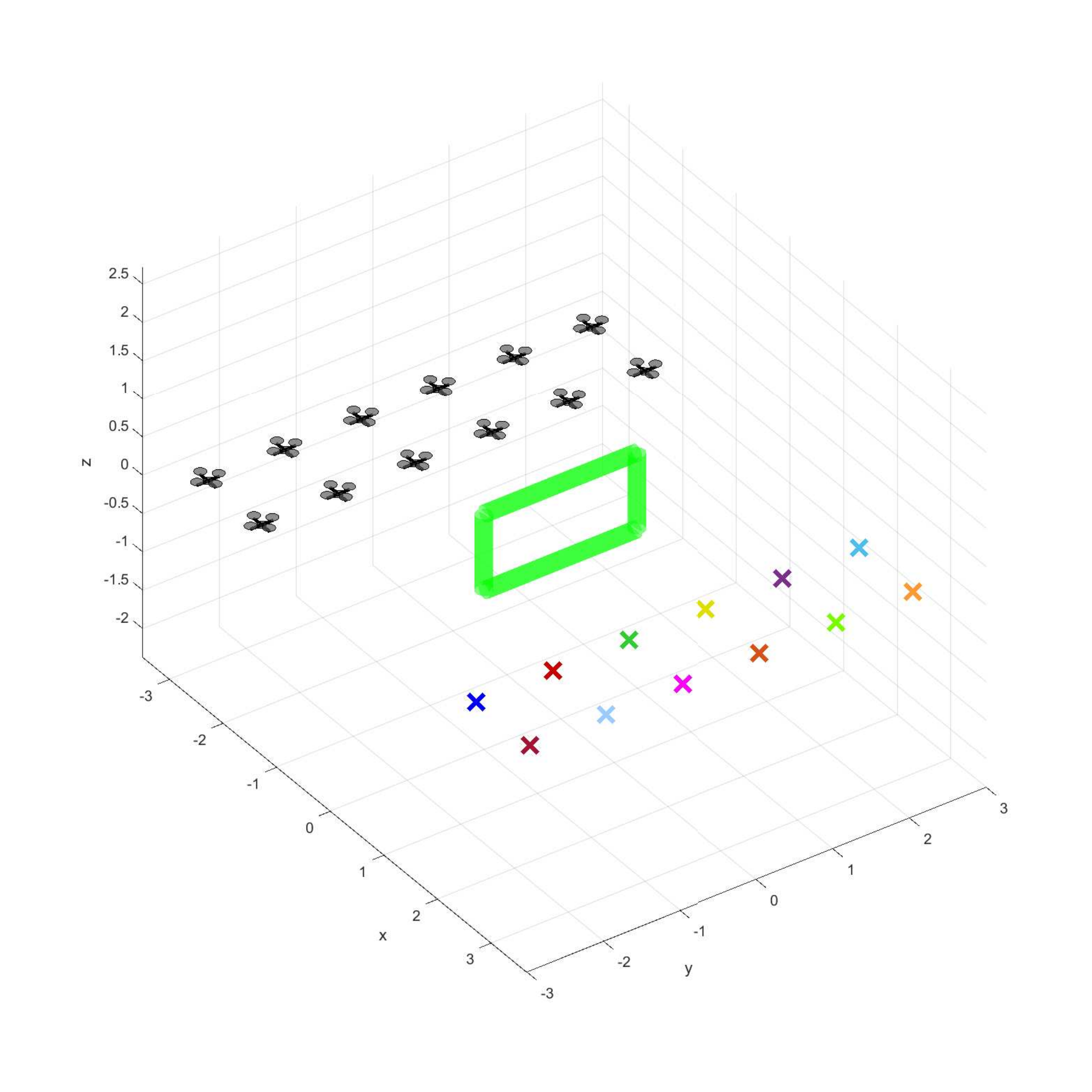}};
    \node[align=center, text=NavyBlue] (c) at (5.1, 0.42) {$k = 0$};
\end{tikzpicture}
\label{fig_drones_gate_1}}
\hfil
\subfloat{
\begin{tikzpicture}
    \node[anchor=south west,inner sep=0] at (0,0){    \includegraphics[width=0.315\textwidth, trim={1.9cm 2.2cm 2.0cm 2.4cm},clip]{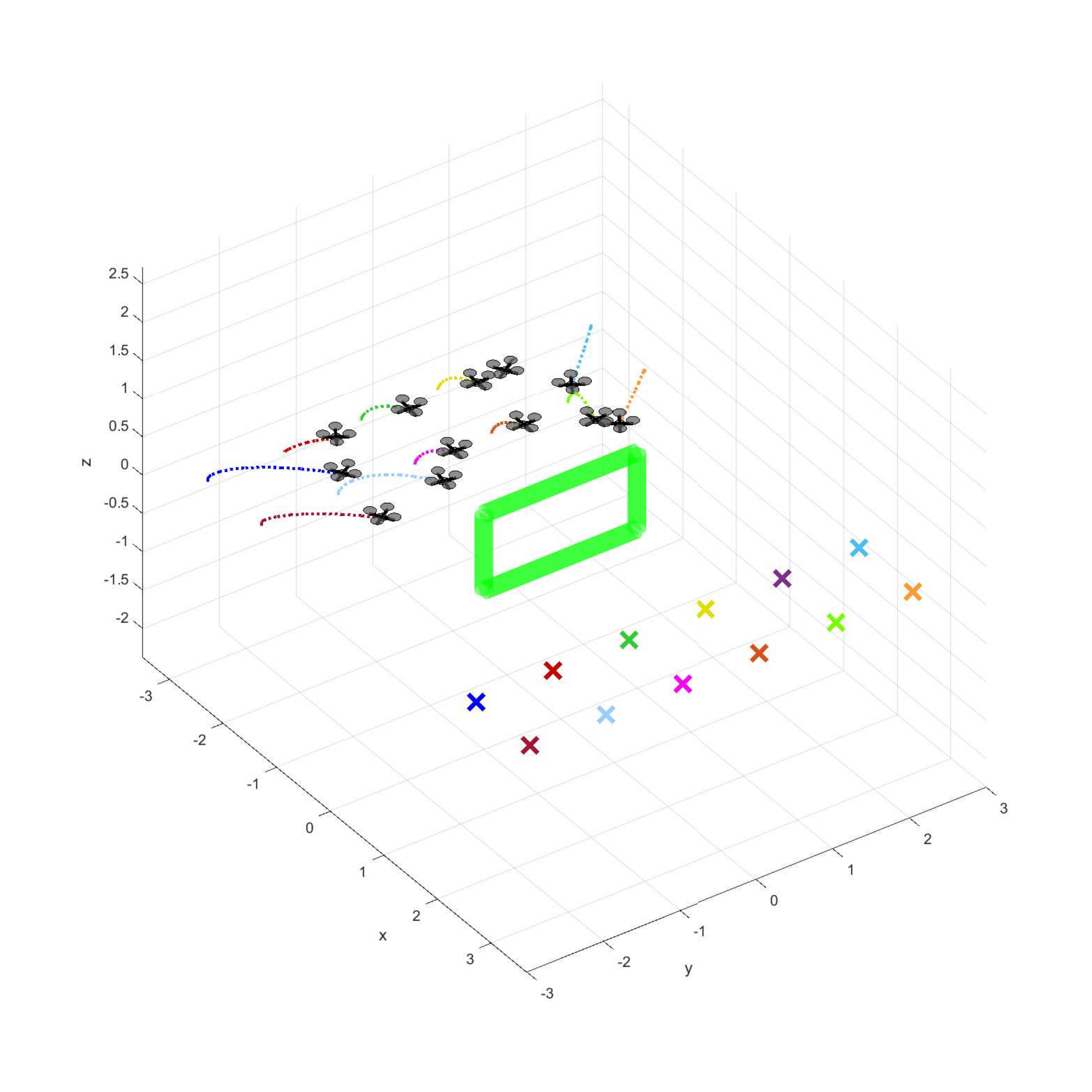}};
    \node[align=center, text=NavyBlue] (c) at (5.03, 0.42) {$k = 15$};
\end{tikzpicture}
\label{fig_drones_gate_2}}
\hfil
\subfloat{
\begin{tikzpicture}
    \node[anchor=south west,inner sep=0] at (0,0){    \includegraphics[width=0.315\textwidth, trim={1.9cm 2.2cm 2.0cm 2.4cm},clip]{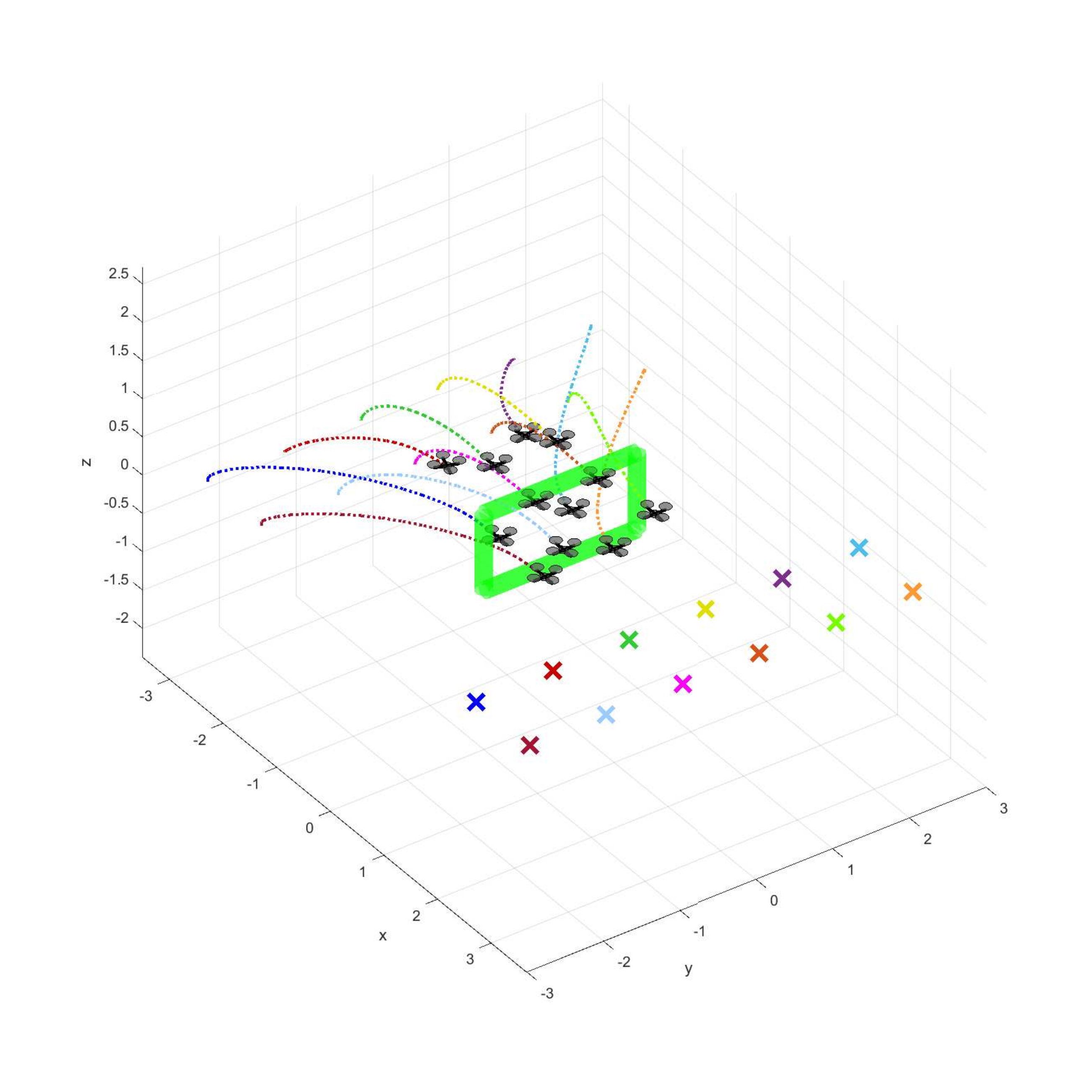}};
    \node[align=center, text=NavyBlue] (c) at (5.03, 0.42) {$k = 30$};
\end{tikzpicture}
\label{fig_drones_gate_3}}
\hfil
\\
\hfil
\subfloat{
\begin{tikzpicture}
    \node[anchor=south west,inner sep=0] at (0,0){    \includegraphics[width=0.315\textwidth, trim={1.9cm 2.2cm 2.0cm 2.4cm},clip]{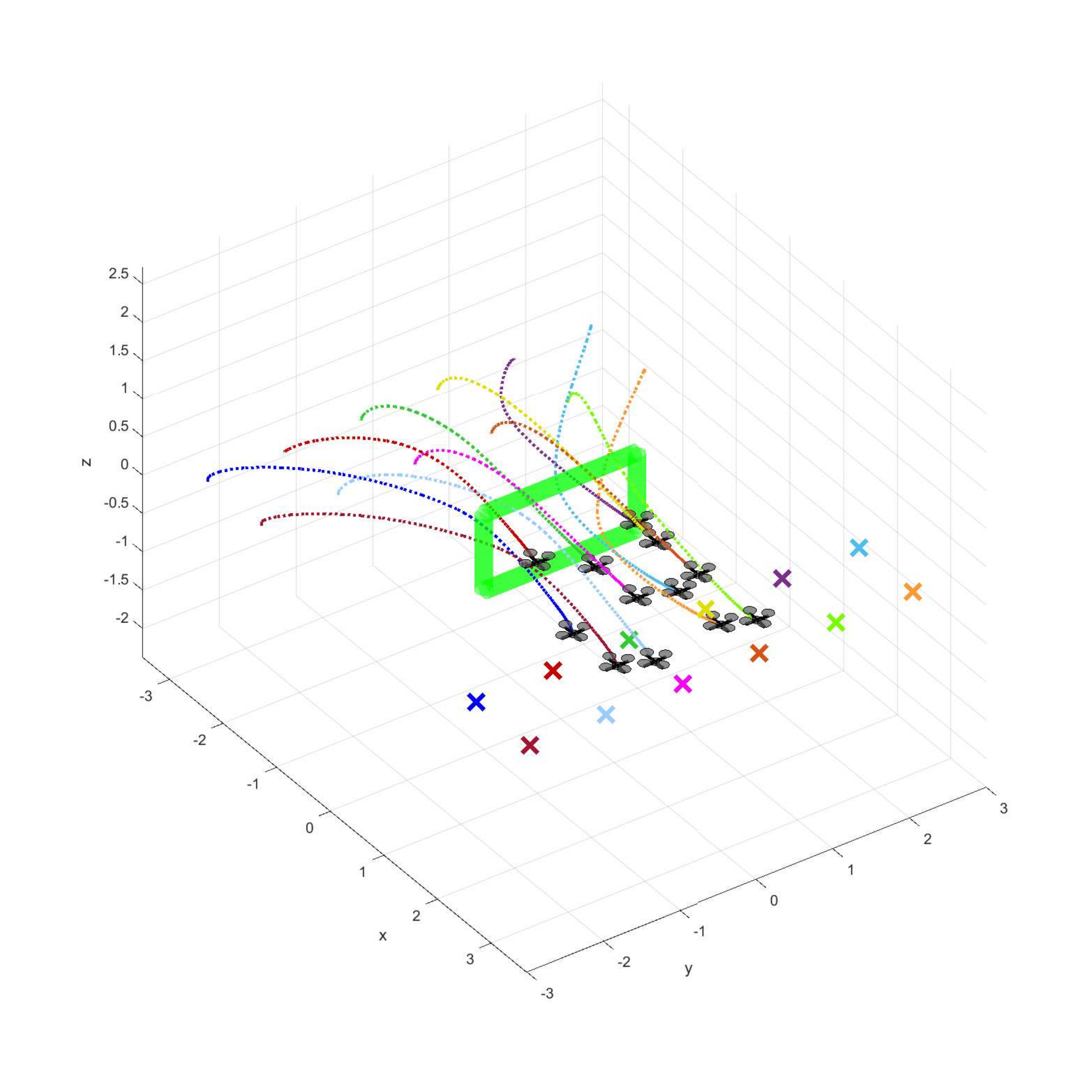}};
    \node[align=center, text=NavyBlue] (c) at (5.03, 0.42) {$k = 70$};
\end{tikzpicture}
\label{fig_drones_gate_4}}
\hfil
\subfloat{
\begin{tikzpicture}
    \node[anchor=south west,inner sep=0] at (0,0){    \includegraphics[width=0.315\textwidth, trim={1.9cm 2.2cm 2.0cm 2.4cm},clip]{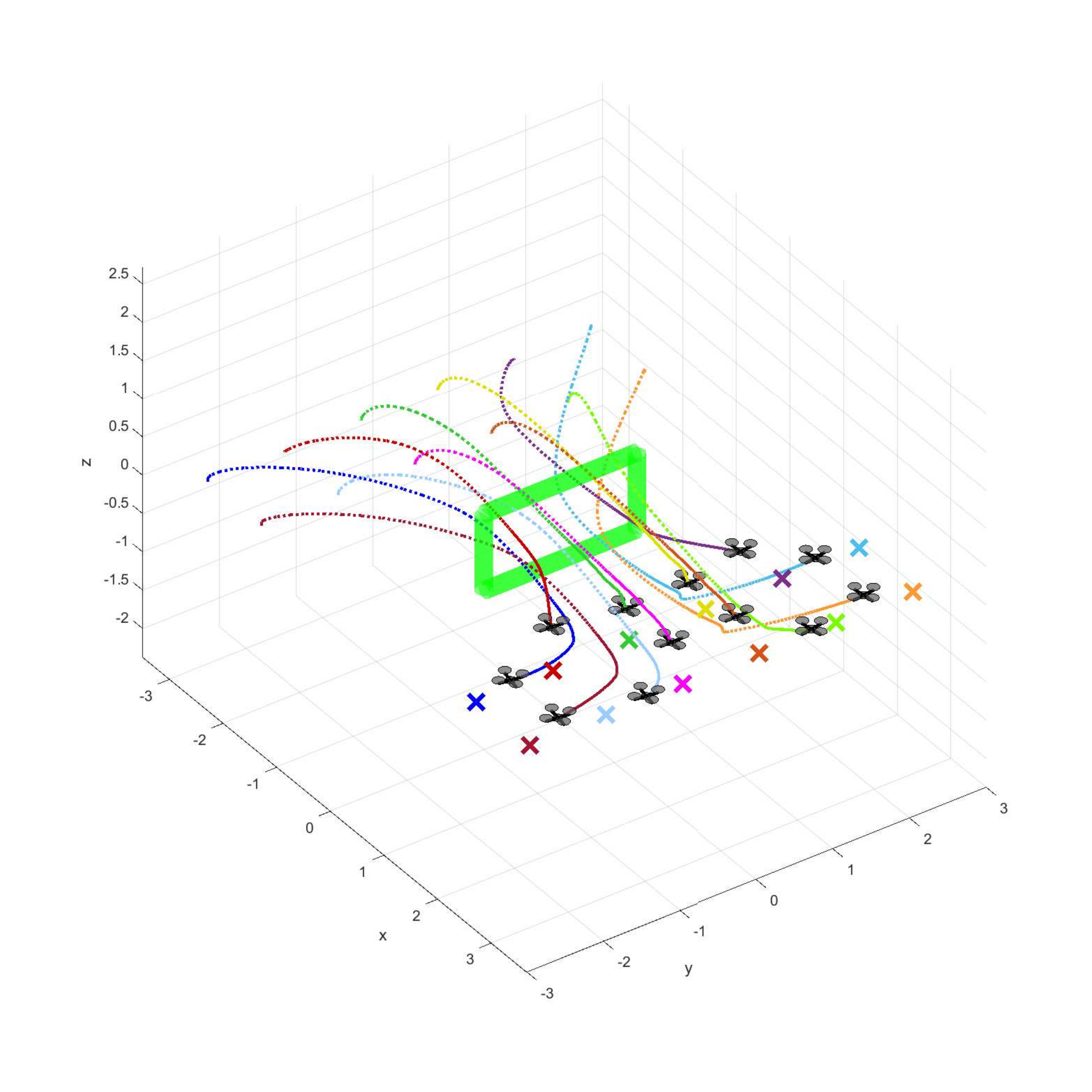}};
    \node[align=center, text=NavyBlue] (c) at (4.96, 0.42) {$k = 130$};
\end{tikzpicture}
\label{fig_drones_gate_5}}
\hfil
\subfloat{
\begin{tikzpicture}
    \node[anchor=south west,inner sep=0] at (0,0){    \includegraphics[width=0.315\textwidth, trim={1.9cm 2.2cm 2.0cm 2.4cm},clip]{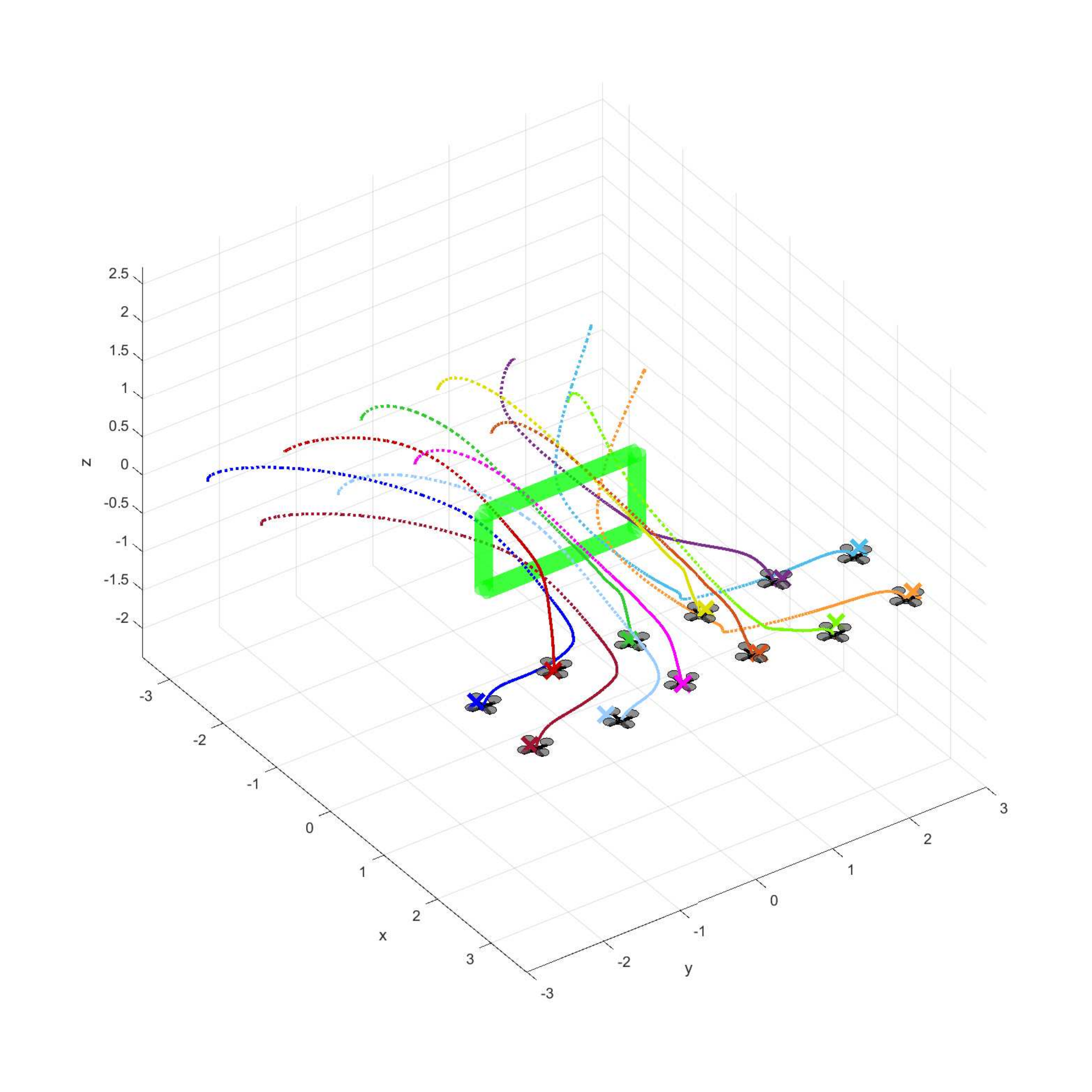}};
    \node[align=center, text=NavyBlue] (c) at (4.96, 0.42) {$k = 250$};
\end{tikzpicture}
\label{fig_drones_gate_6}}
\caption{Multi-UAV ``gate'' task with MD-DDP including $12$ drones: Snapshots at different time instants.}
\label{fig_drones_gate}
\end{figure*}

The first task (Fig. \ref{fig_drones_gate}) requires 12 drones to reach their targets while passing through the green ``gate''. Given a time horizon $K=250$, the additional constraint is given by
\begin{subequations}
\begin{alignat}{2}
\mathrm{y}_{\text{gate}}^{\text{l}} &\leq \mathrm{y}_{i,k} \leq \mathrm{y}_{\text{gate}}^{\text{h}}, 
\ && k \in \llbracket k_1, k_2 \rrbracket,
\label{drone gate constraint y}
\\
\mathrm{z}_{\text{gate}}^{\text{l}} &\leq \mathrm{z}_{i,k} \leq \mathrm{z}_{\text{gate}}^{\text{h}}, 
\ && k \in \llbracket k_1, k_2 \rrbracket,
\label{drone gate constraint z}
\end{alignat}
\end{subequations}
%
with $\mathrm{y}_{\text{gate}}^{\text{l}} = -1.0$m, $\mathrm{y}_{\text{gate}}^{\text{h}} = 1.0$m, $\mathrm{z}_{\text{gate}}^{\text{l}} = -0.5$m, $\mathrm{z}_{\text{gate}}^{\text{h}} = 0.5$m, 
$k_1 = 30$ and $k_2 = 100$. As illustrated in Fig. \ref{fig_drones_gate}, with MD-DDP all drones reach their goals while passing through the gate and avoiding collisions. In the supplementary video, results for both MD-DDP and ND-DDP are provided.

\begin{figure*}[!t]
\centering
\hfil
\subfloat{
\begin{tikzpicture}
    \node[anchor=south west,inner sep=0] at (0,0){    \includegraphics[width=0.42\textwidth, trim={2.1cm 4.4cm 1.9cm 4.4cm},clip]{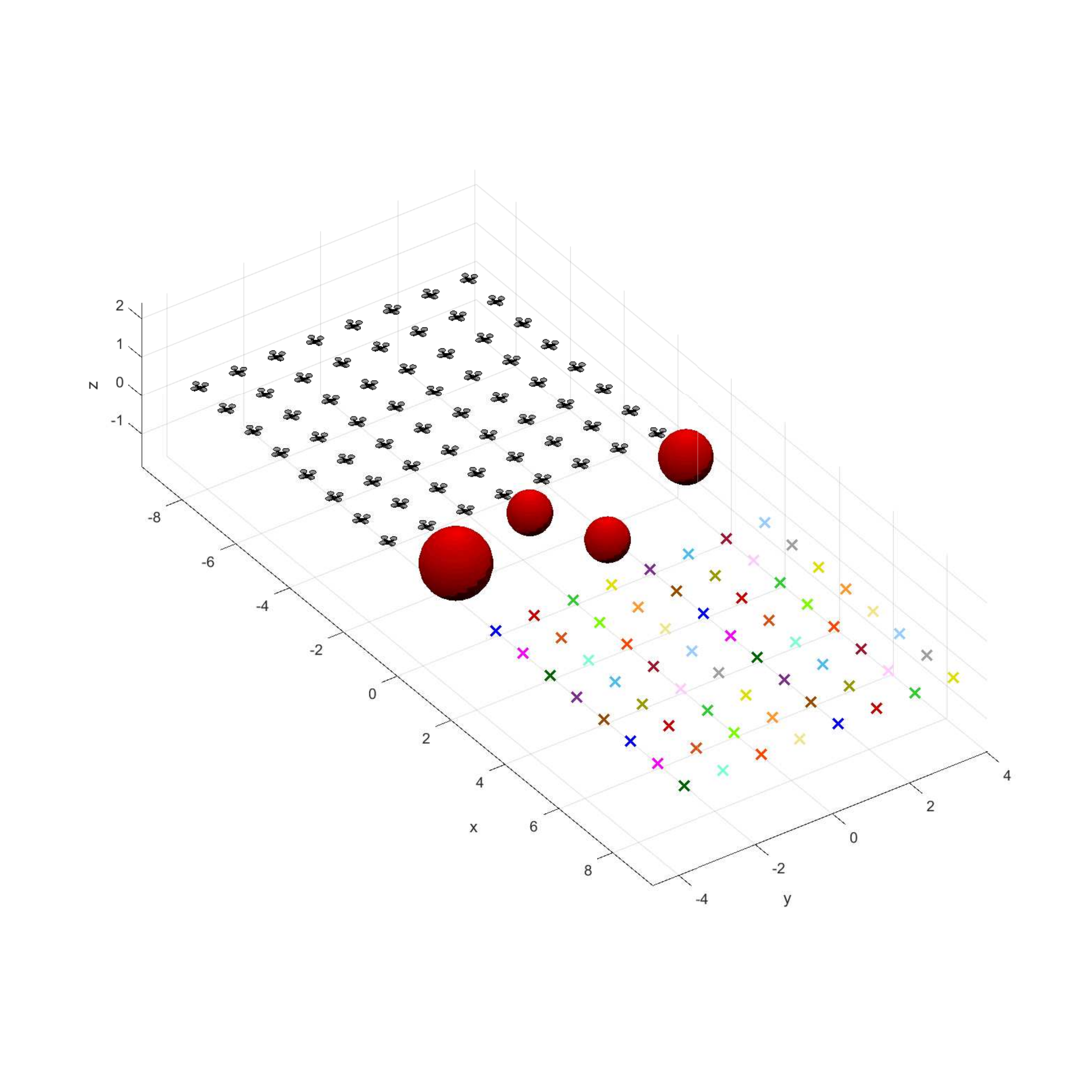}};
    \node[align=center, text=NavyBlue] (c) at (7.2, 0.5) {$k = 0$};
\end{tikzpicture}
\label{fig_drones_formation_1}}
\hfil
\subfloat{
\begin{tikzpicture}
    \node[anchor=south west,inner sep=0] at (0,0){    \includegraphics[width=0.42\textwidth, trim={2.1cm 4.4cm 1.9cm 4.4cm},clip]{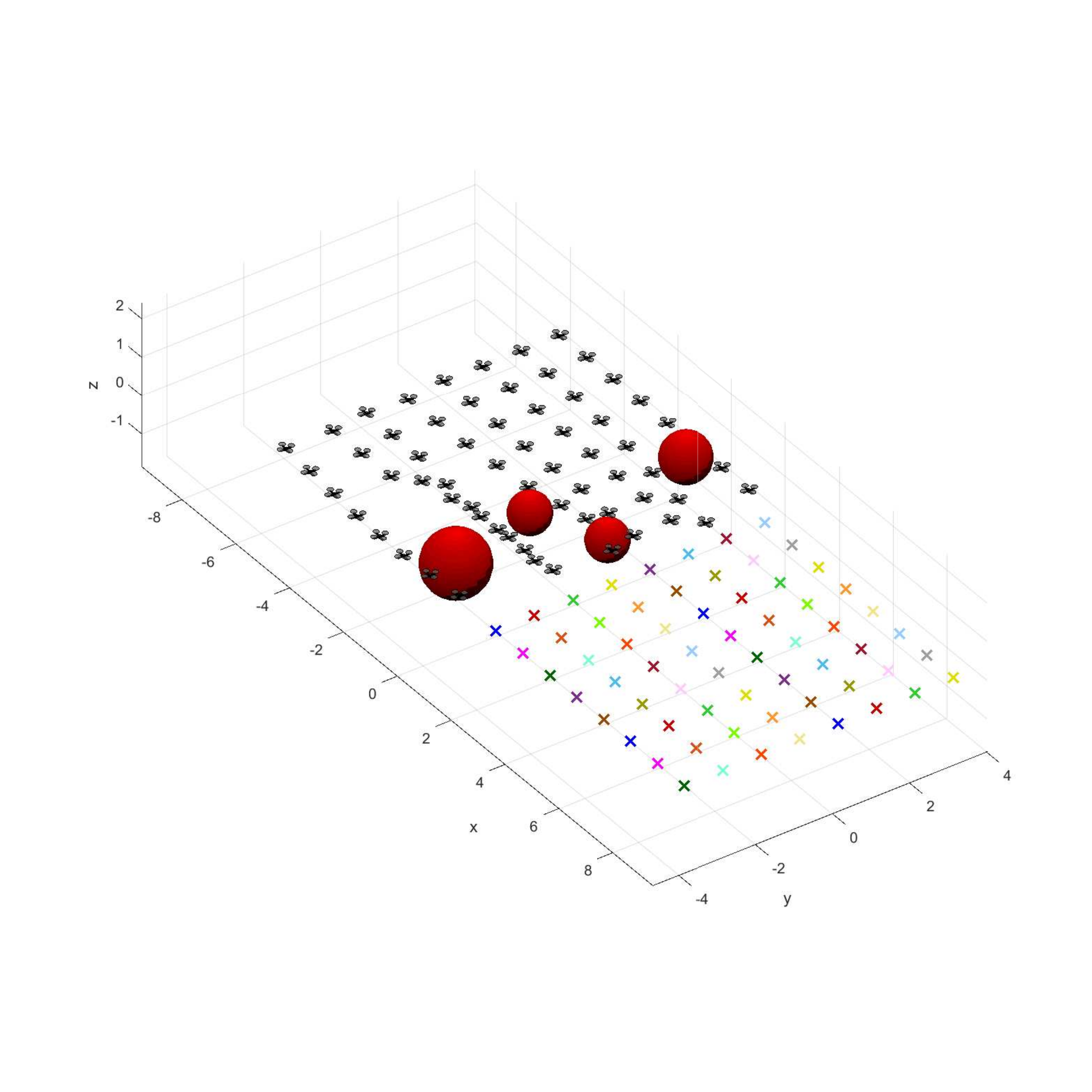}};
    \node[align=center, text=NavyBlue] (c) at (7.13, 0.5) {$k = 40$};
\end{tikzpicture}
\label{fig_drones_formation_2}}
\hfil
\\
\hfil
\subfloat{
\begin{tikzpicture}
    \node[anchor=south west,inner sep=0] at (0,0){    \includegraphics[width=0.42\textwidth, trim={2.1cm 4.2cm 1.9cm 4.4cm},clip]{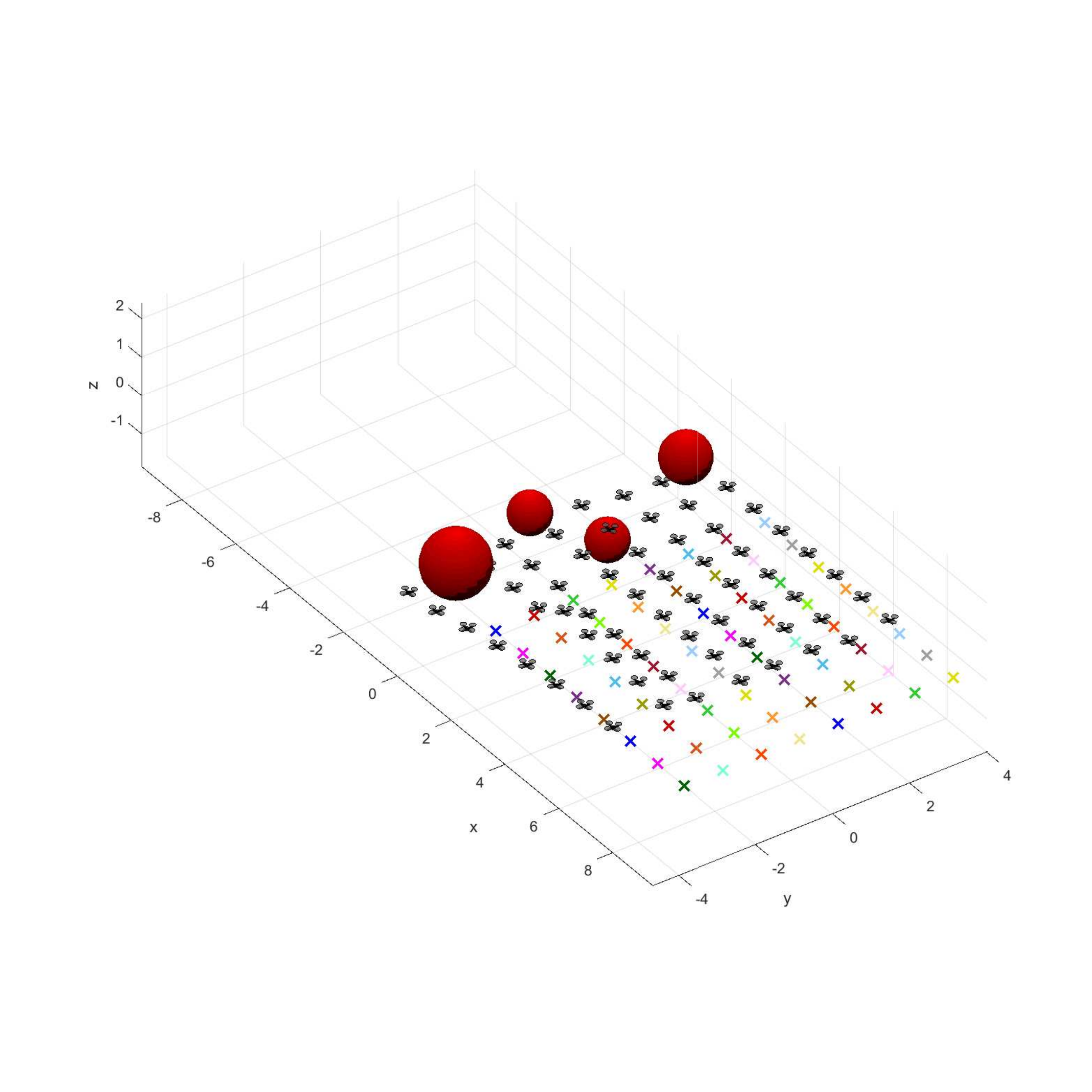}};
    \node[align=center, text=NavyBlue] (c) at (7.06, 0.5) {$k = 120$};
\end{tikzpicture}
\label{fig_drones_formation_3}}
\hfil
\subfloat{
\begin{tikzpicture}
    \node[anchor=south west,inner sep=0] at (0,0){    \includegraphics[width=0.42\textwidth, trim={2.1cm 4.2cm 1.9cm 4.4cm},clip]{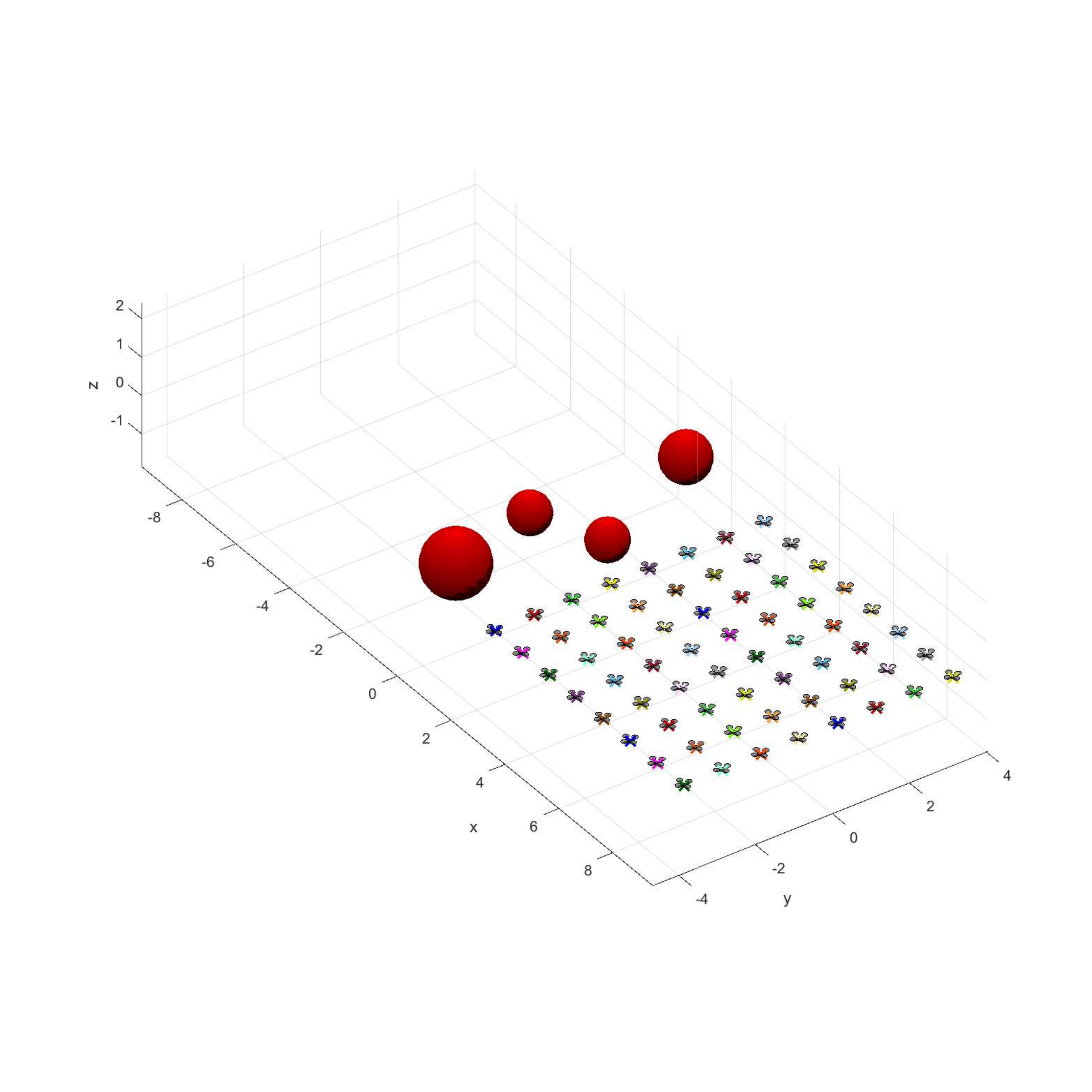}};
    \node[align=center, text=NavyBlue] (c) at (7.06, 0.5) {$k = 400$};
\end{tikzpicture}
\label{fig_drones_formation_4}}
\hfil
\caption{Multi-UAV formation task with MD-DDP including $64$ drones: Snapshots at different time instants.}
\label{fig_drones_formation}
\end{figure*}

Subsequently, a task where a large-scale team of drones must move from an initial square formation to a target one, is presented. Except for avoiding each other, the drones must also avoid obstacles in between, which is enforced with constraints \eqref{obs avoidance constraints} in 3D. In Fig. \ref{fig_drones_formation}, a task with 64 drones using MD-DDP is shown, where all drones successfully reach their targets while staying safe. The video includes results with 64 drones using ND-DDP and with 64 and 256 drones using MD-DDP. 

\subsection{Ablative Analysis}

Here, we show how the improvements proposed in Section \ref{sec:improvements} (Nesterov acceleration and decentralized penalty parameters adaptation), can increase the performance of the algorithms. For brevity, only a comparison for MD-DDP is presented, as the same patterns are observed for ND-DDP as well. The 256 cars formation task is used for all tests. Since we wish to investigate whether the improvements could allow for earlier termination, the algorithms are no longer terminated at maximum ADMM iterations, but once the total primal and dual residuals norms reach below the following thresholds
%
\begin{equation}
\| r^{\text{pri}}_{b,\text{total}} \|_2  \leq \epsilon_b^{\text{pri}}, \quad
\| r^{\text{dual}}_{\text{total}} \|_2 \leq \epsilon_b^{\text{dual}}, \quad b = 1,2,3
\label{ablative analysis term criteria}
\end{equation}
where 
$r^{\text{pri}}_{b,\text{total}} = \big[ r_{b,1}^{\text{pri}}; \dots; r_{b,M}^{\text{pri}} \big] $, 
$r^{\text{dual}}_{b,\text{total}} = \big[ r_{b,1}^{\text{dual}}; \dots; r_{b,M}^{\text{dual}} \big] $. The expressions for all ``per-agent'' residuals $r_{b,i}^{\text{pri}}, \ r_{b,i}^{\text{dual}}, \ i \in \llbracket 1,M \rrbracket$, are provided in Appendix \ref{sec: PPA details}. The threshold values are set to $\epsilon_1^{\text{pri}} = 5$, $\epsilon_{\{2,3\}}^{\text{pri}} = 10$ and $\epsilon_1^{\text{dual}} = 50$, $\epsilon_{\{2,3\}}^{\text{dual}} = 10^3$.

\begin{table}[]
\centering
\begin{tabular}{@{}cccc@{}}
\hline
Method & $\eta$ & Iterations & +/- \% \\
\hline 
Vanilla MD-DDP & - & 127 & - \\
\hline
\multicolumn{1}{c}{\multirow{8}{*}{Nesterov accelerated MD-DDP}} & 0.05 & 119 & -6.3\% \\
 & 0.1 & 116 & -8.7\% \\ 
& 0.15 & 108 & -15.0\% \\
& \textbf{0.2} & \textbf{101} & \textbf{-20.5\%} \\ 
& 0.25 & 104 & -18.1\% \\ 
& 0.3 & 113 & -11.0\% \\
& 0.35 & 122 & -3.9\% \\ 
& 0.4 & 136 & +7.1\% \\ 
\hline
\end{tabular}
\caption{Comparison of ADMM iterations between vanilla MD-DDP and Nesterov accelerated MD-DDP for different values of the parameter $\eta$.}
\label{tab: nesterov}
\end{table}

Initially, a comparison between vanilla and Nesterov accelerated MD-DDP is presented. Table \ref{tab: nesterov} shows the amount of ADMM iterations required for the criteria \eqref{ablative analysis term criteria} to be met, for different values of $\eta$. For the accelerated algorithm to achieve a computational benefit, the parameter $\eta$ must be within the region $[0.05, 0.35]$. The iterations percentage decrease varies from $-3.9\%$ to $-20.5\%$, with the largest reduction observed for $\eta = 0.2$. Therefore, the results show that Nesterov acceleration can indeed improve performance, with the appropriate tuning of parameter $\eta$. 

\begin{table}[]
\centering
\begin{tabular}{@{}cccc@{}}
\hline
Method & $n_{\text{adapt}}$ & Iterations & +/- \% \\
\hline 
Without adaptation & - & 184 & - \\
\hline
\multicolumn{1}{c}{\multirow{4}{*}{Centralized adaptation}} & 1 & 172 & -6.5\% \\
 & 5 & 159 & -13.6\% \\ 
& 10 & 131 & -28.8\% \\
& 20 & 138 & -25.0\% \\
\hline
\multicolumn{1}{c}{\multirow{4}{*}{Decentralized adaptation}} & 1 & 171 & -7.1\% \\ 
& 5 & 151 & -17.9\% \\ 
& \textbf{10} & \textbf{123} & \textbf{-33.2\%} \\
& 20 & 130 & -29.3\% \\
\hline
\end{tabular}
\caption{Comparison of ADMM iterations between MD-DDP with decentralized, centralized or no adaptation for the penalty parameters.}
\label{tab: ppa}
\end{table}

Subsequently, the advantages of the proposed decentralized penalty parameter adaptation scheme are illustrated. 
MD-DDP using the proposed scheme is compared against MD-DDP without adaptation and with a centralized adaptation scheme as in \cite[Section 3.4.1]{boyd2011distributed} using the total residuals norms of the global problem $\| r^{\text{pri}}_{b,\text{total}} \|_2$, $\| r^{\text{dual}}_{b,\text{total}} \|_2$ in the criteria. We use $\chi^{\text{incr}} = \chi^{\text{decr}} = 2$, $\sigma_1^{\text{incr}} = \sigma_2^{\text{incr}} = 1/200$, $\sigma_3^{\text{incr}} = 1/20$, $\sigma_1^{\text{decr}} = \sigma_2^{\text{decr}} = 1/50$ and $\sigma_3^{\text{decr}} = 1/5$ for both the centralized and decentralized schemes. To verify the adaptation capabilities of the proposed approach, we initialize the penalty parameter matrices with $\bT_i^0 = \bR_i$, $\bP_i^0 = \bQ_i$, $\bM_i^0 = \bdiag(\{\bQ_i\}_{j \in \calN_i})$ which are substantially different than the values used in Section \ref{sec: sim car} after proper tuning.  Table \ref{tab: ppa} shows the required ADMM iterations until convergence for each case, while also varying how frequently we perform an adaptation, i.e., $n_{\text{adapt}}$. The results indicate that both the centralized and decentralized schemes will adapt the parameters such that convergence is achieved faster than without adaptation. The best performance is achieved using the decentralized scheme with $n_{\text{adapt}} = 10$, while in all cases, it performs either better or equally well with the centralized one. Therefore, the proposed scheme not only maintains the decentralized nature of the algorithms, but also demonstrates a favorable performance against the equivalent centralized one. 

\subsection{Scalability Comparison with Other Methods}

Next, the scalability of ND-DDP and MD-DDP is compared against centralized AL-DDP and centralized/decentralized SQP. All simulations were in Matlab R2021b using a laptop computer with an 11th Gen Intel(R) Core(TM) i7-11800H @ 2.30GHz and a 32GB RAM memory. For the SQP methods, the NLP solver \texttt{fmincon} of the Matlab Optimization Toolbox is used. In the centralized methods, Problem \ref{multi-agent optimal control problem 1} is directly provided to the solvers. For decentralized SQP, we solve the Step 1 problems of ND-DDP with SQP instead of AL-DDP. The computational times for decentralized methods are measured by considering the ``slowest'' agent at every step. ND-DDP and decentralized SQP are terminated after $N=50$ ADMM iterations and MD-DDP after $N=200$ iterations. 

\begin{table*}[]
\centering
\begin{tabular}{@{}cccccccccccc@{}}
\hline 
Method & $M=2$ & $M=4$ & $M=8$ & $M=16$ & $M=32$ & $M=64$ & $M=256$ & $M=1024$ & $M=4096$ \\
\hline 
ND-DDP (Proposed) & 36s & 1m 11s & 2m 58s & 6m 4s & 7m 35s & 8m 18s & 10m 38s & - & - \\
MD-DDP (Proposed) & 40s & 48s & \textbf{1m 15s} & \textbf{1m 41s} & \textbf{1m 49s} & \textbf{1m 56s} & \textbf{2m 2s} & \textbf{2m 24s} & \textbf{2m 46s} \\
Centralized AL-DDP \cite{aoyama2020constrainedDDPicra} & \textbf{11s} & \textbf{34s} & 2m 3s & 5m 2s & 25m 56s & 2h 47m & - & - & - \\
Decentralized SQP  & 8m 28s & 21m 51s & 42m 37s & 1h 47m & - & - & - & - & -\\
Centralized SQP  & 2m 8s & 7m 30s & 31m 45s & 2h 32m & - & - & - & - & -\\
\hline
\end{tabular}
\caption{Computational times comparison between ND-DDP, MD-DDP, Centralized AL-DDP, Decentralized SQP and Centralized SQP for an increasing number of agents $M$ in the multi-car formation task. }
\label{tab: comp times}
\end{table*}

\begin{figure*}[!t]
\centering
\hfil
\subfloat{
\begin{tikzpicture}
    \node[anchor=south west,inner sep=0] at (0,0){    \includegraphics[width=0.31\textwidth, trim={8cm 0cm 9cm 0.5cm},clip]{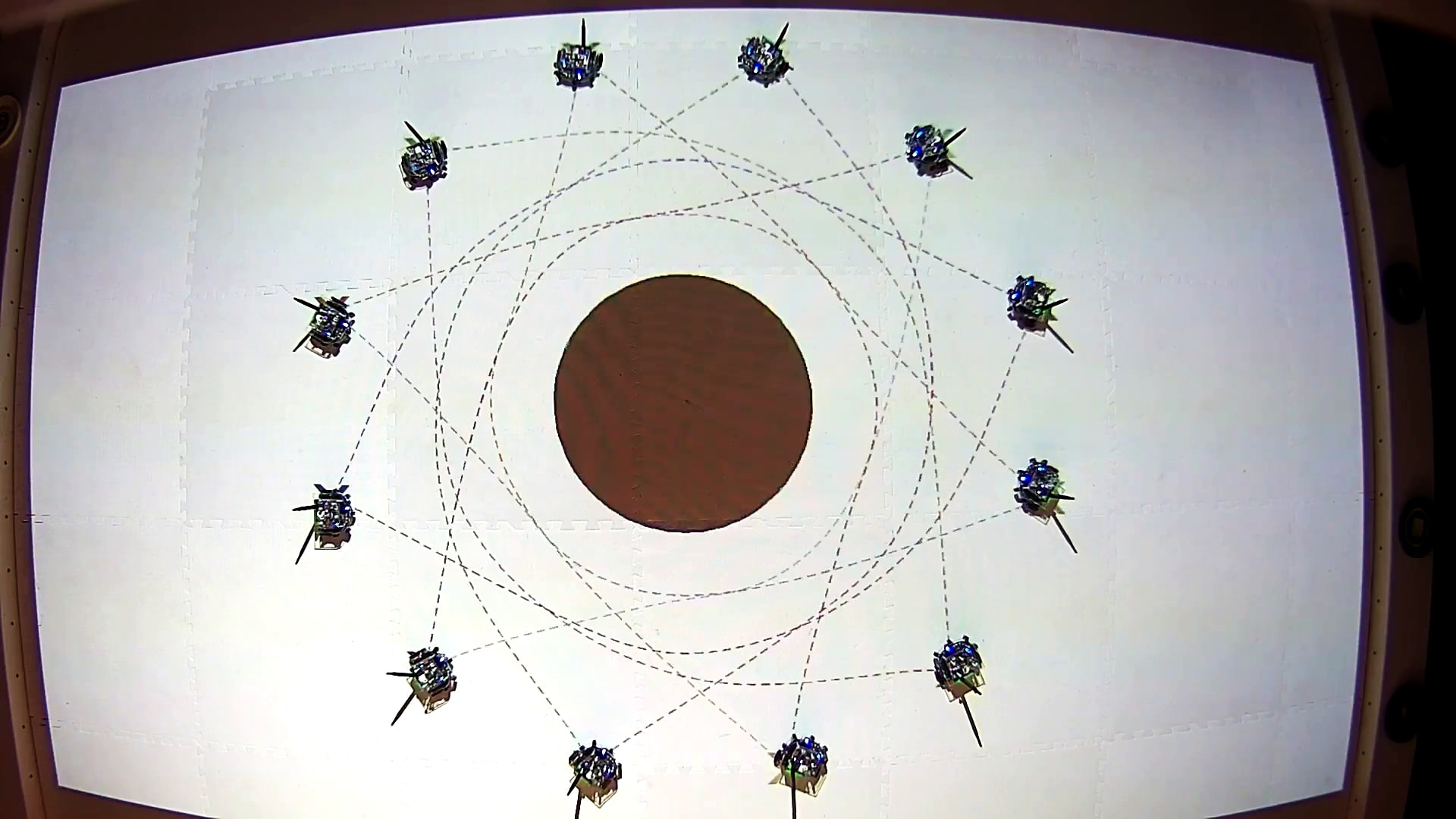}};
    \node[align=center, text=NavyBlue] (c) at (5, 0.3) {$k = 0$};
\end{tikzpicture}
\label{fig_rob_swapp_1}}
\hfil
\subfloat{
\begin{tikzpicture}
    \node[anchor=south west,inner sep=0] at (0,0){    \includegraphics[width=0.31\textwidth, trim={8cm 0cm 9cm 0.5cm},clip]{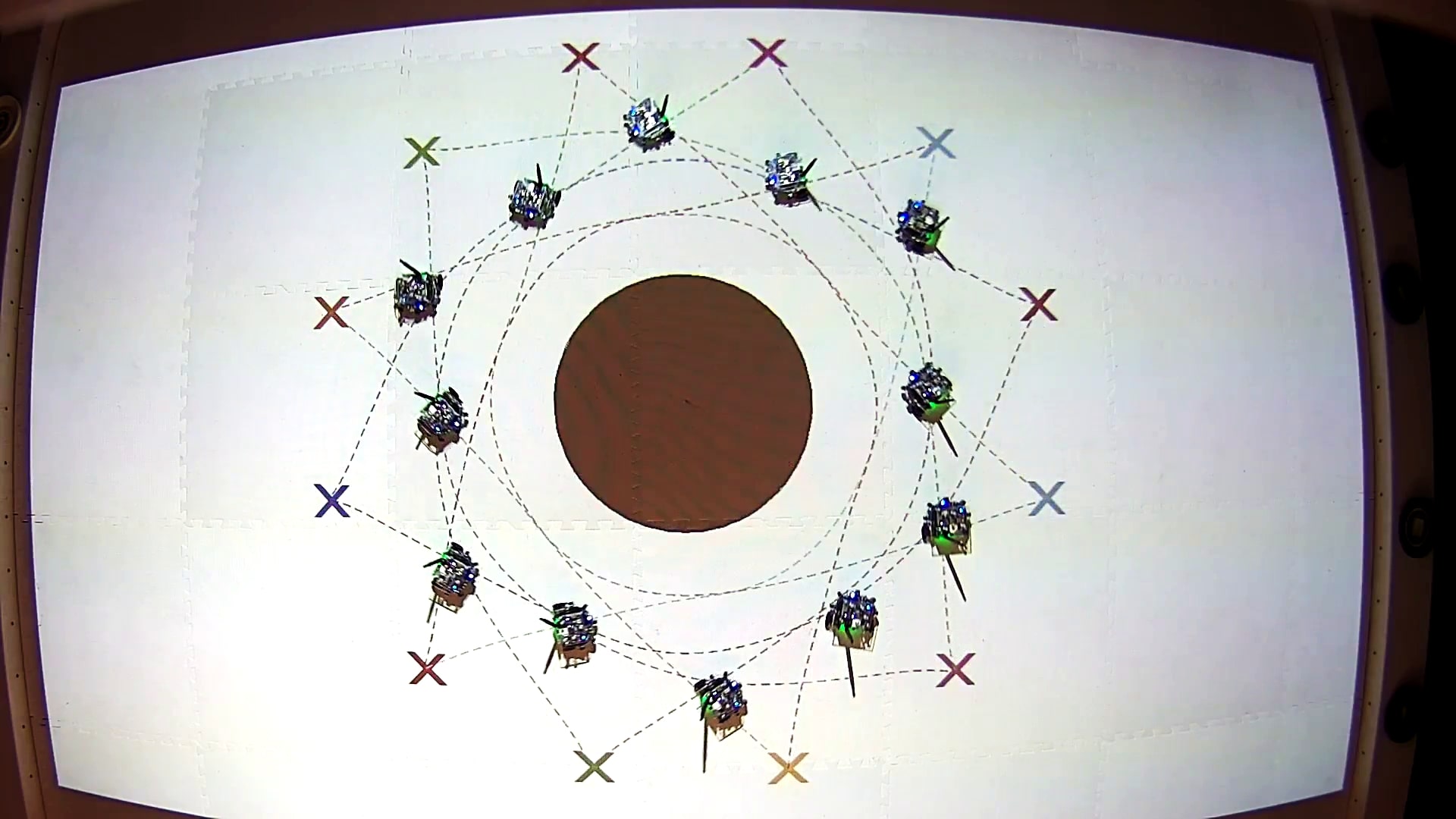}};
    \node[align=center, text=NavyBlue] (c) at (4.92, 0.3) {$k = 60$};
\end{tikzpicture}
\label{fig_rob_swapp_2}}
\hfil
\subfloat{
\begin{tikzpicture}
    \node[anchor=south west,inner sep=0] at (0,0){    \includegraphics[width=0.31\textwidth, trim={8cm 0cm 9cm 0.5cm},clip]{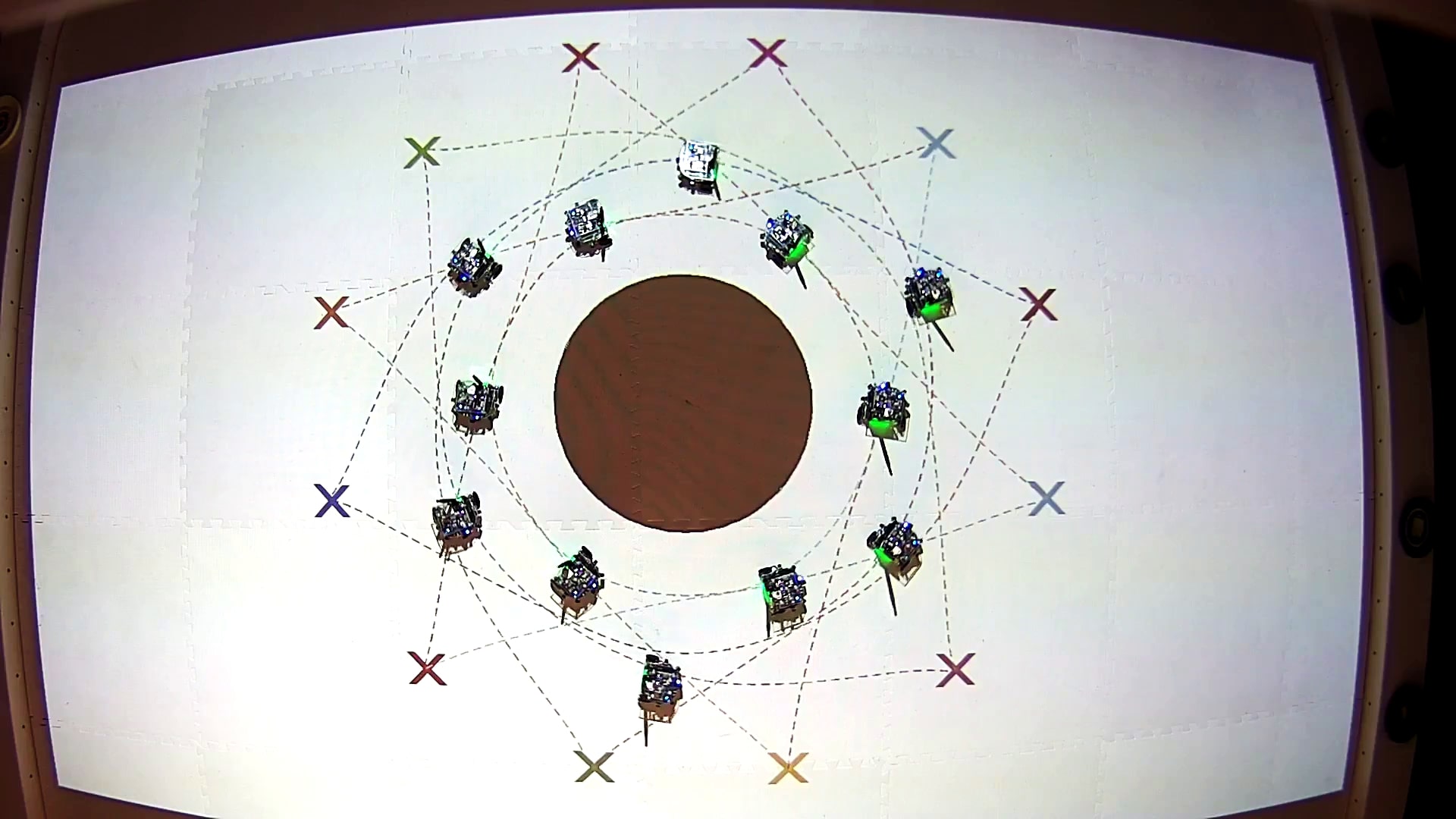}};
    \node[align=center, text=NavyBlue] (c) at (4.84, 0.3) {$k = 150$};
\end{tikzpicture}
\label{fig_rob_swapp_3}}
\hfil
\\
\hfil
\subfloat{
\begin{tikzpicture}
    \node[anchor=south west,inner sep=0] at (0,0){    \includegraphics[width=0.31\textwidth, trim={8cm 0cm 9cm 0.5cm},clip]{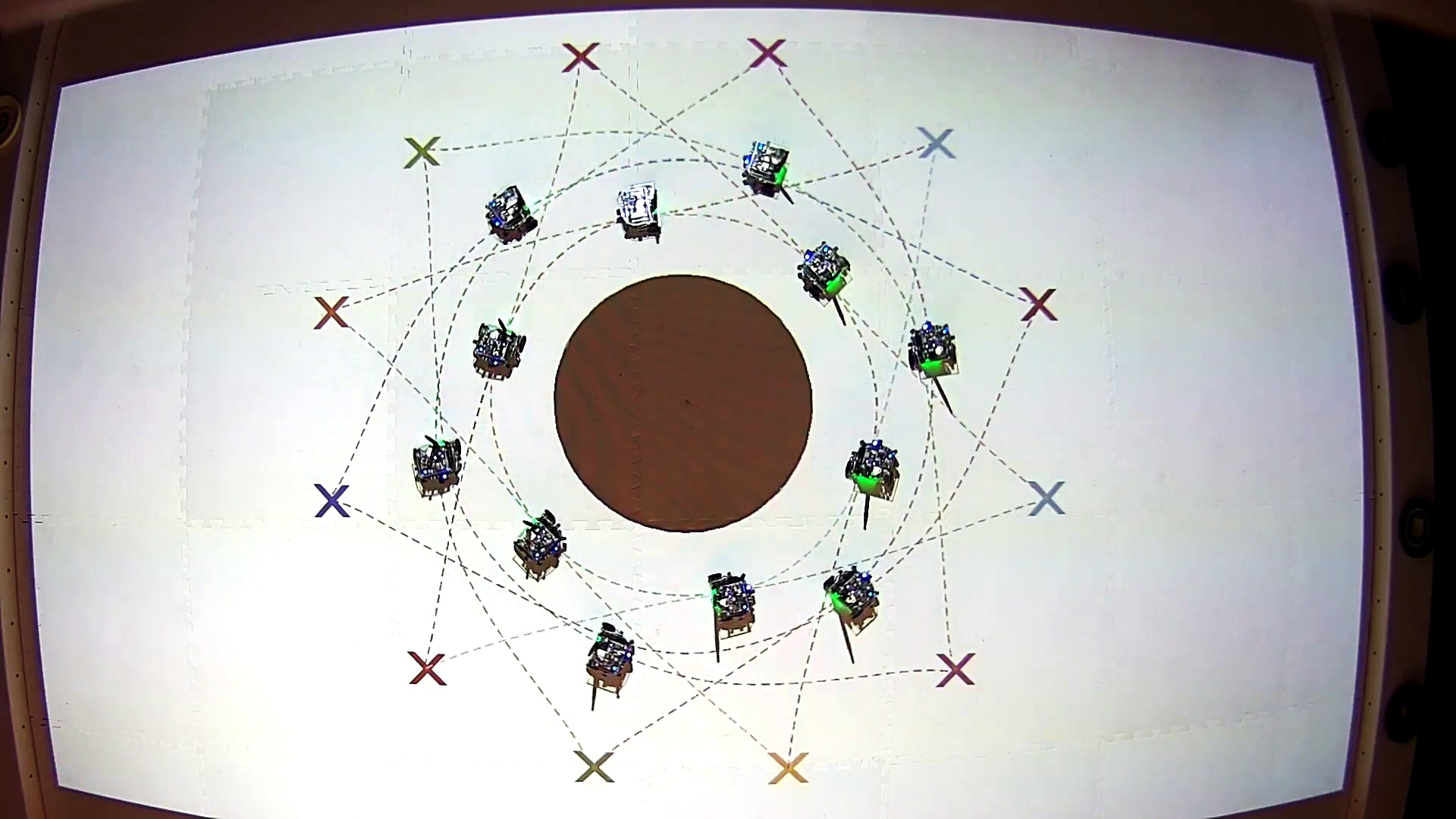}};
    \node[align=center, text=NavyBlue] (c) at (4.84, 0.3) {$k = 300$};
\end{tikzpicture}
\label{fig_rob_swapp_4}}
\hfil
\subfloat{
\begin{tikzpicture}
    \node[anchor=south west,inner sep=0] at (0,0){    \includegraphics[width=0.31\textwidth, trim={8cm 0cm 9cm 0.5cm},clip]{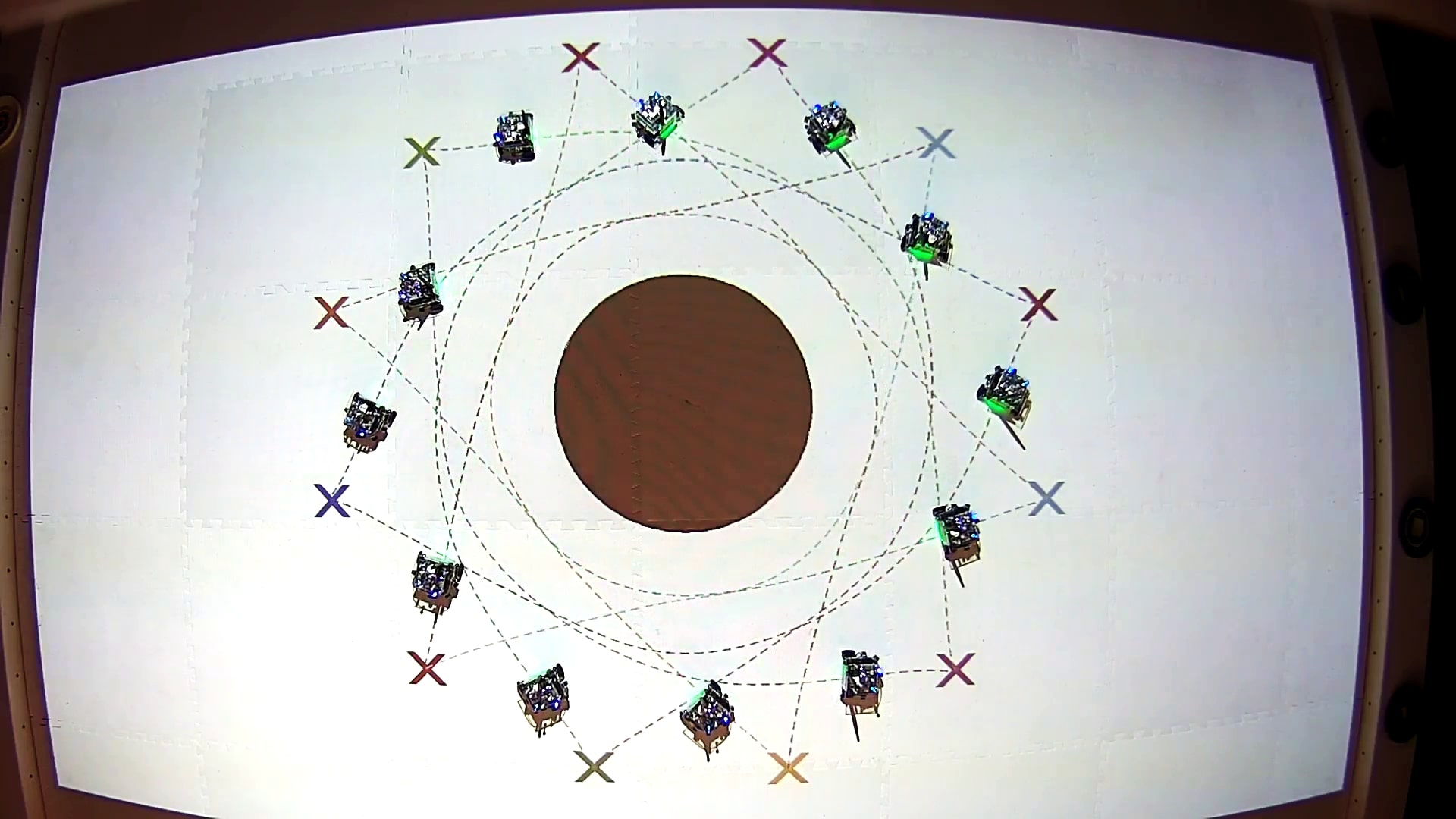}};
    \node[align=center, text=NavyBlue] (c) at (4.84, 0.3) {$k = 450$};
\end{tikzpicture}
\label{fig_rob_swapp_5}}
\hfil
\subfloat{
\begin{tikzpicture}
    \node[anchor=south west,inner sep=0] at (0,0){    \includegraphics[width=0.31\textwidth, trim={8cm 0cm 9cm 0.5cm},clip]{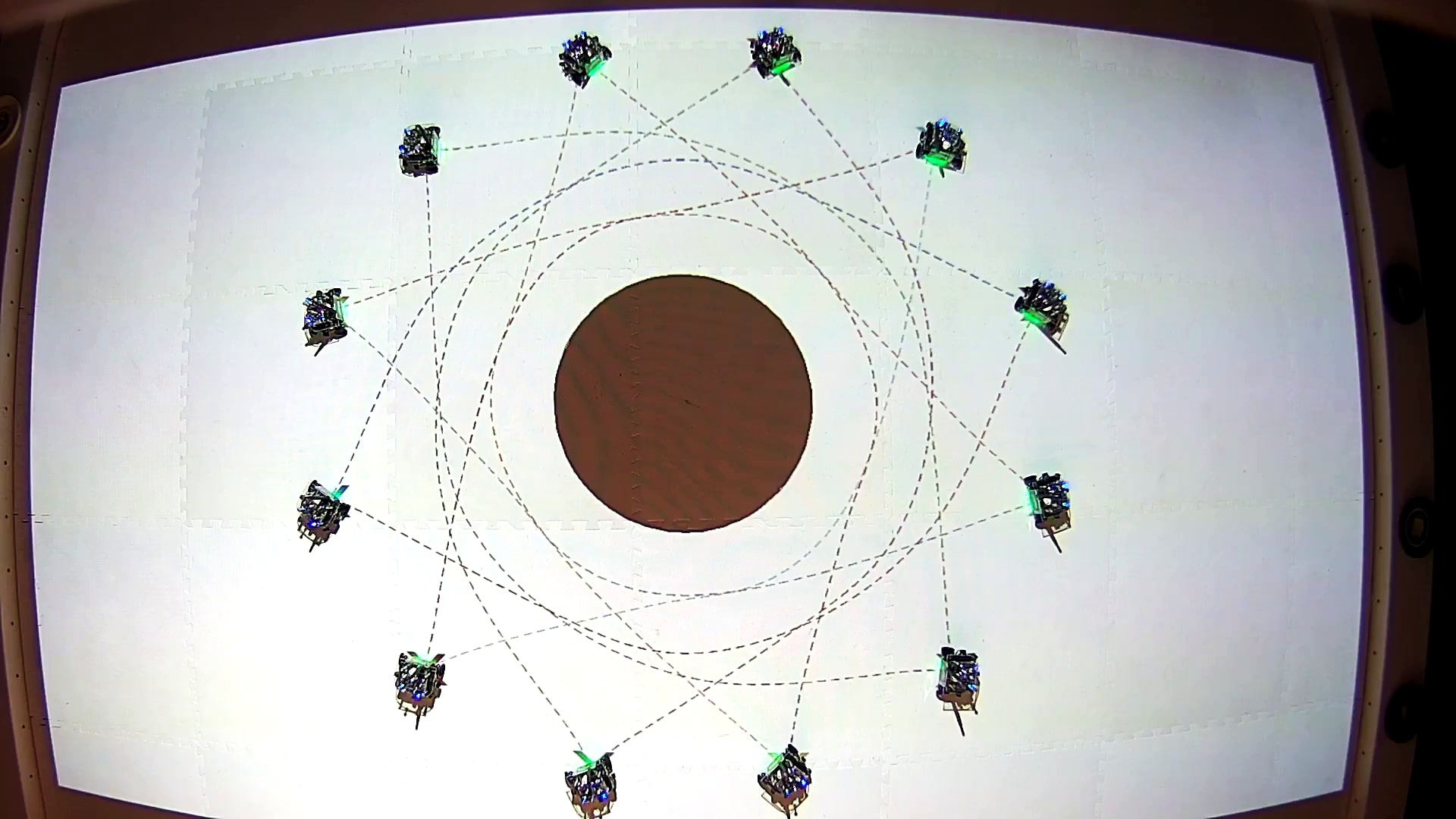}};
    \node[align=center, text=NavyBlue] (c) at (4.84, 0.3) {$k = 900$};
\end{tikzpicture}
\label{fig_rob_swapp_6}}
\hfil
\caption{Hardware experiment: ``Circle Swapping'' task with MD-DDP (12 robots). Snapshots at different time instants.}
\label{fig_rob_swapp}
\end{figure*}
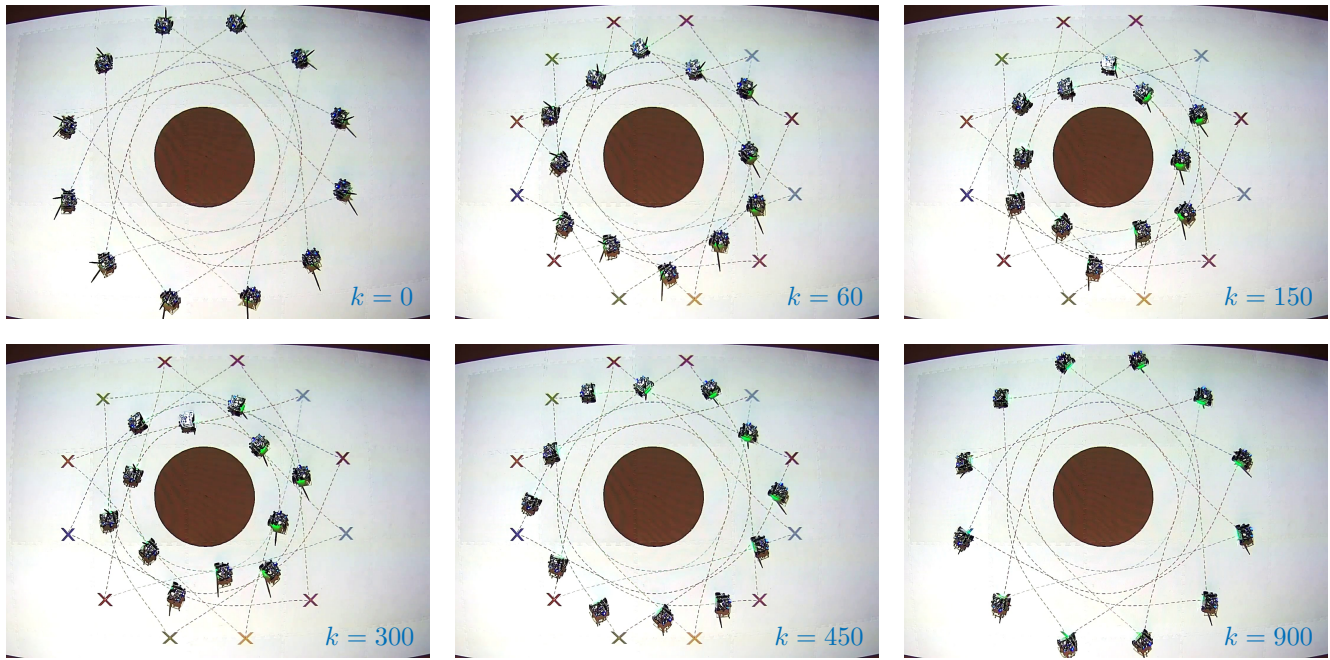

\begin{figure*}[!t]
\centering
\hfil
\subfloat{
\begin{tikzpicture}
    \node[anchor=south west,inner sep=0] at (0,0){    \includegraphics[width=0.31\textwidth, trim={8cm 0cm 9cm 0.5cm},clip]{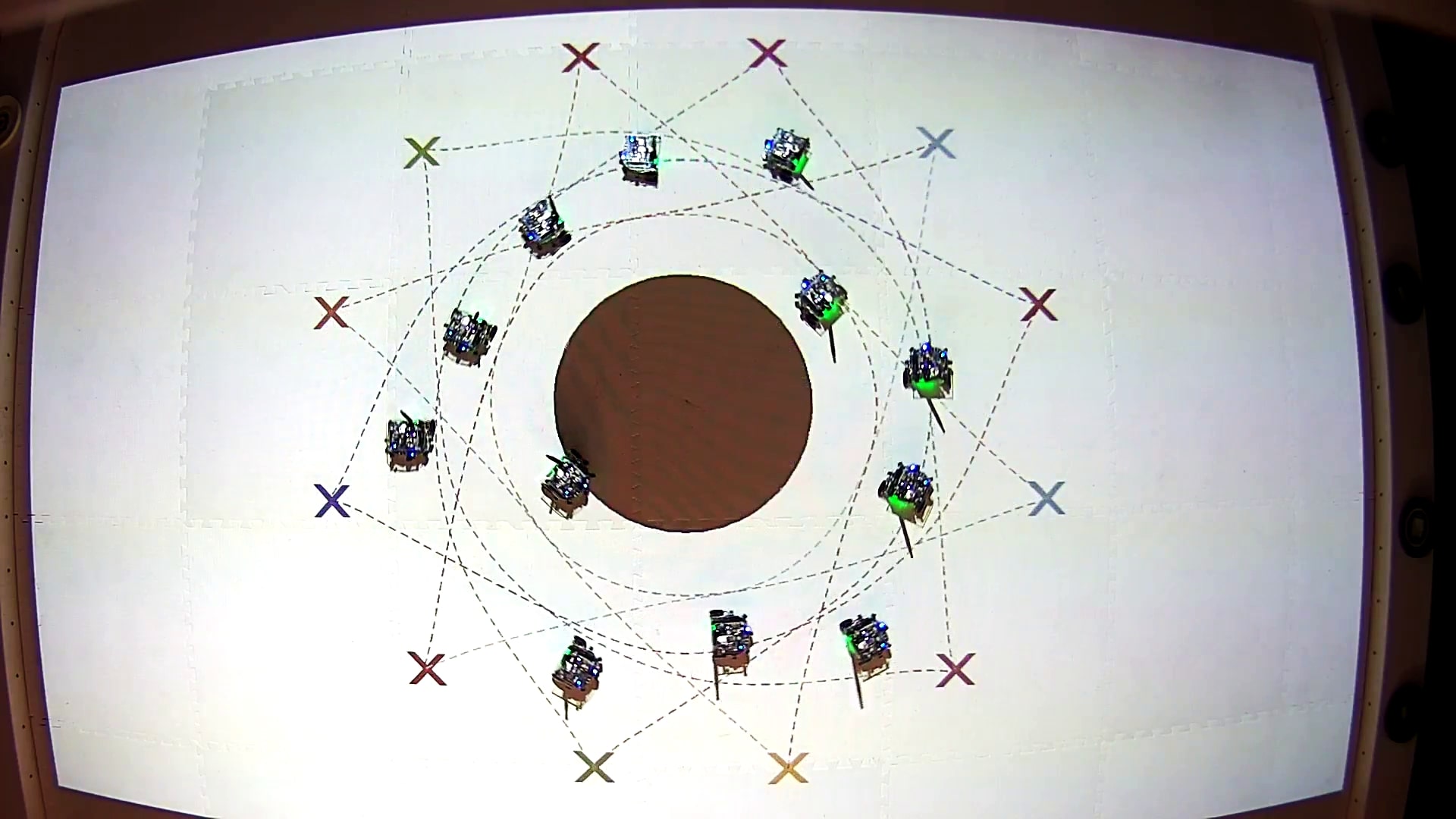}};
    \node[align=center, text=NavyBlue] (c) at (5, 0.3) {$k = 150$};
\end{tikzpicture}
\label{fig_rob_swapp_openloop_1}}
\hfil
\subfloat{
\begin{tikzpicture}
    \node[anchor=south west,inner sep=0] at (0,0){    \includegraphics[width=0.31\textwidth, trim={8cm 0cm 9cm 0.5cm},clip]{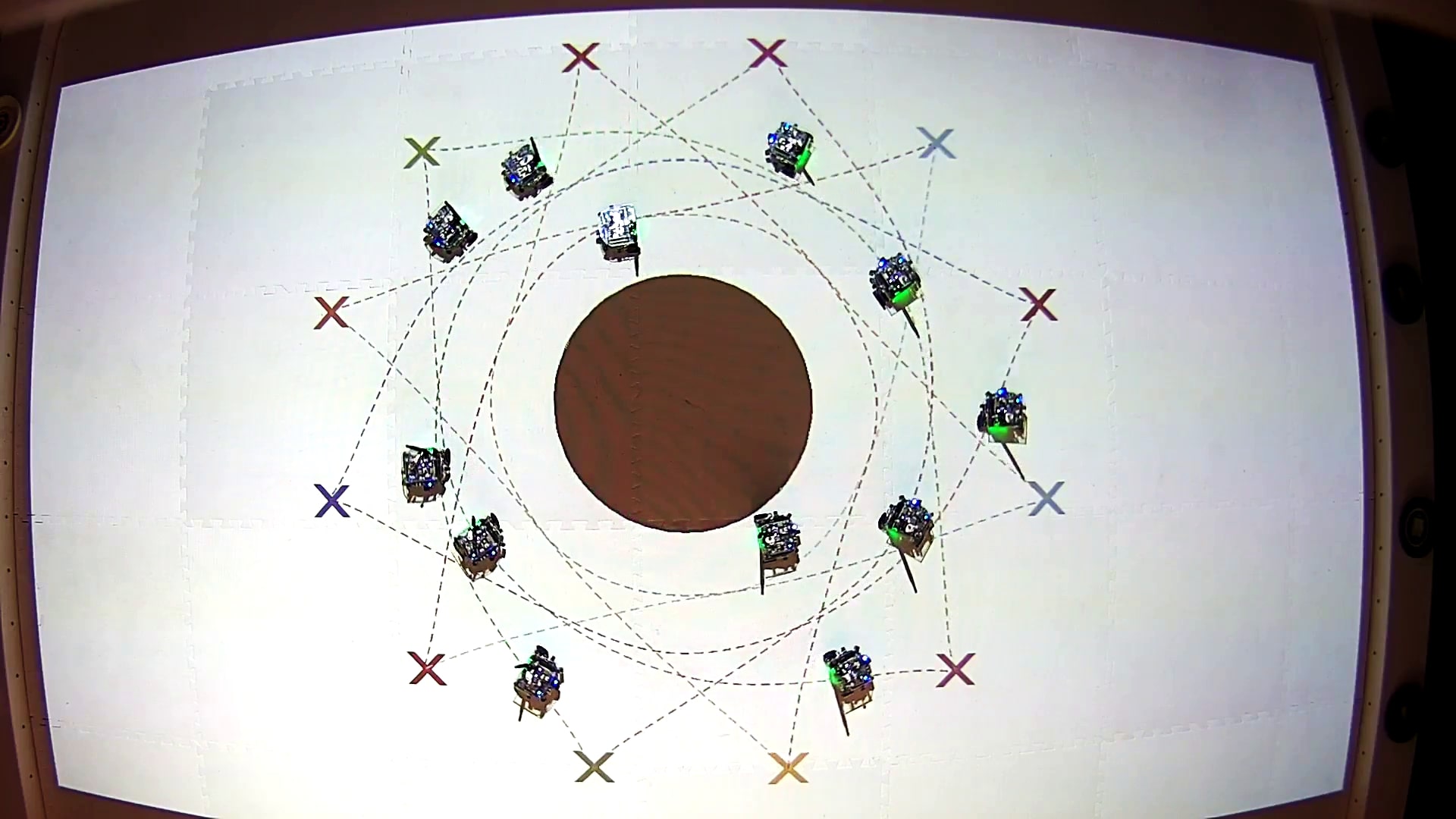}};
    \node[align=center, text=NavyBlue] (c) at (5, 0.3) {$k = 300$};
\end{tikzpicture}
\label{fig_rob_swapp_openloop_2}}
\hfil
\subfloat{
\begin{tikzpicture}
    \node[anchor=south west,inner sep=0] at (0,0){    \includegraphics[width=0.31\textwidth, trim={8cm 0cm 9cm 0.5cm},clip]{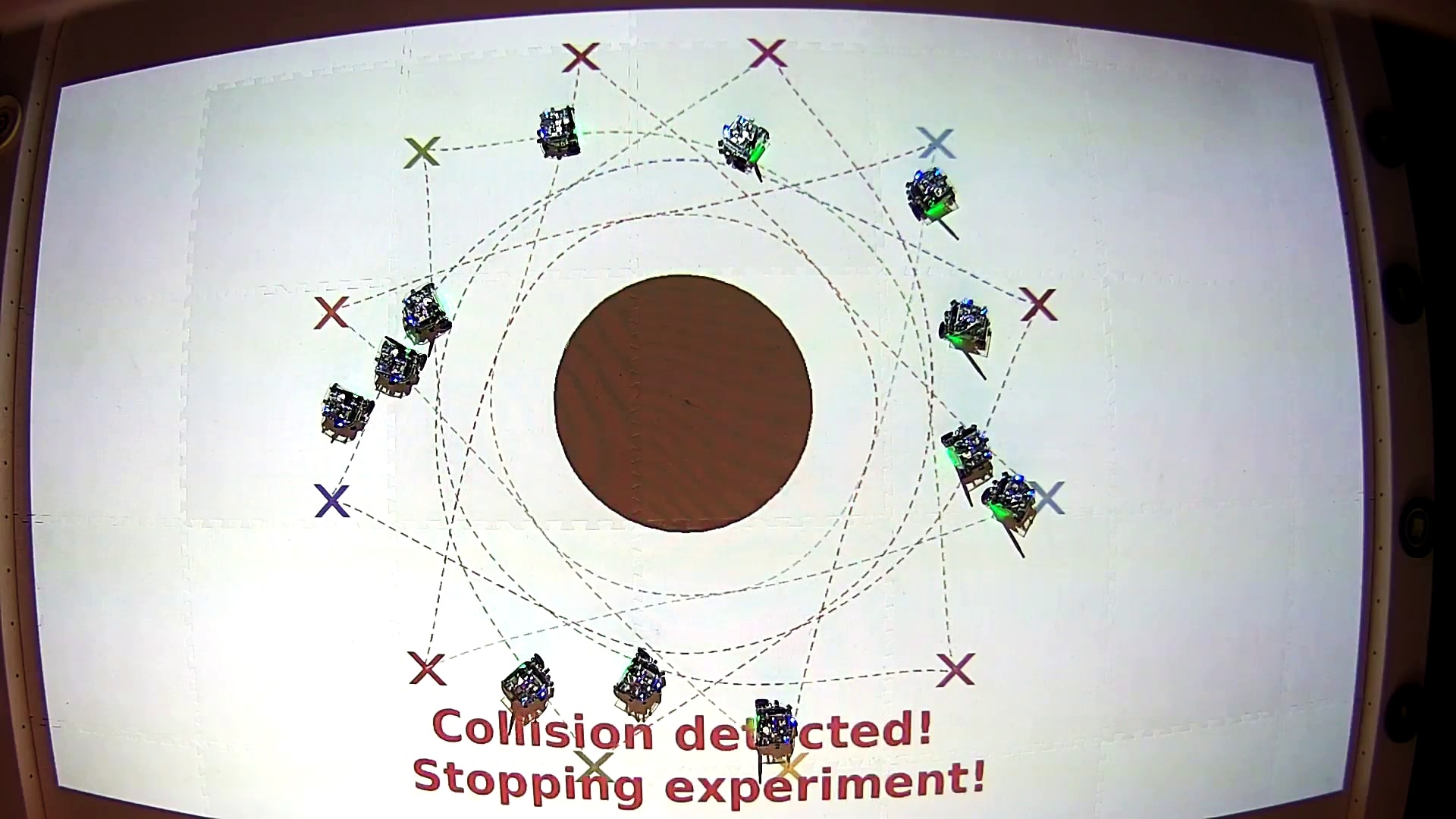}};
    \node[align=center, text=NavyBlue] (c) at (5, 0.3) {$k = 450$};
\end{tikzpicture}
\label{fig_rob_swapp_openloop_3}}
\hfil
\caption{Hardware experiment: Swapping task with MD-DDP (12 robots) using open-loop policies. Without any feedback, the robots end up colliding and the experiment is terminated.}
\label{fig_rob_swapp_openloop}
\end{figure*}
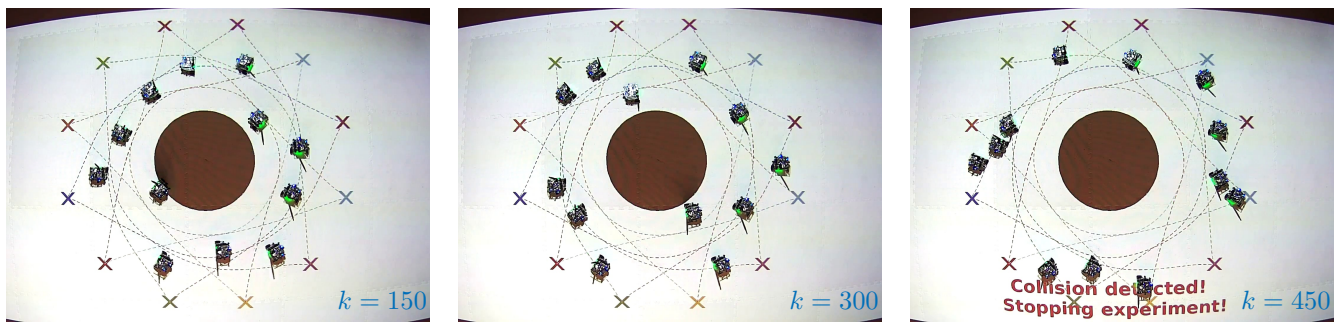

All algorithms were tested on the multi-car formation task for an increasing number of agents $M$. The computational times of all methods are displayed in Table \ref{tab: comp times}. For $M=2, 4, 8$, we used $|\calN_i| = 2, 3, 5$, respectively, while for $M \geq 16$, we set $|\calN_i| = 9$. As $M$ increases, MD-DDP and ND-DDP demonstrate a superior scalability against the other methods. As predicted by its computational complexity (Remark \ref{MD-DDP comp complexity}), the computational demands of MD-DDP do not increase significantly as $M$ grows. The same applies for ND-DDP once the neighborhood size $|\calN_i|$ (and thus $\tilde{q}_i$) remains fixed. The SQP methods present significantly higher computational demands which is mainly attributed to the fact that all states/controls are stacked into one high-dimensional vector for the entire time horizon $K$. On the other hand, DDP scales linearly w.r.t. $K$, which explains its favorable performance for long time horizons. Nevertheless, centralized AL-DDP still demonstrates an inferior scalability compared to ND-DDP and MD-DDP, since it has a cubic complexity in $M$ (see Remark \ref{ND-DDP comp complexity}). Finally, note that the Step 2 computations in MD-DDP were performed sequentially here (and not in parallel; see Remark \ref{MDDDP remark high par}), so the capabilities of the method were not fully exploited.

\section{Hardware Experiments}
\label{sec:Hardware_Experiments}

To illustrate their applicability on real systems, ND-DDP and MD-DDP were employed on the Robotarium platform \cite{wilson2020robotarium} for controlling a multi-robot team. All robots have unicycle dynamics with states $\bx_{i,k} = \begin{bmatrix}
\mathrm{x}_{i,k} & \mathrm{y}_{i,k} & \theta_{i,k}
\end{bmatrix}\T \in \Rb^3
$ 
and control inputs 
$\bu_{i,k} = \begin{bmatrix}
v_{i,k} & \omega_{i,k}
\end{bmatrix}\T \in \Rb^2
$, 
where $(\mathrm{x}_{i,k}, \mathrm{y}_{i,k})$ is the 2D position, $\theta_{i,k}$ is the angle and $v_{i,k}, \ \omega_{i,k}$ are the linear and angular velocities of the $i$-th robot at time $k$. The time step and time horizon are set to $dt=0.033$s and $K=900$. Constraints on the controls of the robots appear through bounding their wheel speeds, 
\begin{equation}
- \bd_{\text{max}} \leq \bC \bu_i  \leq \bd_{\text{max}},  
\end{equation}
with 
\begin{equation}
\bC = \frac{1}{2R} \begin{bmatrix} 2  & L \\ 2 & -L  \end{bmatrix}, \quad \bd_{\text{max}} = \begin{bmatrix} v_{\text{wheel, max}} \\ v_{\text{wheel, max}}
\end{bmatrix},
\end{equation}
where $R = 0.016$m and $L = 0.11$m are the wheel radius and axle length of each robot, respectively, while $v_{\text{wheel, max}} = 12.5$ \text{rad/s} is the maximum wheel speed. Moreover, the robots are subject to the collision avoidance and connectivity maintenance constraints \eqref{obs avoidance constraints}-\eqref{conn maint constraints} with $d_o = 0.15$m, $d_{\text{col}} = 0.2$m and $d_{\text{con}} = 0.5$m. Each robot's cost is quadratic as in \eqref{quadratic agent cost} with $\bQ_i = \diag (100,100,0)$, $\bR_i = \diag(100, 10)$ and $\bQ_i^{\text{f}} = \diag(300, 300, 30)$. To enhance the robustness of the proposed methods against model uncertainty, we utilize the feedback terms provided by DDP and apply the following closed-loop policies 

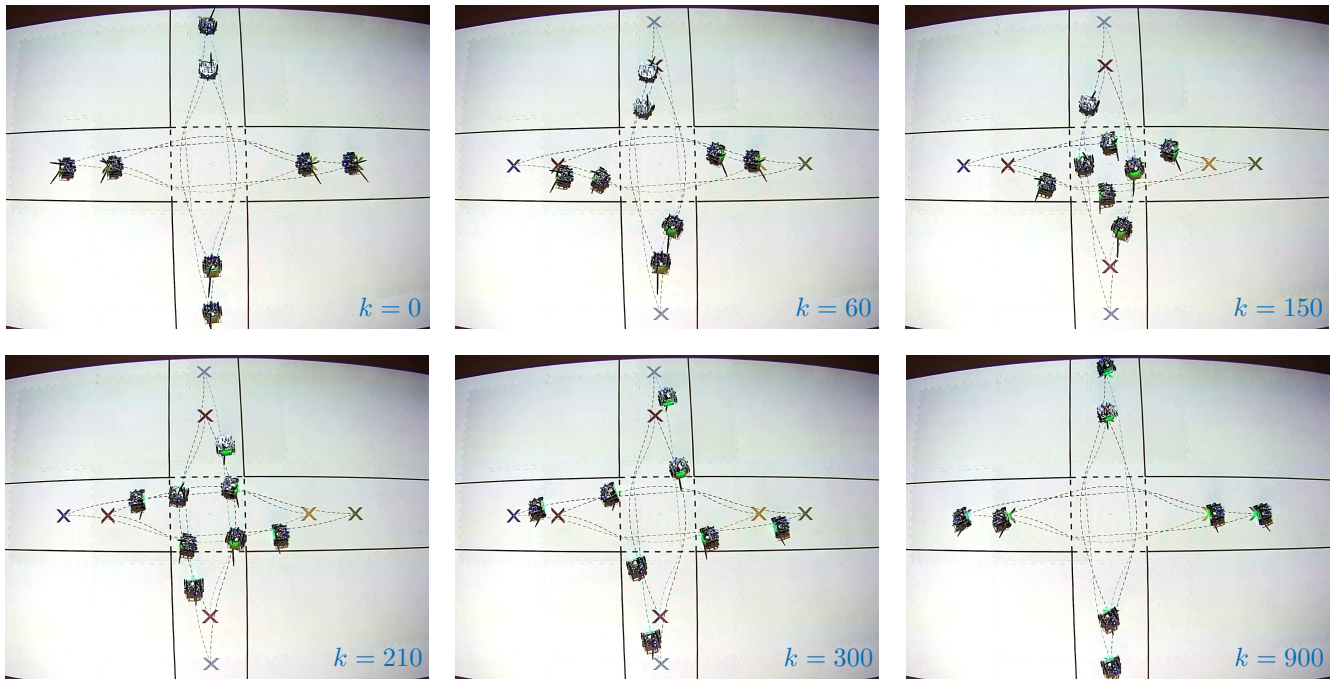
\begin{figure*}[!t]
\centering
\hfil
\subfloat{
\begin{tikzpicture}
    \node[anchor=south west,inner sep=0] at (0,0){    \includegraphics[width=0.31\textwidth, trim={8cm 0cm 10cm 0cm},clip]{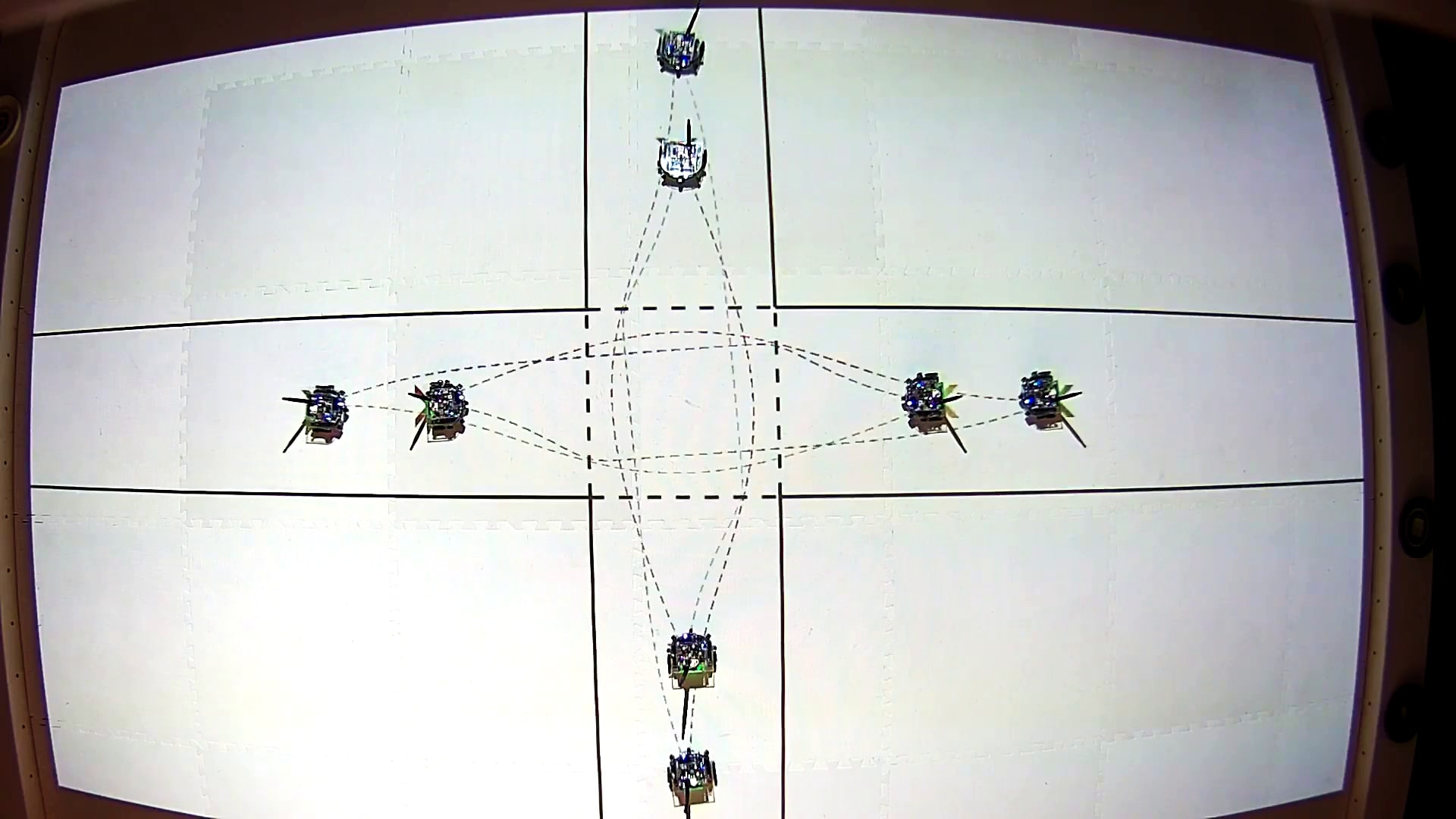}};
    \node[align=center, text=NavyBlue] (c) at (5.1, 0.3) {$k = 0$};
\end{tikzpicture}
\label{fig_rob_inter_1}}
\hfil
\subfloat{
\begin{tikzpicture}
    \node[anchor=south west,inner sep=0] at (0,0){    \includegraphics[width=0.31\textwidth, trim={8cm 0cm 10cm 0cm},clip]{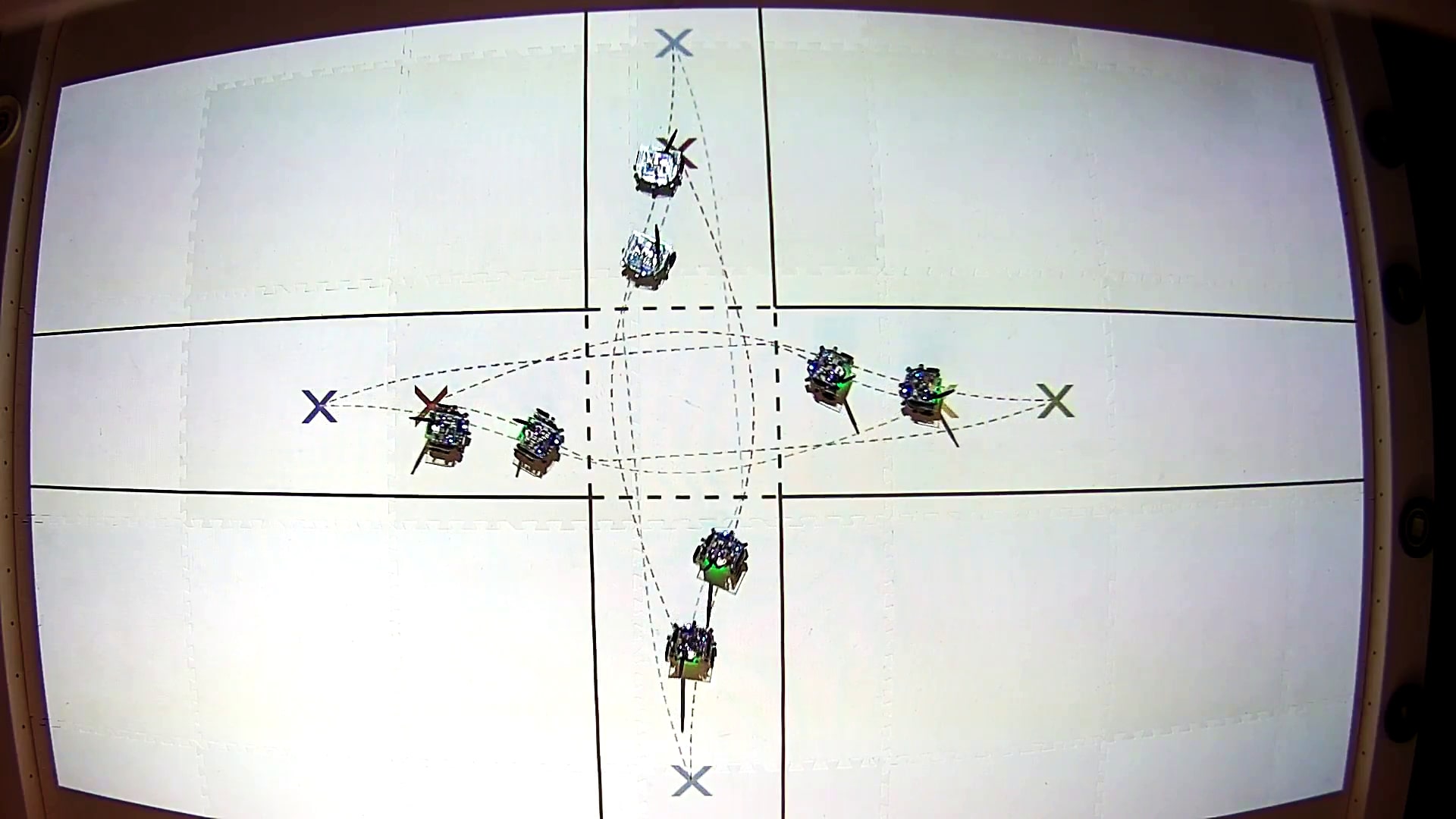}};
    \node[align=center, text=NavyBlue] (c) at (5.03, 0.3) {$k = 60$};
\end{tikzpicture}
\label{fig_rob_inter_2}}
\hfil
\subfloat{
\begin{tikzpicture}
    \node[anchor=south west,inner sep=0] at (0,0){    \includegraphics[width=0.31\textwidth, trim={8cm 0cm 10cm 0cm},clip]{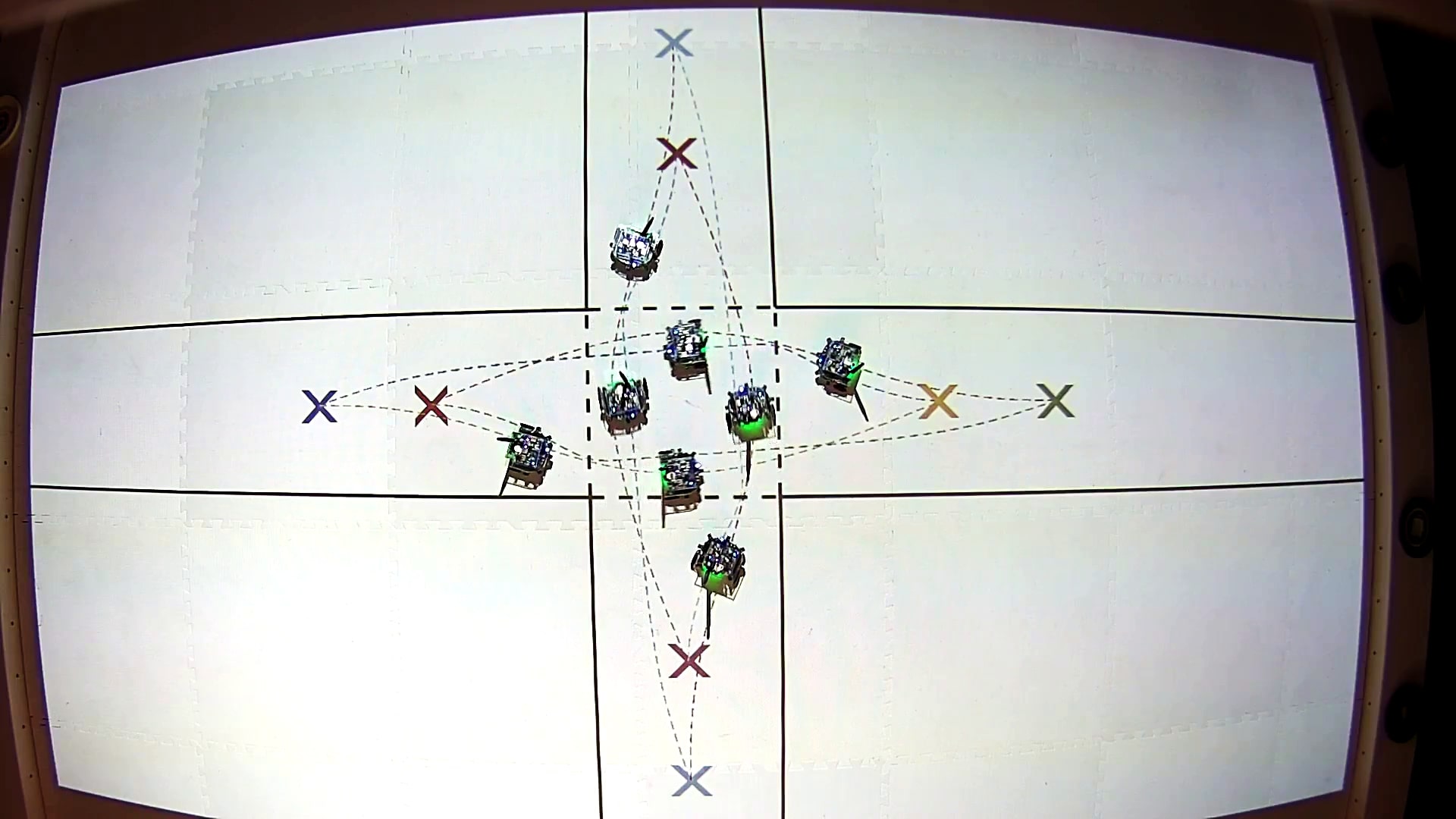}};
    \node[align=center, text=NavyBlue] (c) at (4.96, 0.3) {$k =150$};
\end{tikzpicture}
\label{fig_rob_inter_3}}
\hfil
\\
\hfil
\subfloat{
\begin{tikzpicture}
    \node[anchor=south west,inner sep=0] at (0,0){    \includegraphics[width=0.31\textwidth, trim={8cm 0cm 10cm 0cm},clip]{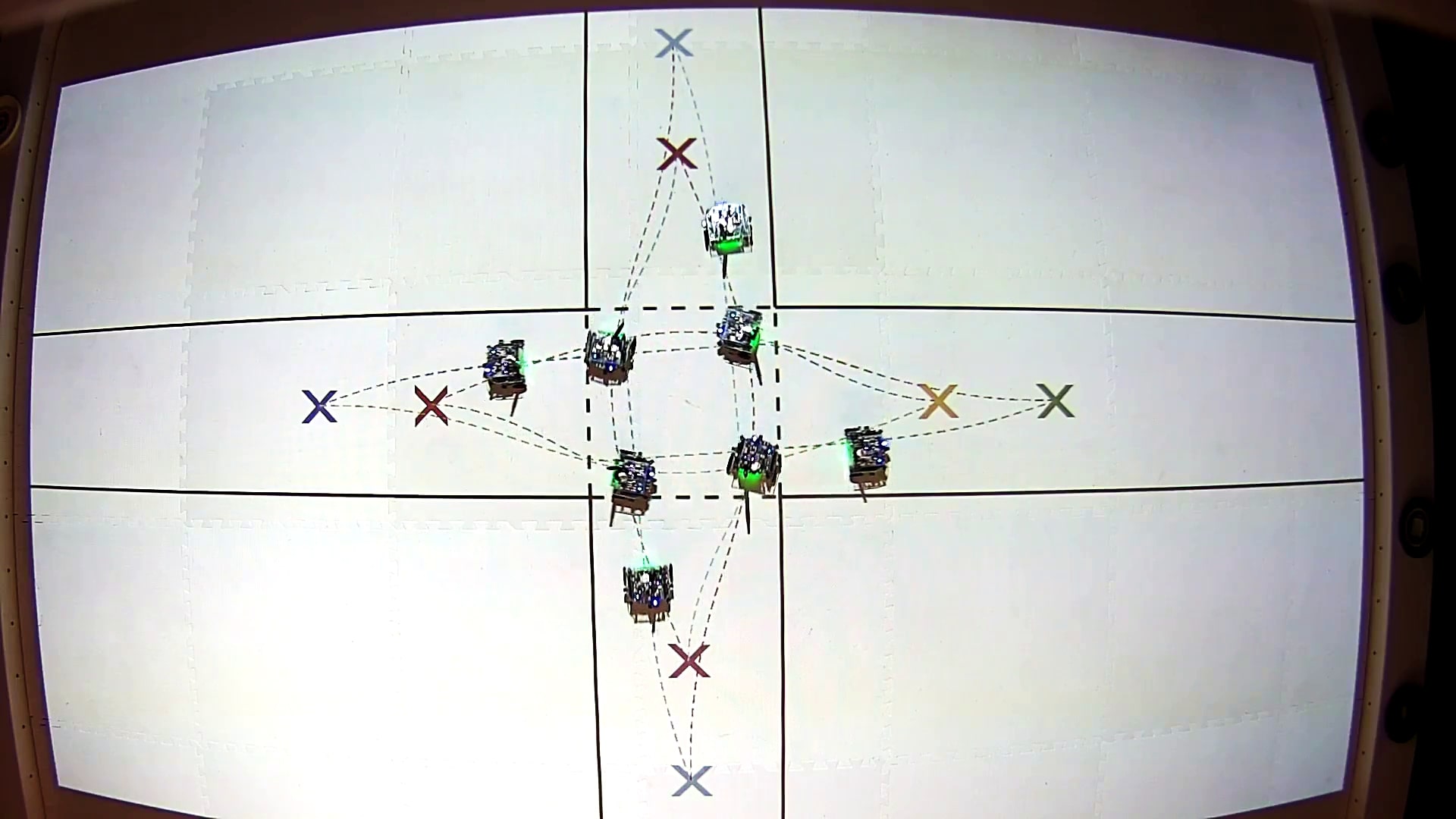}};
    \node[align=center, text=NavyBlue] (c) at (4.96, 0.3) {$k = 210$};
\end{tikzpicture}
\label{fig_rob_inter_4}}
\hfil
\subfloat{
\begin{tikzpicture}
    \node[anchor=south west,inner sep=0] at (0,0){    \includegraphics[width=0.31\textwidth, trim={8cm 0cm 10cm 0cm},clip]{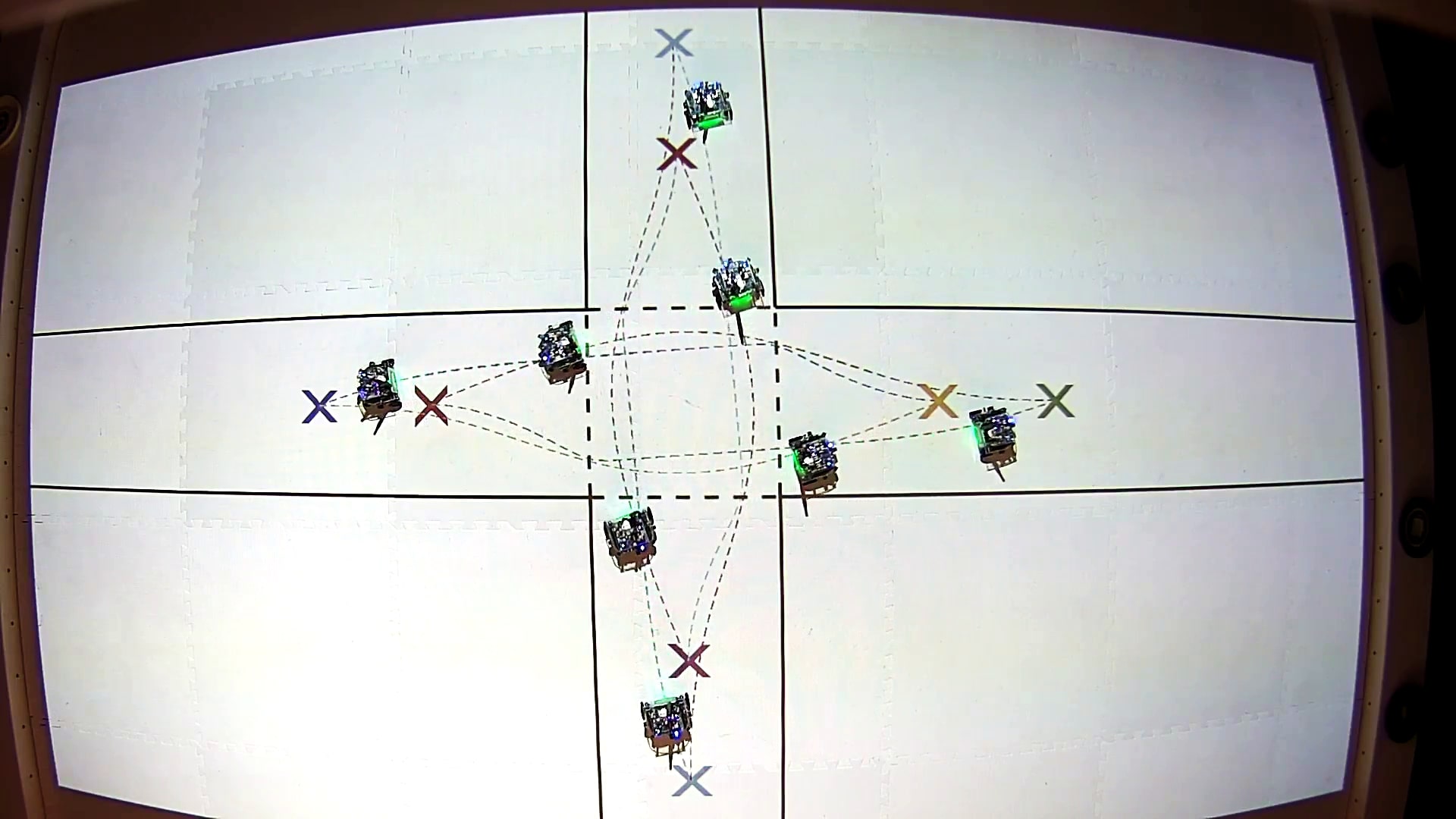}};
    \node[align=center, text=NavyBlue] (c) at (4.96, 0.3) {$k = 300$};
\end{tikzpicture}
\label{fig_rob_inter_5}}
\hfil
\subfloat{
\begin{tikzpicture}
    \node[anchor=south west,inner sep=0] at (0,0){    \includegraphics[width=0.31\textwidth, trim={8cm 0cm 10cm 0cm},clip]{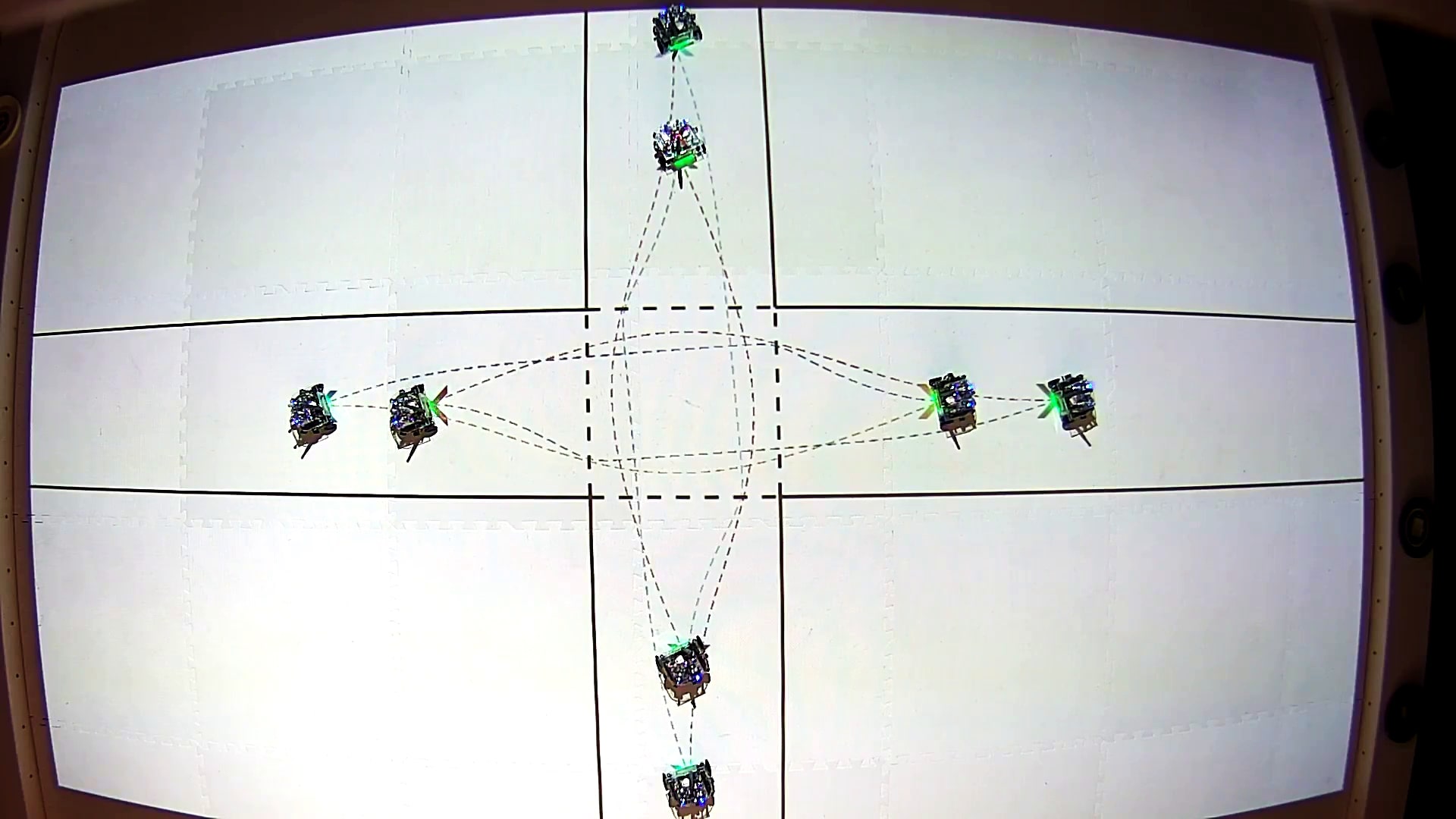}};
    \node[align=center, text=NavyBlue] (c) at (4.96, 0.3) {$k = 900$};
\end{tikzpicture}
\label{fig_rob_inter_6}}
\hfil
\caption{Hardware experiment: ``Intersection'' task with MD-DDP (8 robots). Snapshots at different time instants.}
\label{fig_rob_inter}
\end{figure*}
\begin{equation}
\bu_{i,k} = \bu_{i,k}^* + \bK_{i,k} (\bx_{i,k}^{\text{meas}} - \bx_{i,k}^*)
\label{robotarium closed loop}
\end{equation}
where $\bx_{i,k}^*, \bu_{i,k}^*$ are the computed optimal states and controls, $\bx_{i,k}^{\text{meas}}$ are the measured states and $\bK_{i,k}$ are the computed feedback gains of the $i$-th robot for time $k$. The methods are tested for the same tasks as in Section \ref{sec: sim car}. Results for both ND-DDP and MD-DDP are provided in the supplementary video, while the figures displayed in the main paper correspond to the latter. 

The first task that is presented is the ``circle swapping'' one with 12 robots (Fig. \ref{fig_rob_swapp}). All robots successfully reach to their targets while avoiding collisions with each other and the obstacle. In Fig. \ref{fig_rob_swapp_openloop}, the experiment is repeated using the open-loop policies $\bu_{i,k} = \bu_{i,k}^*$, instead of the closed-loop ones \eqref{robotarium closed loop}. Without any use of feedback, the robots are unable to follow their optimal state trajectories, which leads to unsafe behavior as collisions occur. This highlights the advantage of DDP methods for providing feedback control policies explicitly, in addition to the optimal control and state sequences.

\begin{figure*}[!t]
\centering
\hfil
\subfloat{
\begin{tikzpicture}
    \node[anchor=south west,inner sep=0] at (0,0){    \includegraphics[width=0.315\textwidth, trim={3cm 0cm 5.3cm 0.7cm},clip]{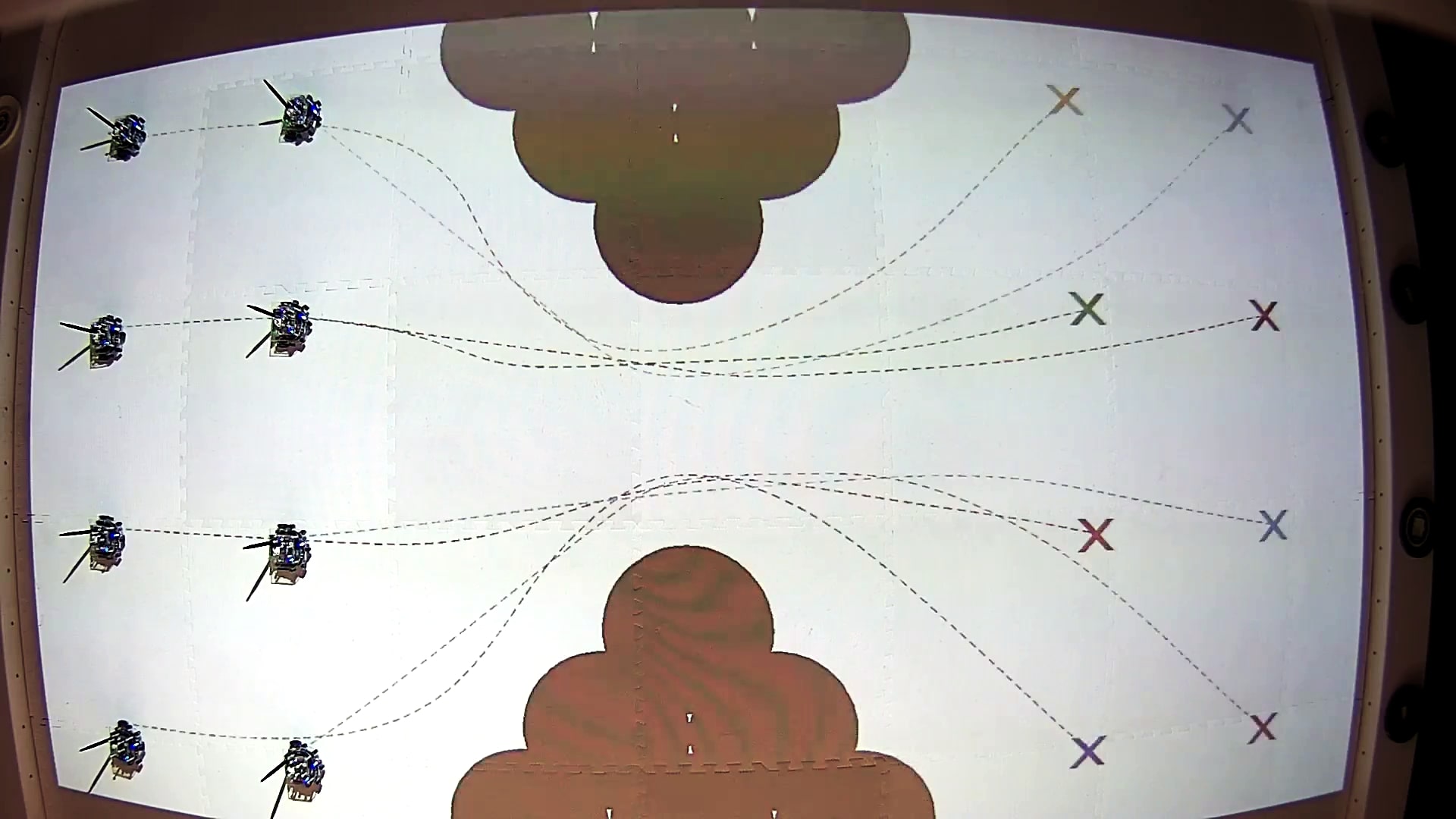}};
    \node[align=center, text=NavyBlue] (c) at (4.66, 0.20) {$k = 0$};
\end{tikzpicture}
\label{fig_rob_bottle_1}}
\hfil
\subfloat{
\begin{tikzpicture}
    \node[anchor=south west,inner sep=0] at (0,0){    \includegraphics[width=0.315\textwidth, trim={3cm 0cm 5.3cm 0.7cm},clip]{images/robotarium_bottleneck_8agents_MDDDP_t_0_edited4.jpg}};
    \node[align=center, text=NavyBlue] (c) at (4.73, 0.20) {$k = 90$};
\end{tikzpicture}
\label{fig_rob_bottle_2}}
\hfil
\subfloat{
\begin{tikzpicture}
    \node[anchor=south west,inner sep=0] at (0,0){    \includegraphics[width=0.315\textwidth, trim={3cm 0cm 5.3cm 0.7cm},clip]{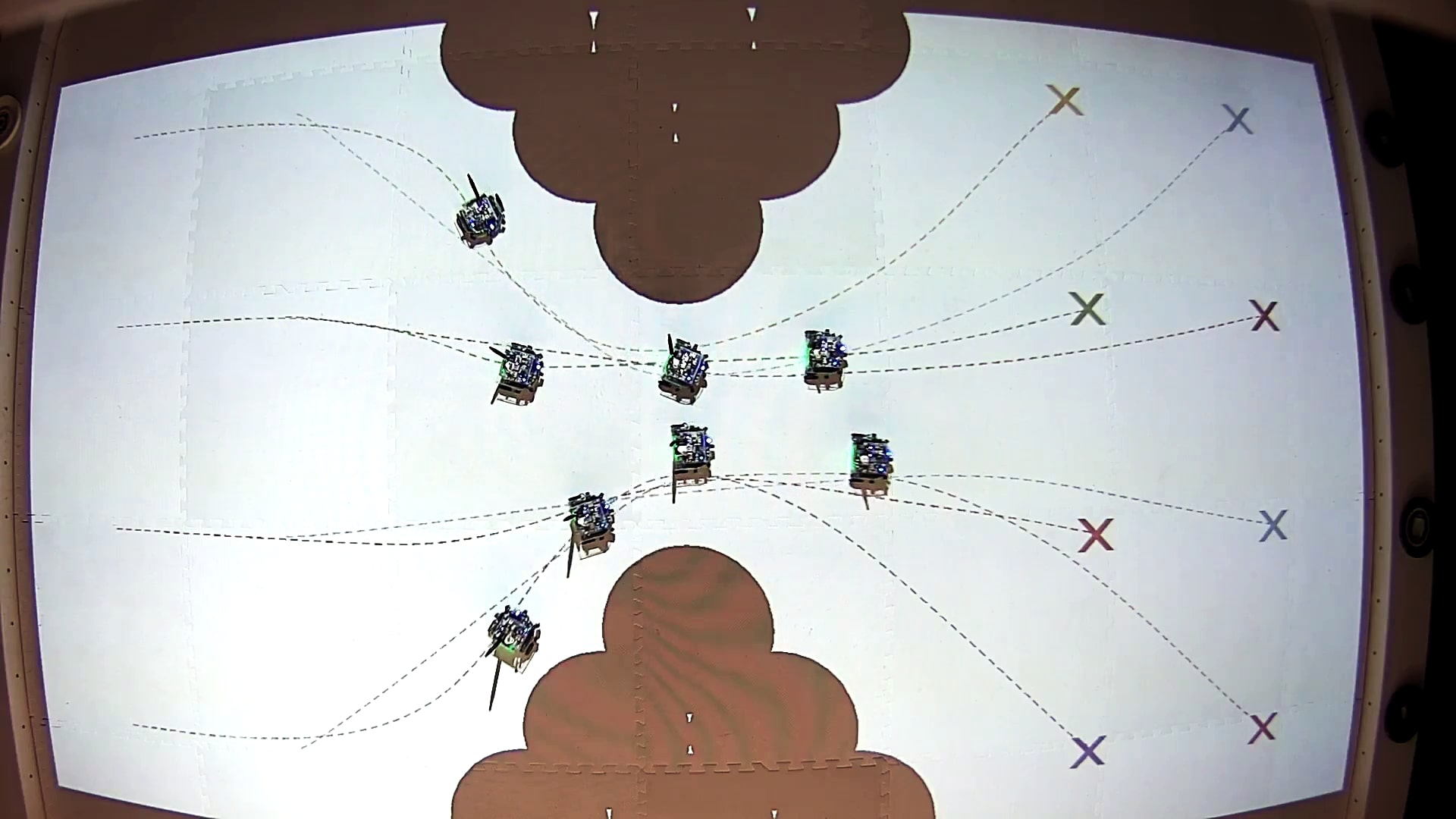}};
    \node[align=center, text=NavyBlue] (c) at (4.8, 0.20) {$k = 240$};
\end{tikzpicture}
\label{fig_rob_bottle_3}}
\hfil
\\
\hfil
\subfloat{
\begin{tikzpicture}
    \node[anchor=south west,inner sep=0] at (0,0){    \includegraphics[width=0.315\textwidth, trim={3cm 0cm 5.3cm 0.7cm},clip]{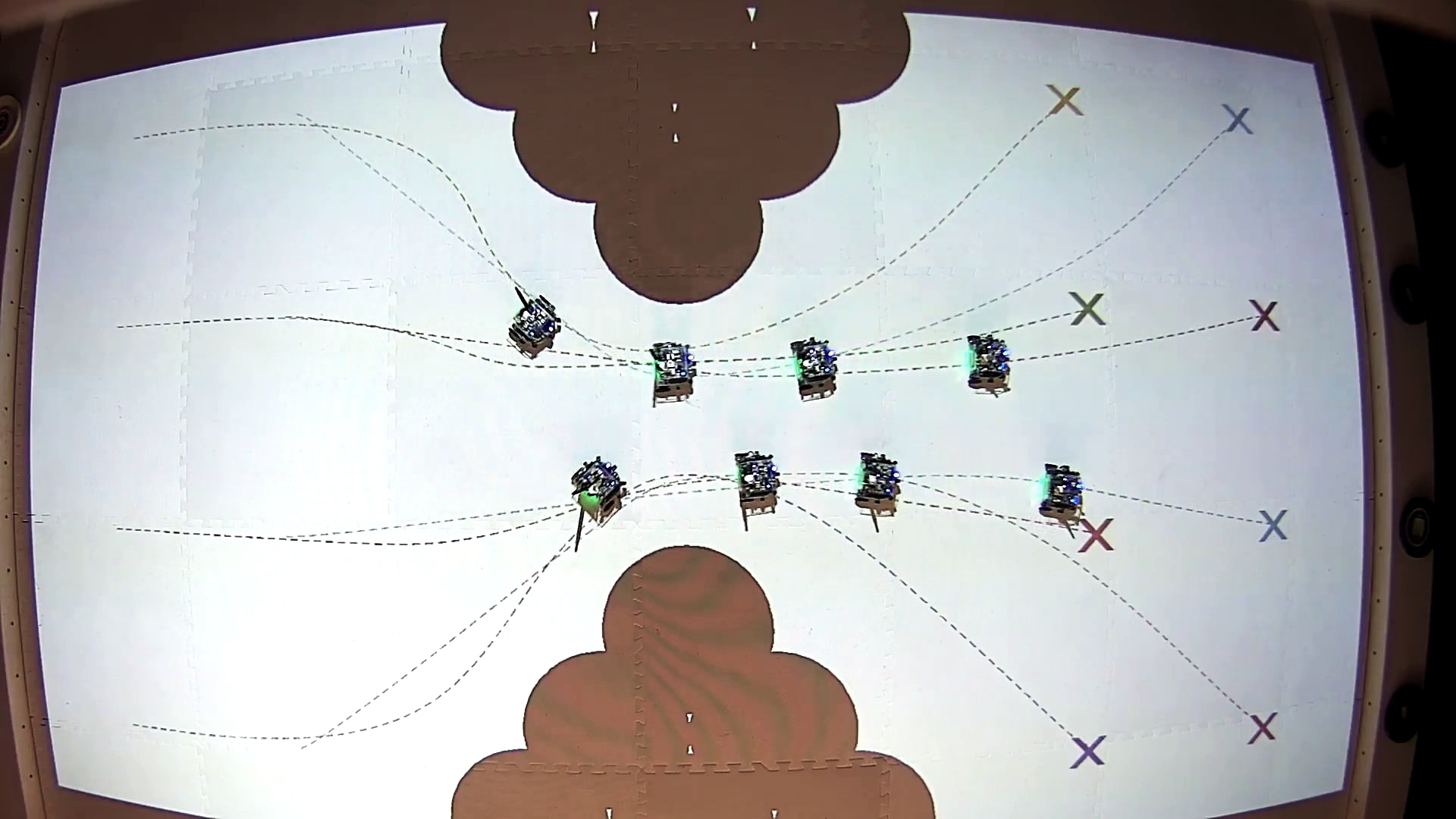}};
    \node[align=center, text=NavyBlue] (c) at (0.8, 0.25) {$k = 330$};
\end{tikzpicture}
\label{fig_rob_bottle_4}}
\hfil
\subfloat{
\begin{tikzpicture}
    \node[anchor=south west,inner sep=0] at (0,0){    \includegraphics[width=0.315\textwidth, trim={3cm 0cm 5.3cm 0.7cm},clip]{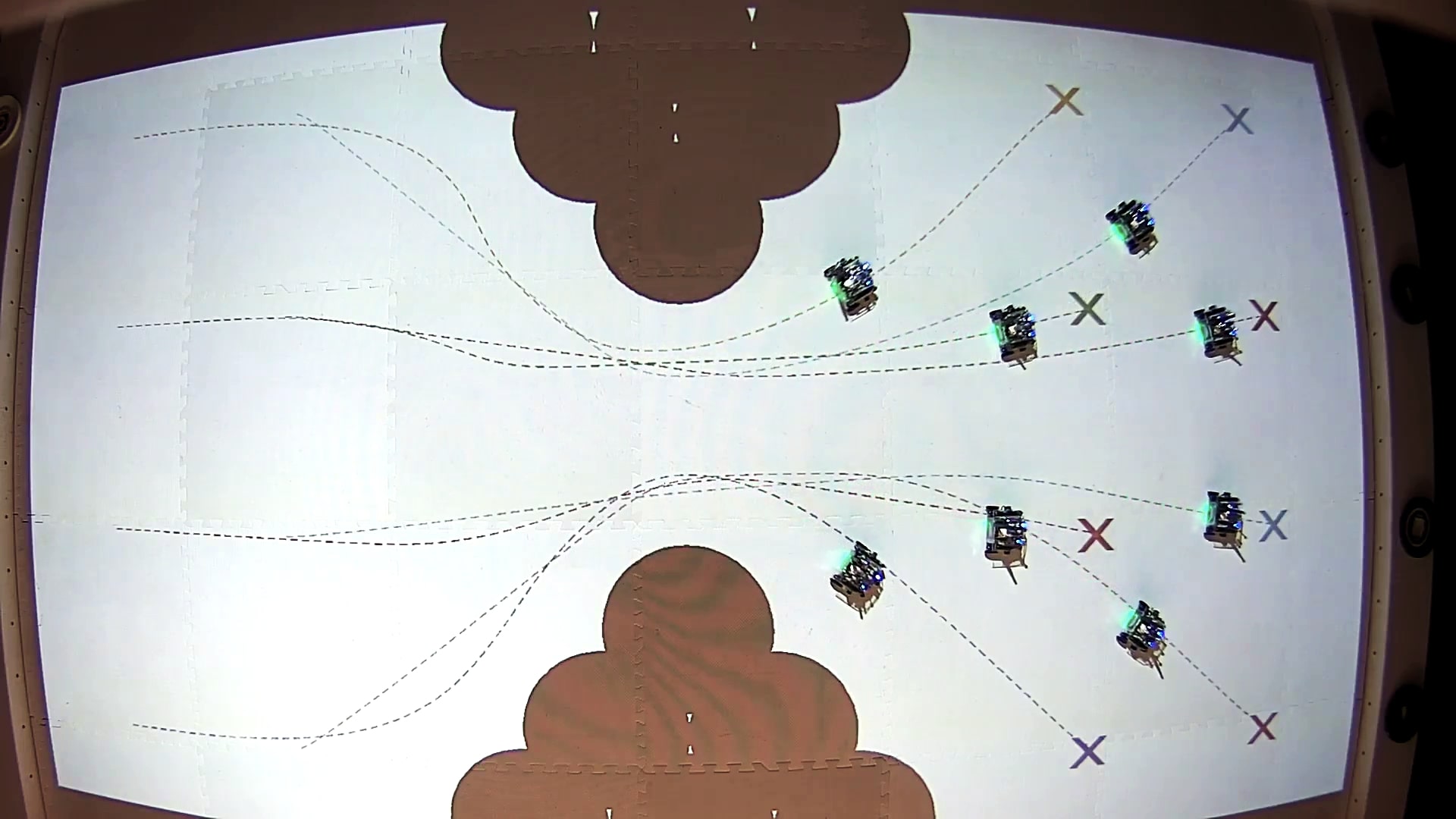}};
    \node[align=center, text=NavyBlue] (c) at (0.8, 0.25) {$k = 510$};
\end{tikzpicture}
\label{fig_rob_bottle_5}}
\hfil
\subfloat{
\begin{tikzpicture}
    \node[anchor=south west,inner sep=0] at (0,0){    \includegraphics[width=0.315\textwidth, trim={3cm 0cm 5.3cm 0.7cm},clip]{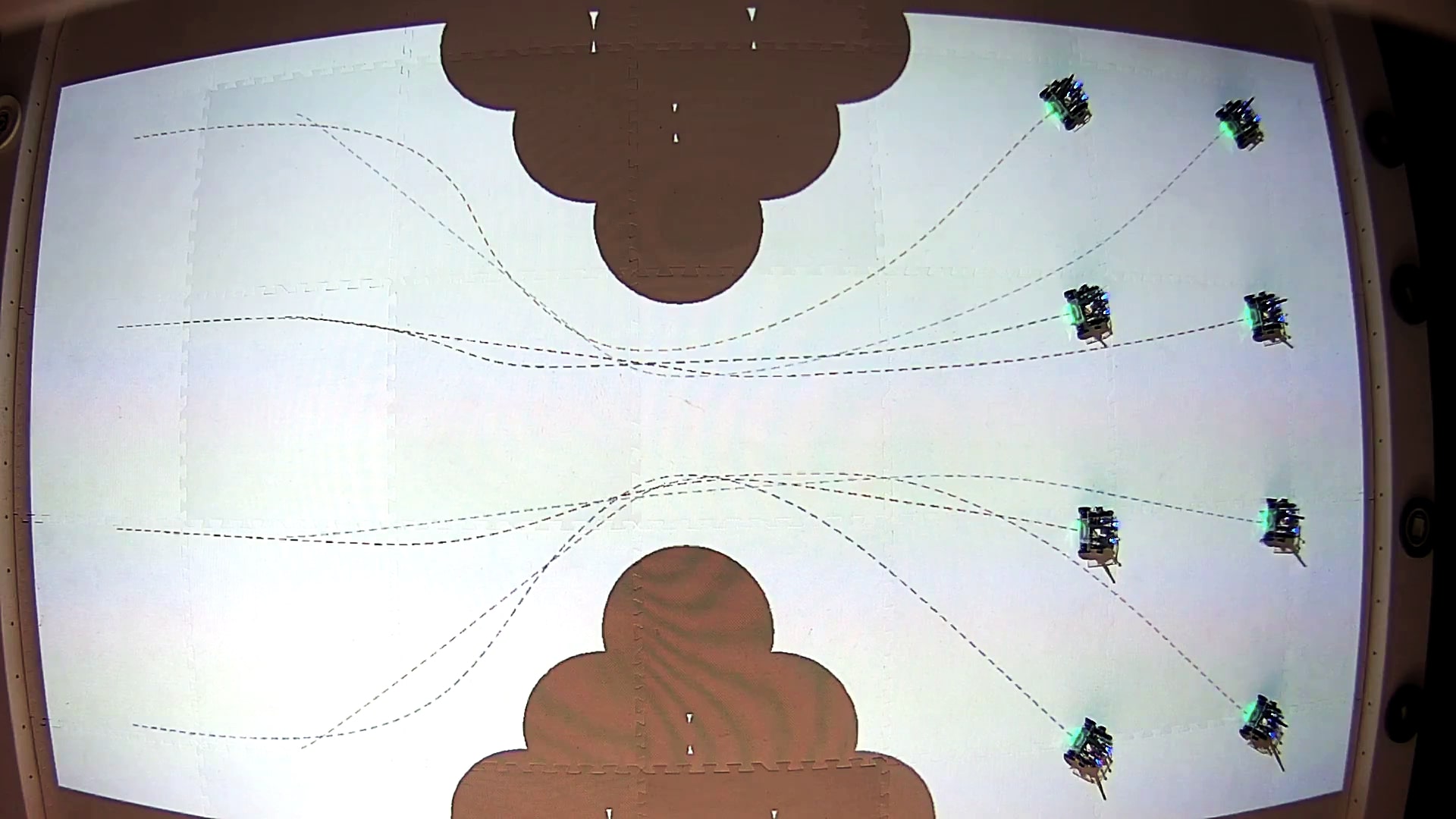}};
    \node[align=center, text=NavyBlue] (c) at (0.8, 0.25) {$k = 900$};
\end{tikzpicture}
\label{fig_rob_bottle_6}}
\hfil
\caption{Hardware experiment: ``Bottleneck'' task with MD-DDP (8 robots). Snapshots at different time instants.}
\label{fig_rob_bottle}
\end{figure*}

Next, we demonstrate 8 robots performing the ``intersection'' task. As shown in Fig. \ref{fig_rob_inter}, the robots successfully cross the intersection while staying safe and within their lanes. In Fig. \ref{fig_rob_bottle}, the ``bottleneck'' task with 8 robots is presented. The robots are able to pass through the gap while avoiding collisions and reach to their goals. Finally, a formation task with 20 robots (maximum available number of robots in the Robotarium) is also shown in Fig. \ref{fig_rob_formation}. All robots are able to reach their desired targets while avoiding the obstacles and each other.

\section{Conclusion}
\label{sec:conclusion}

In this paper, we propose two fully distributed DDP methods based on ADMM. The first one, ND-DDP, is a three-level architecture that utilizes ADMM for consensus, an AL approach for local constraints and DDP as the local trajectory optimizer. The second one, MD-DDP, is a two-level framework that merges the consensus and local constraints levels by treating them both with ADMM, to further improve computational efficiency. Both methods are successfully tested in simulation on a wide variety of multi-car and multi-UAV control problems. To our best knowledge, ND-DDP and MD-DDP are the first fully decentralized DDP methods that exhibit a scalability to multi-robot systems of such a large scale. We also showcase the applicability of the algorithms on hardware systems by successfully employing them on a multi-robot platform.

Future directions include further theoretical and algorithmic investigations on the two methods. From a theoretical perspective, we wish to establish conditions for the convergence of both schemes.
On the algorithmic side, both architectures can be extended to the cases where uncertainty is represented using Gaussian processes \cite{pan2018efficient} or generalized polynomial chaos theory \cite{boutselis2019numerical, aoyama2021receding}. Such extensions will result into a new set of distributed optimization algorithms that are capable of handling epistemic and aleatoric uncertainty. 
Another interesting direction could incorporate a min-max differential game theoretic formulation of DDP \cite{sun2018min}, to address disturbances with an adversarial nature. Finally, generalizations of DDP such as Maximum Entropy DDP \cite{so2021maximum} and Parameterized DDP \cite{oshin2022parameterized} can further expand the capabilities of the ND-DDP and MD-DDP architectures, in terms of avoiding local minima and handling parametric uncertainty, respectively.



{\appendices


\section{MD-DDP Updates with Decentralized Penalty Parameter Adaptation} \label{sec: PPA-MD-DDP}

Here, we provide the MD-DDP updates when the penalty parameter adaptation proposed in Section \ref{sec: arch improvements ppa} is used. 
The Step 1 DDP updates will be given by \eqref{MD-DDP primal update}, but with costs
\begin{align*}
\hat{\ell}_i (\bx_{i,k}, \bu_{i,k}) & = 
\ell_i (\bx_{i,k}, \bu_{i,k}) 
+ \frac{1}{2} \left\| \bx_{i,k} - \tilde{\bx}_{i,k} \right\|_{\bP_i}^2
\nonumber
\\
& \quad 
+ \blambda_{i,k} \T \bx_{i,k}
+ \frac{1}{2} \left\| \bu_{i,k} - \tilde{\bu}_{i,k} \right\|_{\bT_i}^2 
+ \bxi_{i,k} \T \bu_{i,k} ,
\nonumber
\\[0.1cm]
\hat{\phi}_i (\bx_{i,K}) & = 
\phi_i (\bx_{i,K}) 
+ \frac{1}{2} \left\| \bx_{i,K} - \tilde{\bx}_{i,K} \right\|_{\bP_i}^2
+ \blambda_{i,K} \T \bx_{i,K}.
\nonumber
\end{align*}
In Step 2, the safe state updates will be provided by
\begin{align}
\tilde{\bx}_{i,k}^{\text{a}, n+1} = \argmin & \frac{1}{2} \left\| \bx_{i,k} - \tilde{\bx}_{i,k} \right\|_{\bP_i}^2
- \blambda_{i,k} \T \tilde{\bx}_{i,k}
\nonumber
\\
& + \frac{1}{2} \left\| \tilde{\bx}_{i,k}^{\text{a}} - \bz_{i,k}^{\text{a}} \right\|_{\bM_i}^2
+ \by_i \T \tilde{\bx}_i^{\text{a}}
\nonumber
\\[0.1cm]
\mathrm{s.t.} \quad \bg_{i,k}(\tilde{\bx}_{i,k}) & \leq 0, \ 
\bh_{i,k}^{\text{a}}(\tilde{\bx}_{i,k}^{\text{a}}) \leq 0,
\end{align}
while the safe control updates will take the following form
\begin{align}
& \tilde{\bu}_{i,k}^{n+1} = \argmin \frac{1}{2} \left\| \bu_{i,k} - \tilde{\bu}_{i,k} \right\|_{\bT_i}^2
- \bxi_{i,k} \T \tilde{\bu}_{i,k}
\nonumber
\\[0.1cm]
& ~~~~~~~~~~~~ \mathrm{s.t.} \quad \bb_{i,k}(\tilde{\bu}_{i,k}) \leq 0.
\end{align}
The Step 3 global updates will be given by
\begin{equation}
\bz_{i,k}^{n+1} 
= \Big( \sum_{j \in \calP_i}  \bM_j \Big)^{-1} \sum_{j \in \calP_i} 
\bM_j \tilde{\bx}_{j,k}^{i,n+1}.
\end{equation}
Finally, the dual updates will be
\begin{align}
\bxi_{i,k}^{n+1} & = \bxi_{i,k}^n + \bT_i (\bu_{i,k}^{n+1} - \tilde{\bu}_{i,k}^{n+1}), 
\\
\blambda_{i,k}^{n+1} & = \blambda_{i,k}^n + \bP_i (\bx_{i,k}^{n+1} - \tilde{\bx}_{i,k}^{n+1}), 
\\
\by_{i,k}^{n+1} & = \by_{i,k}^n + \bM_i (\tilde{\bx}_{i,k}^{\text{a},n+1} - \bz_{i,k}^{\text{a},n+1}).
\end{align}
For $\bT_i = \diag(\tau, \dots, \tau)$, $\bP_i = \diag(\rho, \dots, \rho)$ and $\bM_i = \diag(\mu, \dots, \mu)$, the vanilla MD-DDP updates are recovered.

\begin{figure*}[!t]
\centering
\hfil
\subfloat{
\begin{tikzpicture}
    \node[anchor=south west,inner sep=0] at (0,0){    \includegraphics[width=0.315\textwidth, trim={3cm 0cm 5.3cm 0.7cm},clip]{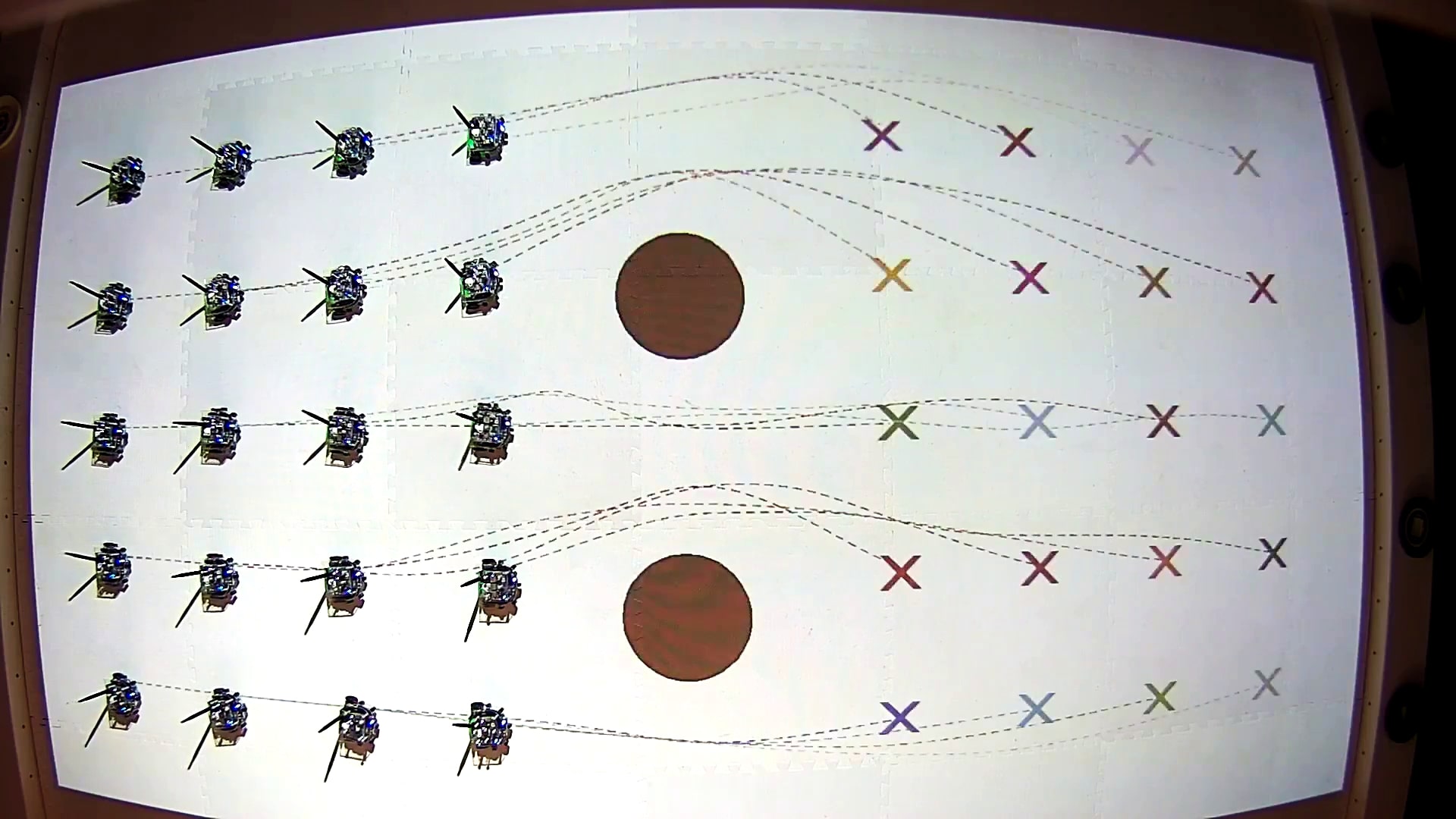}};
    \node[align=center, text=NavyBlue] (c) at (5.1, 0.25) {$k = 0$};
\end{tikzpicture}
\label{fig_rob_formation_1}}
\hfil
\subfloat{
\begin{tikzpicture}
    \node[anchor=south west,inner sep=0] at (0,0){    \includegraphics[width=0.315\textwidth, trim={3cm 0cm 5.3cm 0.7cm},clip]{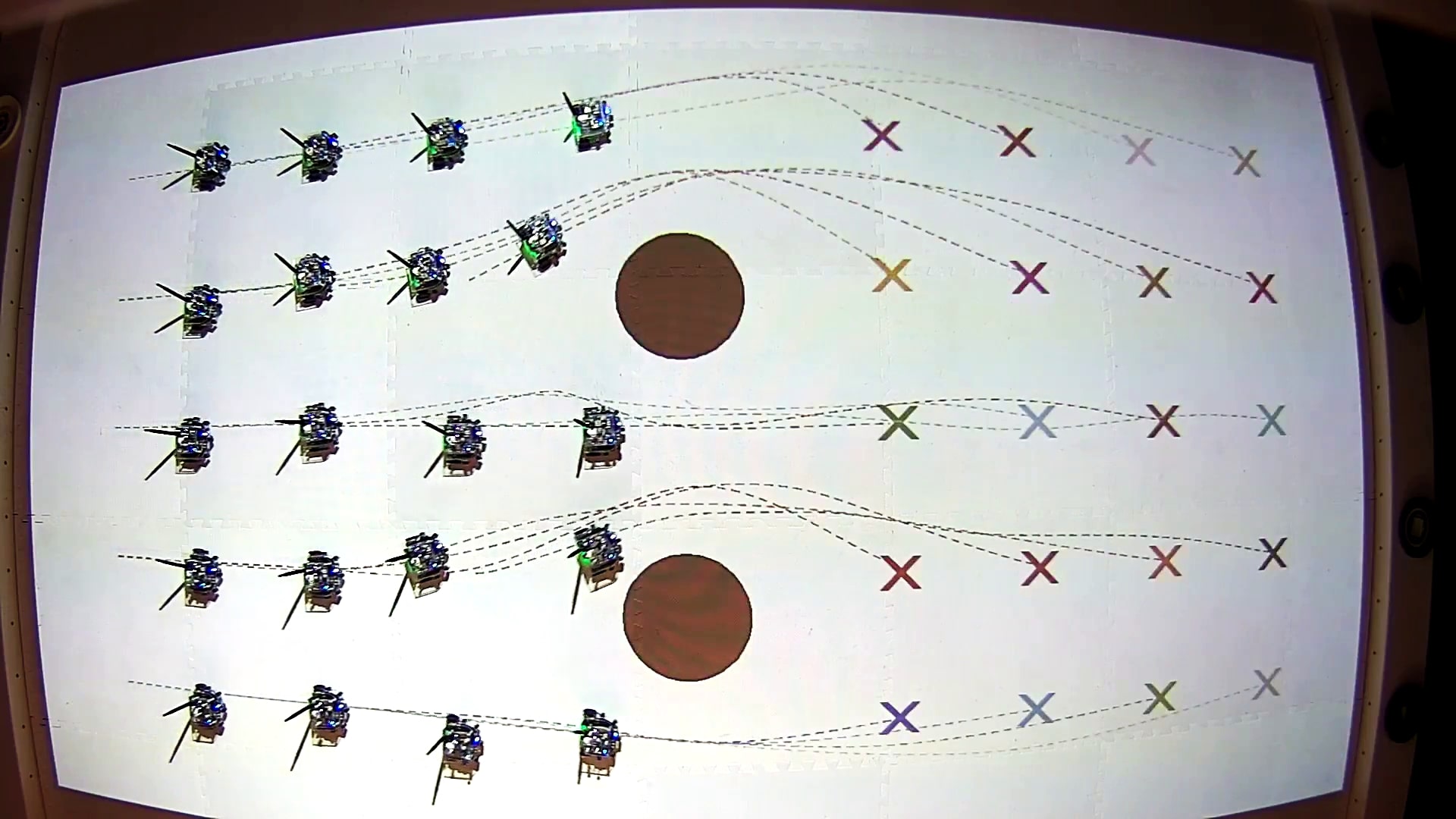}};
    \node[align=center, text=NavyBlue] (c) at (5.0, 0.25) {$k = 60$};
\end{tikzpicture}
\label{fig_rob_formation_2}}
\hfil
\subfloat{
\begin{tikzpicture}
    \node[anchor=south west,inner sep=0] at (0,0){    \includegraphics[width=0.315\textwidth, trim={3cm 0cm 5.3cm 0.7cm},clip]{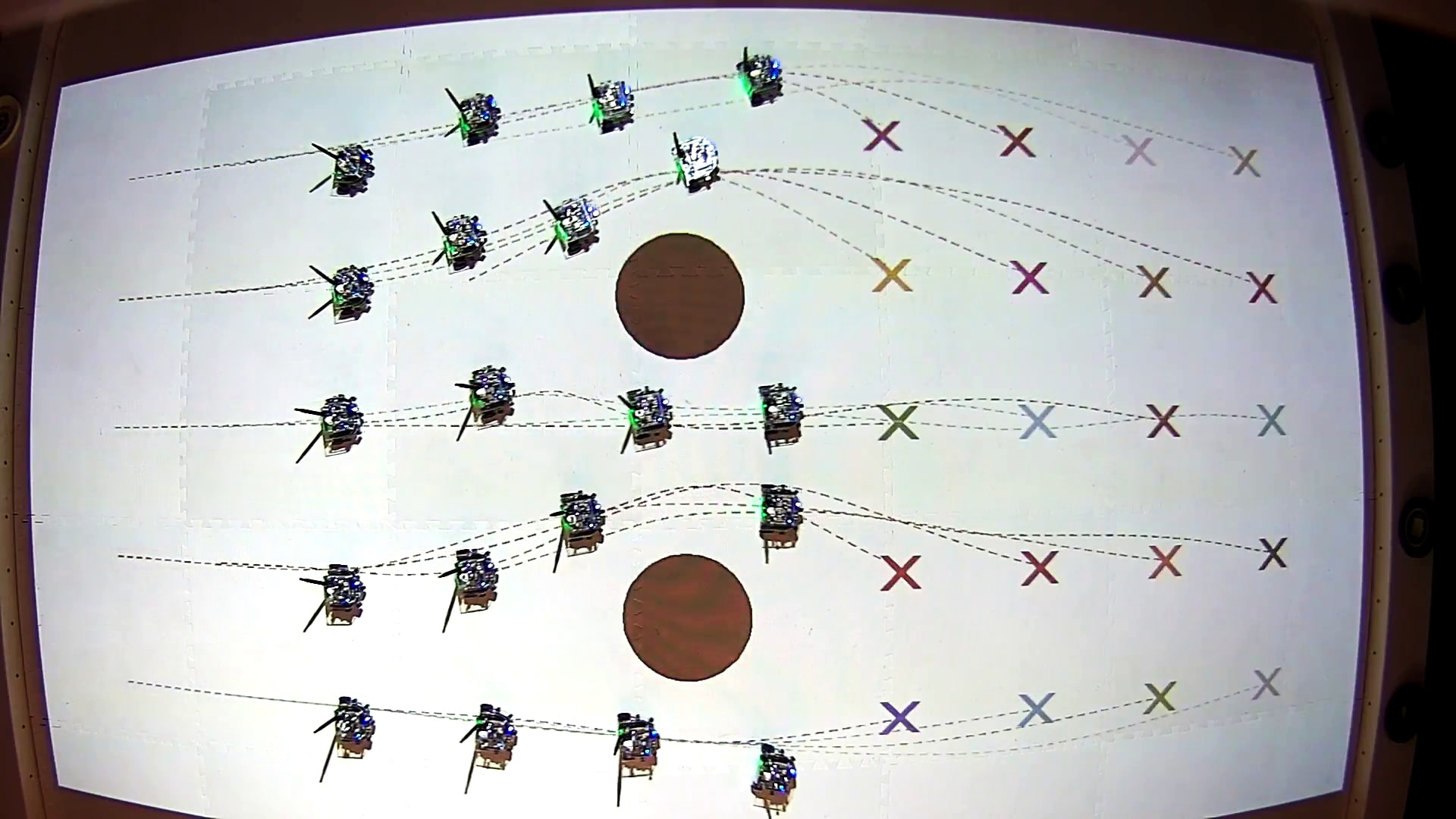}};
    \node[align=center, text=NavyBlue] (c) at (4.9, 0.25) {$k = 120$};
\end{tikzpicture}
\label{fig_rob_formation_3}}
\hfil
\\
\hfil
\subfloat{
\begin{tikzpicture}
    \node[anchor=south west,inner sep=0] at (0,0){    \includegraphics[width=0.315\textwidth, trim={3cm 0cm 5.3cm 0.7cm},clip]{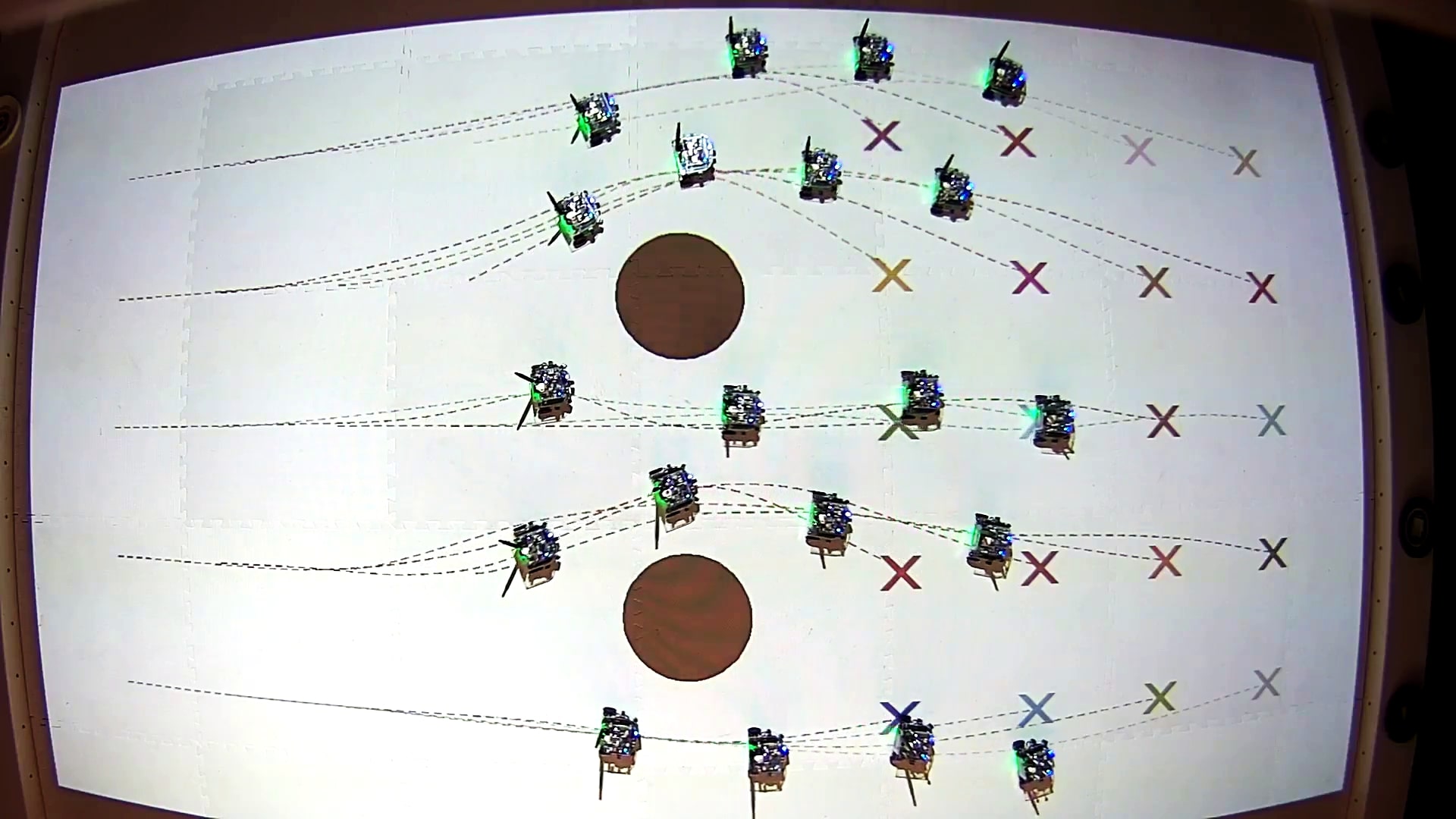}};
    \node[align=center, text=NavyBlue] (c) at (0.8, 0.25) {$k = 240$};
\end{tikzpicture}
\label{fig_rob_formation_4}}
\hfil
\subfloat{
\begin{tikzpicture}
    \node[anchor=south west,inner sep=0] at (0,0){    \includegraphics[width=0.315\textwidth, trim={3cm 0cm 5.3cm 0.7cm},clip]{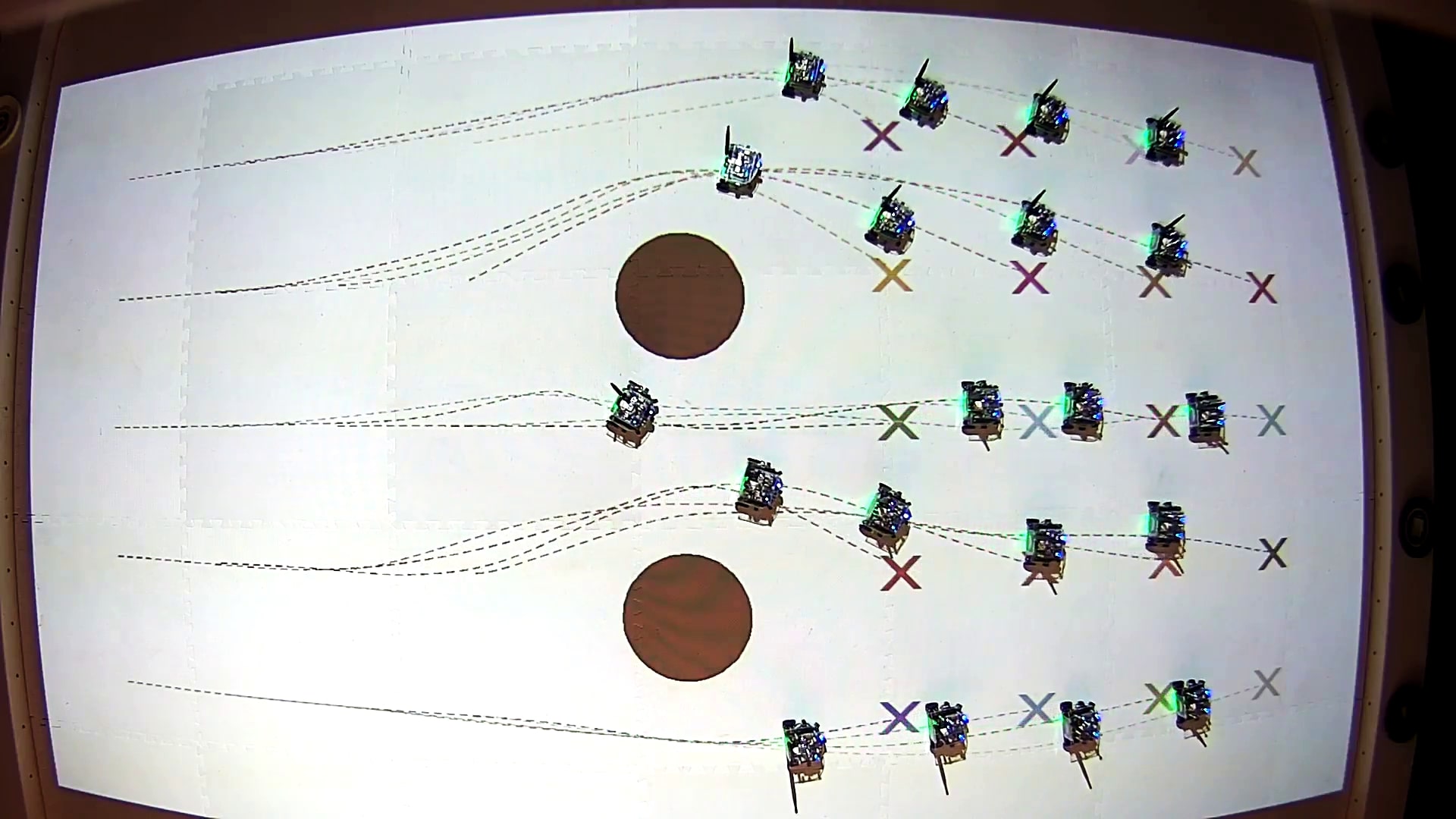}};
    \node[align=center, text=NavyBlue] (c) at (0.8, 0.25) {$k = 360$};
\end{tikzpicture}
\label{fig_rob_formation_5}}
\hfil
\subfloat{
\begin{tikzpicture}
    \node[anchor=south west,inner sep=0] at (0,0){    \includegraphics[width=0.315\textwidth, trim={3cm 0cm 5.3cm 0.7cm},clip]{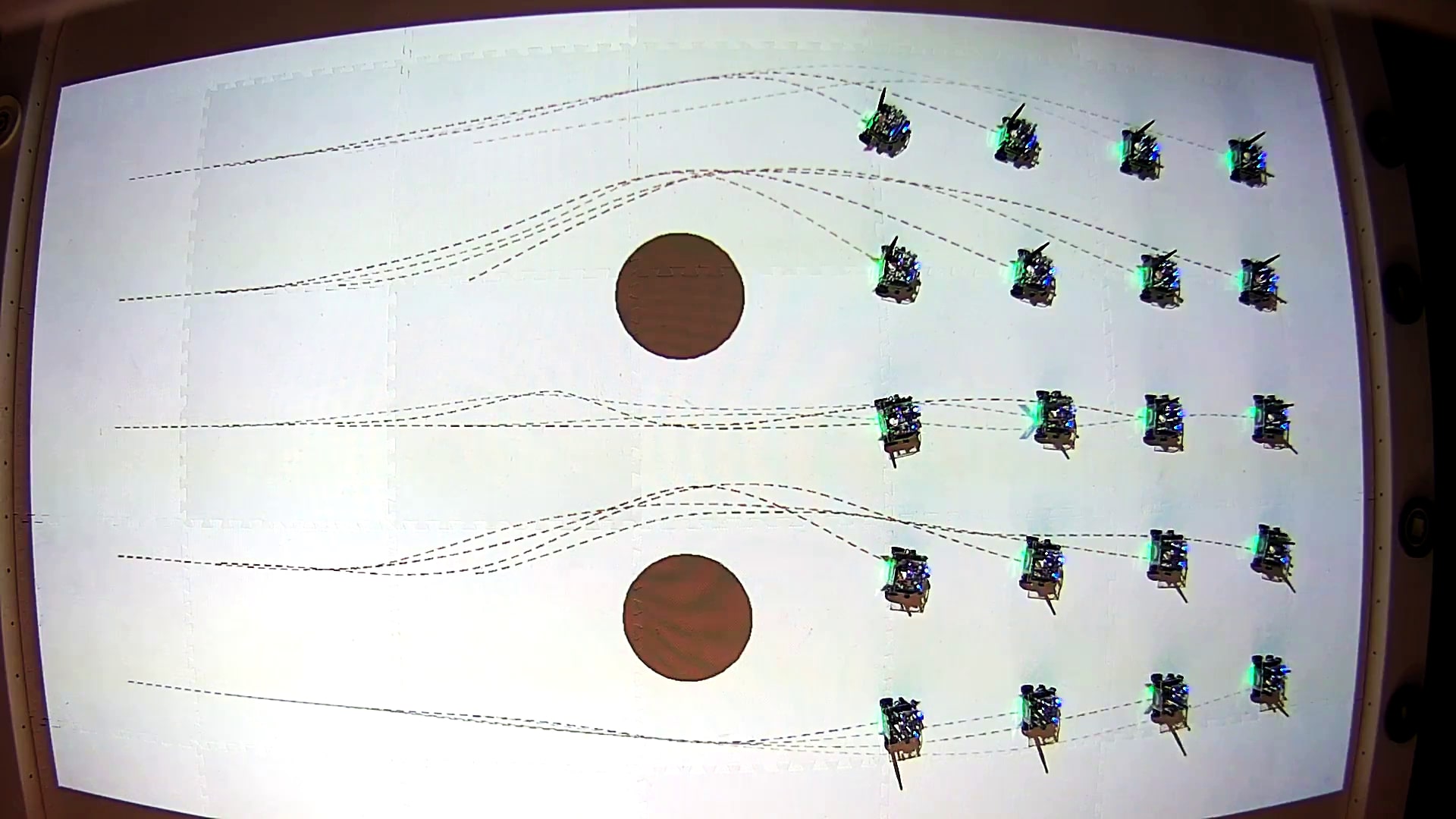}};
    \node[align=center, text=NavyBlue] (c) at (0.8, 0.25) {$k = 900$};
\end{tikzpicture}
\label{fig_rob_formation_6}}
\hfil
\caption{Hardware experiment: Formation task with MD-DDP (20 robots). Snapshots at different time instants.}
\label{fig_rob_formation}
\end{figure*}
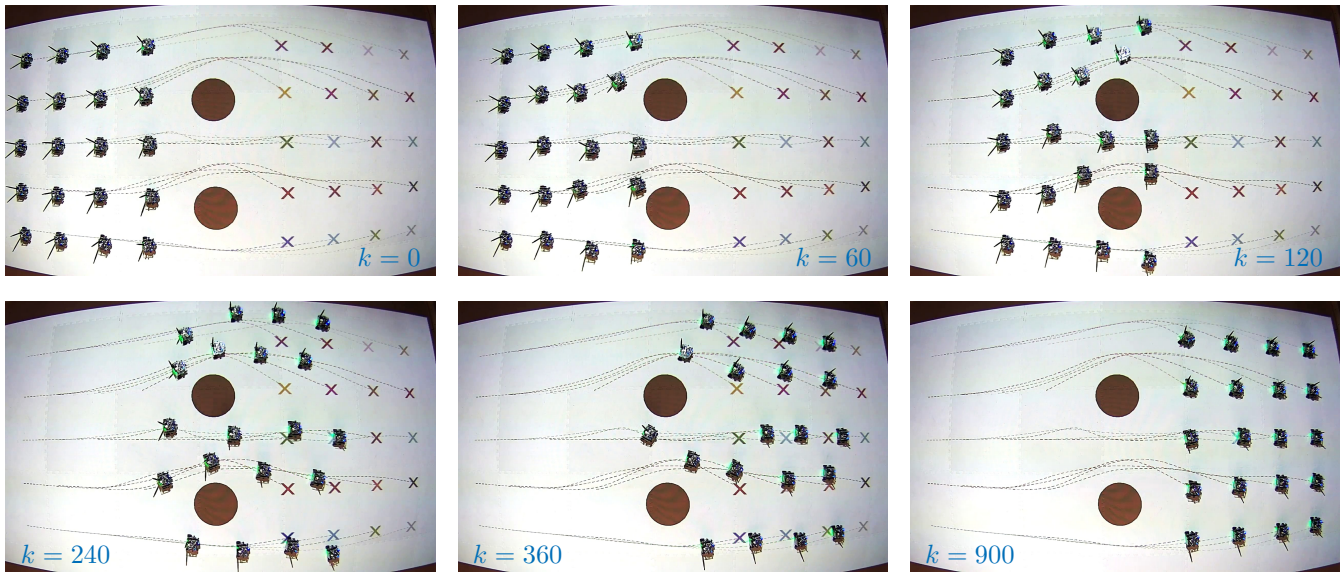

\section{Penalty Parameters Adaptation Details} \label{sec: PPA details}

Detailed expressions for the adaptation rules of the scheme presented in Section \ref{sec: arch improvements ppa} are provided here. First, the ``per-agent'' primal and dual residuals of the MD-DDP algorithm at iteration $n$ are given by
%
%
\begin{alignat*}{2}
& r_{1,i}^{\text{pri},n} = \bu_i^n - \tilde{\bu}_i^n , \quad 
~~~&&r_{1,i}^{\text{dual},n} = \bar{\bT}_i(\tilde{\bu}_i^{n} - \tilde{\bu}_i^{n-1})  , 
\\
& r_{2,i}^{\text{pri},n} = \bx_i^n - \tilde{\bx}_i^n, \quad
\quad 
~~&&r_{2,i}^{\text{dual},n} = \bar{\bP}_i(\tilde{\bx}_i^{n} - \tilde{\bx}_i^{n-1})
\\ 
& r_{3,i}^{\text{pri},n} = \tilde{\bx}_i^{\text{a},n} - \bz_i^{\text{a},n} , \quad  
&&r_{3,i}^{\text{dual},n} = \bar{\bM}_i(\bz_i^{\text{a},n} - \bz_i^{\text{a},n-1}),
\end{alignat*}
where $\bar{\bT}_i = \bdiag(\bT_i, \dots, \bT_i) \in \Rb^{K q_i \times K q_i}$, $\bar{\bP}_i = \bdiag(\bP_i, \dots, \bP_i) \in \Rb^{(K+1) p_i \times (K+1) p_i}$ and $\bar{\bM}_i = \bdiag(\bM_i, \dots, \bM_i) \in \Rb^{(K+1) \tilde{p}_i \times (K+1) \tilde{p}_i}$. For a detailed explanation on how such expressions are derived for ADMM methods, see \cite[Section 3.3]{boyd2011distributed}. Given the residuals, we propose the following update rules 
\begin{equation}
a_{b,i}^{n+1} = 
\begin{cases} 
\chi^{\text{incr}} a_{b,i}^n &\mbox{if} 
\ \| r_{b,i}^{\text{pri},n} \|_2 > \sigma_b^{\text{incr}} \| r_{b,i}^{\text{dual},n} \|_2
\\[0.1cm]
a_{b,i}^n/\chi^{\text{decr}}  & \mbox{if} 
\ \| r_{b,i}^{\text{pri},n} \|_2 < \sigma_b^{\text{decr}} \| r_{b,i}^{\text{dual},n} \|_2 
\\[0.1cm]
a_{b,i}^n & \mbox{otherwise}, 
\end{cases} 
\end{equation}
for $b = 1,2,3$ with $\sigma_b^{\text{incr}} > \sigma_b^{\text{decr}} > 0$, $\chi^{\text{incr}} > 1$ and $\chi^{\text{decr}} > 1$. Lower and upper bounds $a_{i,b,\text{min}}^n$, $a_{i,b,\text{max}}^n \ \forall b \in \{1,2,3\}$ can also be specified. The rules are of similar form with \cite[Section 3.4.1]{boyd2011distributed}, \cite{he2000alternating} but in a ``per-agent'' format.

}

 
%

\bibliographystyle{IEEEtran}

\bibliography{references}

\begin{thebibliography}{10}
\providecommand{\url}[1]{#1}
\csname url@samestyle\endcsname
\providecommand{\newblock}{\relax}
\providecommand{\bibinfo}[2]{#2}
\providecommand{\BIBentrySTDinterwordspacing}{\spaceskip=0pt\relax}
\providecommand{\BIBentryALTinterwordstretchfactor}{4}
\providecommand{\BIBentryALTinterwordspacing}{\spaceskip=\fontdimen2\font plus
\BIBentryALTinterwordstretchfactor\fontdimen3\font minus
  \fontdimen4\font\relax}
\providecommand{\BIBforeignlanguage}[2]{{%
\expandafter\ifx\csname l@#1\endcsname\relax
\typeout{** WARNING: IEEEtran.bst: No hyphenation pattern has been}%
\typeout{** loaded for the language `#1'. Using the pattern for}%
\typeout{** the default language instead.}%
\else
\language=\csname l@#1\endcsname
\fi
#2}}
\providecommand{\BIBdecl}{\relax}
\BIBdecl

\bibitem{chalaki_malikopoulos2021time}
B.~Chalaki and A.~A. Malikopoulos, ``Time-optimal coordination for connected
  and automated vehicles at adjacent intersections,'' \emph{IEEE Transactions
  on Intelligent Transportation Systems}, pp. 1--16, 2021.

\bibitem{xiao_cassandras2021decentralized}
W.~Xiao and C.~G. Cassandras, ``Decentralized optimal merging control for
  connected and automated vehicles with safety constraint guarantees,''
  \emph{Automatica}, vol. 123, p. 109333, 2021.

\bibitem{kabore2021distributed}
K.~M. Kabore and S.~Güler, ``Distributed formation control of drones with
  onboard perception,'' \emph{IEEE/ASME Transactions on Mechatronics}, pp.
  1--11, 2021.

\bibitem{yu_dimarogonas2022distributed}
P.~Yu and D.~V. Dimarogonas, ``Distributed motion coordination for multirobot
  systems under {LTL} specifications,'' \emph{IEEE Transactions on Robotics},
  vol.~38, no.~2, pp. 1047--1062, 2022.

\bibitem{Tomlin_AirTrafficManagement2011}
W.~Zhang, M.~Kamgarpour, D.~Sun, and C.~J. Tomlin, ``Decentralized flight path
  planning for air traffic management,'' in \emph{Proceedings of the 2011
  American Control Conference}, 2011, pp. 2137--2142.

\bibitem{turgut2008self}
A.~E. Turgut, H.~{\c{C}}elikkanat, F.~G{\"o}k{\c{c}}e, and E.~{\c{S}}ahin,
  ``Self-organized flocking in mobile robot swarms,'' \emph{Swarm
  Intelligence}, vol.~2, no.~2, pp. 97--120, 2008.

\bibitem{zhu_ferrari2021adaptive}
P.~Zhu, C.~Liu, and S.~Ferrari, ``Adaptive online distributed optimal control
  of very-large-scale robotic systems,'' \emph{IEEE Transactions on Control of
  Network Systems}, vol.~8, no.~2, pp. 678--689, 2021.

\bibitem{abdulghafoor_bakolas2022multi}
A.~Abdulghafoor and E.~Bakolas, ``Multi-agent distributed optimal control for
  tracking large-scale multi-target systems in dynamic environments,'' 2022.

\bibitem{zhou2022anovel}
Z.~Zhou and H.~Xu, ``A novel mean-field-game-type optimal control for very
  large-scale multiagent systems,'' \emph{IEEE Transactions on Cybernetics},
  vol.~52, no.~6, pp. 5197--5208, 2022.

\bibitem{riegger2016centralized}
L.~Riegger, M.~Carlander, N.~Lidander, N.~Murgovski, and J.~Sj{\"o}berg,
  ``Centralized mpc for autonomous intersection crossing,'' in \emph{2016 IEEE
  19th international conference on intelligent transportation systems
  (ITSC)}.\hskip 1em plus 0.5em minus 0.4em\relax IEEE, 2016, pp. 1372--1377.

\bibitem{diehl2009nonlinearmpc}
M.~Diehl, H.~J. Ferreau, and N.~Haverbeke, \emph{Efficient Numerical Methods
  for Nonlinear MPC and Moving Horizon Estimation}.\hskip 1em plus 0.5em minus
  0.4em\relax Berlin, Heidelberg: Springer Berlin Heidelberg, 2009, pp.
  391--417.

\bibitem{mombaur2009using}
K.~Mombaur, ``Using optimization to create self-stable human-like running,''
  \emph{Robotica}, vol.~27, no.~3, pp. 321--330, 2009.

\bibitem{von1992direct}
O.~Von~Stryk and R.~Bulirsch, ``Direct and indirect methods for trajectory
  optimization,'' \emph{Annals of operations research}, vol.~37, no.~1, pp.
  357--373, 1992.

\bibitem{jacobson1970differential}
D.~H. Jacobson and D.~Q. Mayne, \emph{Differential dynamic programming}.\hskip
  1em plus 0.5em minus 0.4em\relax Elsevier Publishing Company, 1970, no.~24.

\bibitem{morimoto2003minimax}
J.~Morimoto, G.~Zeglin, and C.~Atkeson, ``Minimax differential dynamic
  programming: application to a biped walking robot,'' in \emph{Proceedings
  2003 IEEE/RSJ International Conference on Intelligent Robots and Systems
  (IROS 2003) (Cat. No.03CH37453)}, vol.~2, 2003, pp. 1927--1932 vol.2.

\bibitem{kumar2016optimal}
V.~Kumar, E.~Todorov, and S.~Levine, ``Optimal control with learned local
  models: Application to dexterous manipulation,'' in \emph{2016 IEEE
  International Conference on Robotics and Automation (ICRA)}, 2016, pp.
  378--383.

\bibitem{budhiraja_carpentier2018differential}
R.~Budhiraja, J.~Carpentier, C.~Mastalli, and N.~Mansard, ``Differential
  dynamic programming for multi-phase rigid contact dynamics,'' in \emph{2018
  IEEE-RAS 18th International Conference on Humanoid Robots (Humanoids)}, 2018,
  pp. 1--9.

\bibitem{houghton2022path}
M.~D. Houghton, A.~B. Oshin, M.~J. Acheson, E.~A. Theodorou, and I.~M. Gregory,
  ``Path planning: Differential dynamic programming and model predictive path
  integral control on vtol aircraft,'' in \emph{AIAA SCITECH 2022 Forum}, 2022,
  p. 0624.

\bibitem{todorov2005iLQR}
E.~Todorov and W.~Li, ``A generalized iterative {LQG} method for
  locally-optimal feedback control of constrained nonlinear stochastic
  systems,'' in \emph{Proceedings of the 2005, American Control Conference,
  2005.}, 2005, pp. 300--306 vol. 1.

\bibitem{liao1991convergence}
L.-Z. Liao and C.~Shoemaker, ``Convergence in unconstrained discrete-time
  differential dynamic programming,'' \emph{IEEE Transactions on Automatic
  Control}, vol.~36, no.~6, pp. 692--706, 1991.

\bibitem{yakowitz1984computational}
S.~Yakowitz and B.~Rutherford, ``Computational aspects of discrete-time optimal
  control,'' \emph{Applied Mathematics and Computation}, vol.~15, no.~1, pp.
  29--45, 1984.

\bibitem{liao1992advantages}
L.-z. Liao and C.~A. Shoemaker, ``Advantages of differential dynamic
  programming over newton's method for discrete-time optimal control
  problems,'' Cornell University, Tech. Rep., 1992.

\bibitem{murray1979constrained}
D.~M. Murray and S.~J. Yakowitz, ``Constrained differential dynamic programming
  and its application to multireservoir control,'' \emph{Water Resources
  Research}, vol.~15, no.~5, pp. 1017--1027, 1979.

\bibitem{tassa2014controlddp}
Y.~Tassa, N.~Mansard, and E.~Todorov, ``Control-limited differential dynamic
  programming,'' in \emph{2014 IEEE International Conference on Robotics and
  Automation (ICRA)}, 2014, pp. 1168--1175.

\bibitem{xie2017differential}
Z.~Xie, C.~K. Liu, and K.~Hauser, ``Differential dynamic programming with
  nonlinear constraints,'' in \emph{2017 IEEE International Conference on
  Robotics and Automation (ICRA)}, 2017, pp. 695--702.

\bibitem{pavlov2021interior}
A.~Pavlov, I.~Shames, and C.~Manzie, ``Interior point differential dynamic
  programming,'' \emph{IEEE Transactions on Control Systems Technology},
  vol.~29, no.~6, pp. 2720--2727, 2021.

\bibitem{jallet2022constrained}
W.~Jallet, A.~Bambade, N.~Mansard, and J.~Carpentier, ``Constrained
  differential dynamic programming: A primal-dual augmented lagrangian
  approach,'' 2022.

\bibitem{howell2019altro}
T.~A. Howell, B.~E. Jackson, and Z.~Manchester, ``{ALTRO}: A fast solver for
  constrained trajectory optimization,'' in \emph{2019 IEEE/RSJ International
  Conference on Intelligent Robots and Systems (IROS)}, 2019, pp. 7674--7679.

\bibitem{aoyama2020constrainedDDPicra}
Y.~Aoyama, G.~Boutselis, A.~Patel, and E.~A. Theodorou, ``Constrained
  differential dynamic programming revisited,'' in \emph{2021 IEEE
  International Conference on Robotics and Automation (ICRA)}, 2021, pp.
  9738--9744.

\bibitem{almubarak2022safety}
H.~Almubarak, K.~Stachowicz, N.~Sadegh, and E.~A. Theodorou, ``Safety embedded
  differential dynamic programming using discrete barrier states,'' \emph{IEEE
  Robotics and Automation Letters}, vol.~7, no.~2, pp. 2755--2762, 2022.

\bibitem{kavuncu2021potential}
T.~Kavuncu, A.~Yaraneri, and N.~Mehr, ``{Potential iLQR: A Potential-Minimizing
  Controller for Planning Multi-Agent Interactive Trajectories},'' in
  \emph{Proceedings of Robotics: Science and Systems}, Virtual, July 2021.

\bibitem{so2022multimodal}
O.~So, K.~Stachowicz, and E.~A. Theodorou, ``Multimodal maximum entropy dynamic
  games,'' \emph{arXiv preprint arXiv:2201.12925}, 2022.

\bibitem{wang2020preconditioned}
Y.~Wang and K.~Tsumura, ``Preconditioned distributed trajectory optimization
  algorithm using differential dynamic programming,'' in \emph{2020 59th IEEE
  Conference on Decision and Control (CDC)}, 2020, pp. 2985--2991.

\bibitem{boyd2011distributed}
S.~Boyd, N.~Parikh, E.~Chu, B.~Peleato, and J.~Eckstein, ``Distributed
  optimization and statistical learning via the alternating direction method of
  multipliers,'' \emph{Foundations and Trends® in Machine Learning}, vol.~3,
  no.~1, pp. 1--122, 2011.

\bibitem{saravanos2021distributed}
A.~D. Saravanos, A.~Tsolovikos, E.~Bakolas, and E.~Theodorou, ``{Distributed
  Covariance Steering with Consensus ADMM for Stochastic Multi-Agent
  Systems},'' in \emph{Proceedings of Robotics: Science and Systems}, Virtual,
  July 2021.

\bibitem{pereira2022decentralized}
M.~A. Pereira, A.~D. Saravanos, O.~So, and E.~A. Theodorou, ``{Decentralized
  Safe Multi-agent Stochastic Optimal Control using Deep FBSDEs and ADMM},'' in
  \emph{Proceedings of Robotics: Science and Systems}, New York City, NY, USA,
  June 2022.

\bibitem{rey_lygeros2018fully}
F.~Rey, Z.~Pan, A.~Hauswirth, and J.~Lygeros, ``Fully decentralized admm for
  coordination and collision avoidance,'' in \emph{2018 European Control
  Conference (ECC)}, 2018, pp. 825--830.

\bibitem{shorinwa_macschwager2020scalable}
O.~Shorinwa, T.~Halsted, and M.~Schwager, ``Scalable distributed optimization
  with separable variables in multi-agent networks,'' in \emph{2020 American
  Control Conference (ACC)}, 2020, pp. 3619--3626.

\bibitem{halsted2021survey}
T.~Halsted, O.~Shorinwa, J.~Yu, and M.~Schwager, ``A survey of distributed
  optimization methods for multi-robot systems,'' \emph{arXiv preprint
  arXiv:2103.12840}, 2021.

\bibitem{tang2021fast}
W.~Tang and P.~Daoutidis, ``Fast and stable nonconvex constrained distributed
  optimization: the {ELLADA} algorithm,'' \emph{Optimization and Engineering},
  pp. 1--43, 2021.

\bibitem{zhang2021semi}
X.~Zhang, Z.~Cheng, J.~Ma, S.~Huang, F.~L. Lewis, and T.~H. Lee,
  ``Semi-definite relaxation-based {ADMM} for cooperative planning and control
  of connected autonomous vehicles,'' \emph{IEEE Transactions on Intelligent
  Transportation Systems}, vol.~23, no.~7, pp. 9240--9251, 2022.

\bibitem{zhang2021parallel}
X.~Zhang, Z.~Cheng, J.~Ma, L.~Zhao, C.~Xiang, and T.~H. Lee, ``Parallel
  collaborative motion planning with alternating direction method of
  multipliers,'' in \emph{IECON 2021 – 47th Annual Conference of the IEEE
  Industrial Electronics Society}, 2021, pp. 1--6.

\bibitem{park2019distributed}
S.-S. Park, Y.~Min, J.-S. Ha, D.-H. Cho, and H.-L. Choi, ``A distributed {ADMM}
  approach to non-myopic path planning for multi-target tracking,'' \emph{IEEE
  Access}, vol.~7, pp. 163\,589--163\,603, 2019.

\bibitem{wilson2020robotarium}
S.~Wilson, P.~Glotfelter, L.~Wang, S.~Mayya, G.~Notomista, M.~Mote, and
  M.~Egerstedt, ``The robotarium: Globally impactful opportunities, challenges,
  and lessons learned in remote-access, distributed control of multirobot
  systems,'' \emph{IEEE Control Systems Magazine}, vol.~40, no.~1, pp. 26--44,
  2020.

\bibitem{bellman1966dynamic}
R.~Bellman, ``Dynamic programming,'' \emph{Science}, vol. 153, no. 3731, pp.
  34--37, 1966.

\bibitem{tao2011recovering}
M.~Tao and X.~Yuan, ``Recovering low-rank and sparse components of matrices
  from incomplete and noisy observations,'' \emph{SIAM Journal on
  Optimization}, vol.~21, no.~1, pp. 57--81, 2011.

\bibitem{tang2019distributed}
W.~Tang and P.~Daoutidis, ``Distributed nonlinear model predictive control
  through accelerated parallel {ADMM},'' in \emph{2019 American Control
  Conference (ACC)}, 2019, pp. 1406--1411.

\bibitem{goldstein2014fast}
T.~Goldstein, B.~O'Donoghue, S.~Setzer, and R.~Baraniuk, ``Fast alternating
  direction optimization methods,'' \emph{SIAM Journal on Imaging Sciences},
  vol.~7, no.~3, pp. 1588--1623, 2014.

\bibitem{rockafellar1976monotone}
R.~T. Rockafellar, ``Monotone operators and the proximal point algorithm,''
  \emph{SIAM Journal on Control and Optimization}, vol.~14, no.~5, pp.
  877--898, 1976.

\bibitem{he2000alternating}
B.~He, H.~Yang, and S.~Wang, ``Alternating direction method with self-adaptive
  penalty parameters for monotone variational inequalities,'' \emph{Journal of
  Optimization Theory and applications}, vol. 106, no.~2, pp. 337--356, 2000.

\bibitem{song2016fast}
C.~Song, S.~Yoon, and V.~Pavlovic, ``Fast {ADMM} algorithm for distributed
  optimization with adaptive penalty,'' \emph{Proceedings of the AAAI
  Conference on Artificial Intelligence}, vol.~30, no.~1, Feb. 2016.

\bibitem{xu2017adaptiveB}
Z.~Xu, G.~Taylor, H.~Li, M.~A.~T. Figueiredo, X.~Yuan, and T.~Goldstein,
  ``Adaptive consensus {ADMM} for distributed optimization,'' in
  \emph{Proceedings of the 34th International Conference on Machine Learning},
  ser. Proceedings of Machine Learning Research, D.~Precup and Y.~W. Teh, Eds.,
  vol.~70.\hskip 1em plus 0.5em minus 0.4em\relax PMLR, 06--11 Aug 2017, pp.
  3841--3850.

\bibitem{sun2019two}
K.~Sun and X.~A. Sun, ``A two-level distributed algorithm for nonconvex
  constrained optimization,'' \emph{arXiv preprint arXiv:1902.07654}, 2019.

\bibitem{zhang2020improved}
X.~Zhang, J.~Ma, Z.~Cheng, S.~Huang, C.~W. de~Silva, and T.~H. Lee, ``Improved
  hierarchical {ADMM} for nonconvex cooperative distributed model predictive
  control,'' \emph{arXiv preprint arXiv:2011.00463}, 2020.

\bibitem{fortin2000augmented}
M.~Fortin and R.~Glowinski, \emph{Augmented Lagrangian methods: applications to
  the numerical solution of boundary-value problems}.\hskip 1em plus 0.5em
  minus 0.4em\relax Elsevier, 2000.

\bibitem{glowinski1989augmented}
R.~Glowinski and P.~Le~Tallec, \emph{Augmented Lagrangian and
  operator-splitting methods in nonlinear mechanics}.\hskip 1em plus 0.5em
  minus 0.4em\relax SIAM, 1989.

\bibitem{jiang2014alternating}
B.~Jiang, S.~Ma, and S.~Zhang, ``Alternating direction method of multipliers
  for real and complex polynomial optimization models,'' \emph{Optimization},
  vol.~63, no.~6, pp. 883--898, 2014.

\bibitem{magnusson2015distributed}
S.~Magnússon, P.~C. Weeraddana, and C.~Fischione, ``A distributed approach for
  the optimal power-flow problem based on {ADMM} and sequential convex
  approximations,'' \emph{IEEE Transactions on Control of Network Systems},
  vol.~2, no.~3, pp. 238--253, 2015.

\bibitem{erseghe2014distributed}
T.~Erseghe, ``Distributed optimal power flow using {ADMM},'' \emph{IEEE
  Transactions on Power Systems}, vol.~29, no.~5, pp. 2370--2380, 2014.

\bibitem{shen2014augmented}
Y.~Shen, Z.~Wen, and Y.~Zhang, ``Augmented lagrangian alternating direction
  method for matrix separation based on low-rank factorization,''
  \emph{Optimization Methods and Software}, vol.~29, no.~2, pp. 239--263, 2014.

\bibitem{xu2016empirical}
Z.~Xu, S.~De, M.~Figueiredo, C.~Studer, and T.~Goldstein, ``An empirical study
  of {ADMM} for nonconvex problems,'' \emph{arXiv preprint arXiv:1612.03349},
  2016.

\bibitem{augugliaro2012generation}
F.~Augugliaro, A.~P. Schoellig, and R.~D'Andrea, ``Generation of collision-free
  trajectories for a quadrocopter fleet: {A} sequential convex programming
  approach,'' in \emph{2012 IEEE/RSJ International Conference on Intelligent
  Robots and Systems}, 2012, pp. 1917--1922.

\bibitem{luukkonen2011modelling}
T.~Luukkonen, ``Modelling and control of quadcopter,'' \emph{Independent
  research project in applied mathematics, Espoo}, vol.~22, p.~22, 2011.

\bibitem{pan2018efficient}
Y.~Pan, G.~I. Boutselis, and E.~A. Theodorou, ``Efficient reinforcement
  learning via probabilistic trajectory optimization,'' \emph{IEEE Transactions
  on Neural Networks and Learning Systems}, vol.~29, no.~11, pp. 5459--5474,
  2018.

\bibitem{boutselis2019numerical}
G.~I. Boutselis, Y.~Pan, and E.~A. Theodorou, ``Numerical trajectory
  optimization for stochastic mechanical systems,'' \emph{SIAM Journal on
  Scientific Computing}, vol.~41, no.~4, pp. A2065--A2087, 2019.

\bibitem{aoyama2021receding}
Y.~Aoyama, A.~D. Saravanos, and E.~A. Theodorou, ``Receding horizon
  differential dynamic programming under parametric uncertainty,'' in
  \emph{2021 60th IEEE Conference on Decision and Control (CDC)}, 2021, pp.
  3761--3767.

\bibitem{sun2018min}
W.~Sun, Y.~Pan, J.~Lim, E.~A. Theodorou, and P.~Tsiotras, ``Min-max
  differential dynamic programming: Continuous and discrete time
  formulations,'' \emph{Journal of Guidance, Control, and Dynamics}, vol.~41,
  no.~12, pp. 2568--2580, 2018.

\bibitem{so2021maximum}
O.~So, Z.~Wang, and E.~A. Theodorou, ``Maximum entropy differential dynamic
  programming,'' \emph{arXiv preprint arXiv:2110.06451}, 2021.

\bibitem{oshin2022parameterized}
A.~Oshin, M.~Houghton, M.~Acheson, I.~Gregory, and E.~Theodorou,
  ``{Parameterized Differential Dynamic Programming},'' in \emph{Proceedings of
  Robotics: Science and Systems}, New York City, NY, USA, June 2022.

\end{thebibliography}











\newpage

 





\end{document}